\documentclass[preprints,review,accept,moreauthors,pdftex]{Definitions/mdpi}

\newcommand{\ee}{$\epsilon_{\rm e}$}
\newcommand{\epsp}{$\epsilon_{\rm p}$}
\newcommand{\epsrad}{$\epsilon_{\rm rad}$}
\newcommand{\bw}{blast-wave}
\newcommand{\lf}{Lorentz factor}

\extrafloats{100}
\usepackage{enotez}
\usepackage{adjustbox}
\firstpage{1} 
\pubvolume{1}
\issuenum{1}
\articlenumber{0}
\pubyear{2022}
\copyrightyear{2022}
\datereceived{} 
\dateaccepted{} 
\datepublished{} 
\hreflink{https://doi.org/} % If needed use \linebreak
\Title{Gamma-ray bursts afterglow physics and the VHE domain}

\TitleCitation{GRB afterglow and the VHE domain}

\Author{Davide Miceli $^{1,2,*}$\orcidA{}, Lara Nava $^{3,4,*}$\orcidB{}}

\AuthorNames{Davide Miceli, Lara Nava}

\AuthorCitation{Miceli, D.; Nava, L.}
\address{%
$^{1}$ \quad Dipartimento di Fisica e Astronomia, Università di Padova, Via Marzolo 8, 35131 Padova, Italy\\
$^{2}$ \quad Istituto Nazionale di Fisica Nucleare (INFN), Sez. Padova, Via Marzolo 8, 35131 Padova, Italy\\
$^{3}$ \quad INAF—Osservatorio Astronomico di Brera, Via Emilio Bianchi 46, 23807 Merate, Italy\\
$^{4}$ \quad Istituto Nazionale di Fisica Nucleare (INFN), Sez. Trieste, Via A. Valerio 2, 34100 Trieste, Italy
}

\corres{Correspondence: davide.miceli@unipd.it (D.M.); lara.nava@inaf.it (L.N.)}

\abstract{Afterglow radiation in gamma-ray bursts (GRB), extending from the radio band to GeV energies, is produced as a result of the interaction between the relativistic jet and the ambient medium. Although in general the origin of the emission is robustly identified as synchrotron radiation from the shock-accelerated electrons, many aspects remain poorly constrained, such as the role of inverse Compton emission, the particle acceleration mechanism, the properties of the environment and of the GRB jet itself.
The extension of the afterglow emission into the TeV band has been discussed and theorized for years, but has eluded for a long time the observations. Recently the Cherenkov telescopes MAGIC and H.E.S.S. have unequivocally proven that afterglow radiation is produced also above $100$\,GeV, up to at least a few TeV. The accessibility of the TeV spectral window will largely improve with the upcoming facility CTA ({the} Cherenkov Telescope Array).
In this review article, we first revise the current model for afterglow emission in GRBs, its limitations and open issues. Then we describe the recent detections of very high energy emission from GRBs and the origin of this radiation. Implications on the understanding of afterglow radiation and constraints on the physics of the involved processes will be deeply investigated, showing how future observations, especially {by} the CTA Observatory, are expected to give a key contribution in improving our comprehension of such elusive sources.
}


\usepackage{comment}
\begin{document}
\section{Introduction}

Gamma-Ray Bursts (GRBs) are observed as transient sources of radiation displaying a distinctive pattern consisting of two different phases.
The first phase is dominated by emission in the {keV-MeV} energy range, lasting from fractions of a second to several minutes, and reaching isotropic equivalent peak luminosities in the range $L\sim10^{49}-10^{53}$\,erg\,s$^{-1}$.
The bimodal distribution of the prompt emission duration reveals that there are two classes of GRBs, called short and long depending on whether the prompt emission lasts shorter or longer than 2 seconds \cite{kouveliotou93,mazets81}.
The second emission phase, called afterglow, follows the prompt with a delay of tens of seconds, and is detected on a very wide range of frequencies, from $\gamma$-rays to the radio band. The afterglow flux decays smoothly as a power-law in time for weeks or months, and the typical frequency of the radiation moves in time from the X-ray to the radio band. 
Since 2019, the detection of a few long GRBs between 0.3 and 3\,TeV on time-scales from tens of seconds to a few days proved for the first time that GRBs can also be sources of radiation in the TeV band, where they can convey a sizable fraction (20-50\%) of the total energy emitted during the afterglow phase \cite{190114C_discovery_paper,hess_180720B,190829A_hess}.

All this prompt/afterglow emission is identified with radiation produced as a result of the launch of an ultra-relativistic ($\Gamma\sim100-1000$) jet from a newly born compact object. The ejecta undergoes first internal dissipation (through mechanisms such as shocks between different parts of the outflow \cite{rees94} or magnetic reconnection episodes \cite{thompson94,spruit01}). In a second moment, the ejecta undergoes external dissipation \cite{rees92}, triggered by interactions with the ambient medium (e.g., the interstellar medium or the wind of the progenitor's star \cite{chevalier99}). The two different dissipation processes occur at different typical distances from the central engine ($R\sim10^{13-14}$\,cm and $R\sim10^{15-20}$\,cm) and generate two well distinguished emission phases, identified as the prompt and afterglow emission, respectively.

For long GRBs, it is widely believed that the involved energetics and time-scales and the successful launch of a relativistic jet can find justification in the collapsar model \cite{woosley93,paczynski98}. In this model, the core of a massive star collapses into a black hole and the accretion from the surrounding disk powers the launch of two opposite, collimated ($\theta_{jet}\sim5-10^\circ$) outflows. 
A similar scenario applies also to short GRBs, where the black hole originates from the merger of two neutron stars (as recently proven by the association of a short GRB with a gravitational wave signal \cite{abbott}) or a neutron star and a black-hole. 
An alternative model {\cite{usov92,thompson10,metzger_progenitors,dallosso11} }considers a millisecond magnetar (i.e. a rapidly rotating neutron star) as the progenitor of long GRBs (or at least a fraction of them). This model has the advantage of explaining more naturally the detections of late time activity ($10^2-10^3$\,s after the prompt onset) in the form of X-ray flares and plateaus, observed in about one third of the population.

Beside the nature of the progenitor's star, another quite pressing open issue in GRB physics concerns the composition of the jet itself, i.e. the nature of the dominant energy stored in the outflow, which can be either magnetic (in the form of Poynting flux \cite{usov92,thompson94}) or kinetic (i.e. bulk motion of the matter). 
This uncertainty reflects into an uncertainty on the mechanism extracting energy from the jet (i.e. the process converting part of the jet energy into random energy of the particles), which is identified with internal shocks in the latter case, and magnetic reconnection events in the case of a Poynting flux dominated outflows {\cite{usov92}}. While internal shocks in a matter-dominated jet have been considered the mainstream model for a long time, tensions between some model predictions and observations have moved the attention in the last decade on a family of models based on magnetic jets \cite{giannios05,zhang11,lazarian19}. In particular, internal shocks are not an efficient mechanism \cite{lazzati99,sironi15b}, and this is in {contrast} with the evidence that only a relatively small fraction ($10-50\%$) of energy is still in the blast-wave during the afterglow phase, meaning that most of it must have been dissipated and radiated away during the prompt. It must be noted, however, that the estimate of the energy content of the blast during the afterglow is indirect, and {contingent upon} a proper modeling of the afterglow emission \cite{fan06}. 
Investigations that took advantage of GeV emission detected by {LAT (the Large Area Telescope onboard the Fermi satellite)}, reached the conclusion that the blast energy is usually underestimated by studies relying on X-ray emission, and inferred a prompt emission efficiency between 1-10\% \cite{beniamini16}, which can be still consistent with internal shocks.
The nature and efficiency of the dissipation mechanism in the prompt phase are still  matter of intense debate. 
In any case, the radiation is expected to be produced by the accelerated electrons, which efficiently loose energy via synchrotron cooling {\cite{rees94,ghisellini00}}. Inconsistencies between the expected synchrotron spectrum and the observed spectral shape of the prompt emission \cite{preece98,ghisellini00} have called into question also the nature of the radiative process. Recent works have performed major advances towards the comprehension of the radiative mechanism responsible for the prompt emission, supporting the synchrotron interpretation \cite{oganesyan17,oganesyan18,oganesyan19,ravasio18,ravasio19}.

The nature of the afterglow emission is much better understood, at least on its general grounds. The interaction between the jet and the external medium triggers the formation of a forward shock running into the external medium and a reverse shock running into the ejecta {\cite{meszaros93,meszaros97,sari_piran_reverseshock}}. These shocks are responsible for the acceleration of particles and for the deceleration of the outflow, eventually down to non-relativistic velocities{ \cite{BM76,sedov46,taylor50}}.
The observed radiation is the result of synchrotron radiation from electrons accelerated at the forward shock{ \cite{saripiran}}. A contribution from the reverse shock may also be relevant, typically in the radio and optical band {\cite{sari_piran_reverseshock,kobayashi_sari_00}}.
Shock formation and particle acceleration in ultra-relativistic shocks are still not completely understood. Very important progresses have been done in the last decade from the theoretical side (see \cite{vanthieghem20} for a recent review), especially thanks to numerical particle-in-cell (PIC) simulations \cite{sironi_spitkovsky,sironi13}. Ultra-relativistic shocks in a weakly magnetised medium are found to be efficient particle accelerators, with {$\xi_e\sim1\%$} of the electrons being accelerated into a power-law distribution with spectral index {$p\sim2.5$}, carrying about $\epsilon_e\sim10\%$ of the shock-dissipated energy. A strong magnetic turbulence is self-generated by the accelerated particles counter-streaming in the upstream, ahead of the shock, at a level of magnetization $\epsilon_B=0.01-0.1$ {\cite{sironi13}}. PIC simulations however are currently probing time-scales that are orders of magnitude smaller than the dynamical time-scale of the blast-wave. This implies that results from simulations can only be extrapolated to the relevant time-scales, introducing a certain degree of uncertainty and caution in using the results as inputs for the modeling of GRB afterglows. What is still poorly understood, even though dedicated simulations are starting to give important clues \cite{keshet09}, is how the micro-turbulence generated in the shock vicinity evolves (decays) with {time. This} is particularly important for a proper interpretation of the observations, since it is likely that the particles produce synchrotron and {synchrotron-self Compton (SSC)} photons in a region of decayed micro-turbulence, and hence feel a magnetic field with $\epsilon_B<<0.01$.

The afterglow emission, its spectral shape from radio to $\gamma$-rays, and its temporal evolution from seconds to months contain a wealth of (convoluted) information on blast dynamics, particle acceleration, magnetic turbulence generation and decay, and external density in the progenitor's surroundings (up to a parsec scale). Nevertheless, since all these ingredients are poorly constrained from theoretical grounds, they enter the afterglow physics as free, unconstrained model {parameters. The }large degeneracy among different parameters and the small number of observables as compared to model variables are limiting our possibility to extract valuable and robust information from the modeling of the observed afterglow radiation.
To go beyond the state-of-the-art, additional efforts are necessary both on the observational and theoretical sides. 

An interesting opportunity has recently opened on the observational side, thanks to the discovery that GRBs can be sources of TeV radiation associated to the afterglow phase \cite{190114C_discovery_paper,hess_180720B}. The characterization of the TeV {spectra and light curves offers} new observables to further constrain the unknown physics of the afterglow {emission. These observations are expected to impact on our current understanding of the environment where GRBs explode} (and hence on the nature of their progenitors), {of} the physics of ultra-relativistic shocks, and {of} the properties of the jet {(e. g.,} bulk Lorentz factor and energy {content). Constraining the jet properties is mandatory for a} correct estimate of the prompt mechanism efficiency and then {for determining its nature.}
It is then evident how the opening of this completely new energy window in GRBs is expected to boost the studies in a field that has many connections both with the general understanding of the GRB phenomenon and with topics of general interest, such as star formation and evolution, the last stages of massive stars and their environments, and plasma physics under extreme conditions.

{Given the impressive amount of new information that VHE observations are going to bring to the field, it is important to revise what is the state-of-the-art, what are the main issues, and how we can benefit from the few existing and the upcoming observations in the TeV domain.}
This review revisits the present understanding of afterglow radiation, the discovery of very-high energy (VHE, $>100$\,GeV) emission from {GRBs, and} future prospects for the detection of GRBs at VHE with the next generation of Cherenkov telescopes. 

For recent and complete reviews on GRB's phenomenology and theoretical interpretation before the TeV era see \cite{kumar_zhang,zhang_book}. An overview focused on high-energy emission (0.1-100\,GeV) observations and interpretation can be found in \cite{nava18}.

This review is organised as follows. 
Section~\ref{chap:model} presents an overview of the afterglow external shock model, revisiting our common understanding and phenomenological description of i) the dynamics of the blast-wave, ii) shock formation, particle acceleration and self-generation of turbulent magnetic field in the shock proximity, and iii) the main processes shaping the radiative output, on the whole electromagnetic spectrum, from radio to very-high energy $\gamma$-rays. 
In Section~\ref{chap:openquestions} we propose a discussion of the main open issues of the afterglow model, outlining which observations are at odds with model predictions, which observed features are missing in the basic scenario and what are the present limitations that prevent us from extracting valuable information from the modeling of multi-wavelength afterglow radiation. In Section~\ref{Chap:TeVdiscovery} we describe the recent discovery that GRBs can be bright TeV emitters. Each GRB with a firm (or with a hint of) detection by MAGIC or H.E.S.S. is discussed in detail. We present multi-wavelength {observations and review} the proposed interpretations of the detected emission. 
In Section~\ref{chap:discussion}, we compare the general properties of the detected GRBs both among each other and with the general population. We discuss how the TeV emission can help to solve some of the most important issues of the afterglow model.
Finally, in Section~\ref{chap:conclusions}, we discuss the prospects for future studies of TeV emission from GRBs with the next generation of Cherenkov telescopes and their expected impact on GRB physics.

\section{The afterglow model}\label{chap:model}

Afterglow emission refers to all the broad-band radiation observed from a GRB on longer timescales (minutes to months) as compared to the initial prompt radiation detected in hard X-rays {\cite{PK00,saripiran,sariesin}}. 
Its temporal evolution is usually well described with simple decaying power-laws, in contrast with the short-time ($<$\,seconds) variability that characterises the prompt emission {\cite{nousek06,nardini06,zaninoni13,bernardini12}}. 
These major differences place the emission region of afterglow radiation at larger radii ($>10^{15}$\,cm), pinpointing its origin in the processes triggered by the interaction between the jet and the circumburst medium. 

The expansion of the relativistic jet into the external medium is expected to drive two different shocks: the forward shock, running into the external medium, and the reverse shock, running into the jet. The shocked ejecta and the shocked external medium, separated by the contact discontinuity, are both sources of synchrotron radiation from the accelerated electrons{ \cite{sari_piran_95}}. Most of the detected radiation is interpreted as emission from ambient particles energized by the forward shock.
Spectra and lightcurves  are then shaped by the environment where the GRB explodes, which in turn is strictly connected to the nature of the progenitor.
The other player that shapes the properties of afterglow radiation is particle acceleration at relativistic shocks, which is though to proceed via diffusive shock acceleration, but for which the details of the underlying physics remain still poorly constrained.
Moreover, the overall luminosity of the afterglow radiation depends on the energy content of the {blast-wave. Such amount} is determined by how efficiently the prompt mechanism has dissipated and released part of the initial explosion energy. 
{Following these considerations, it} is evident how the study of afterglow radiation impacts on the general understanding of the GRB phenomenon: the progenitor and its environment, the nature and efficiency of the mechanisms responsible for prompt emission, {the properties of the jet,} and the micro-physics of relativistic shocks.

In this section, the physics involved in the afterglow scenario is presented, with a particular focus on the forward shock emission and on the radiative output expected at VHE. This section is organized as follows: we revisit the physics of the jet dynamical evolution in its interaction with the ambient medium ({Section~}\ref{subsec:jet_dynamics}), the particle acceleration mechanism ({Section~}\ref{subsec:fermi_mechanism}), and the resulting radiative output and its spectral shape ({Section~}\ref{subsec:radiation_processes}). 

%-------------------------------------
\subsection{Jet dynamics}
\label{subsec:jet_dynamics}

After the reverse shock has crossed the ejecta, the dynamics of the \bw\ enters a self-similar regime (\cite{BM76}, BM76 hereafter).
In a thin shell approximation, the reverse shock crossing time corresponds to the time when the \bw\ starts decelerating. 
The deceleration of the jet, caused by the collision with the external medium, becomes significant at the radius $R_{dec}$ where the energy transferred to the mass $m$ collected from the external medium ($\sim m(R_{dec})c^2\Gamma_0^2$) is comparable to the initial energy ($E_0=M_0\Gamma_0\,c^2$) carried by the jet.
This deceleration radius is typically of the order of $R_{dec}\sim10^{15}-10^{16}$\,cm, depending on the density of the external medium and on the ejecta {mass $M_0$ and initial }bulk Lorentz factor {$\Gamma_0$}. 
Before reaching this radius, the ejecta {expands with constant velocity (coasting phase).}

Most analytic estimates of the afterglow evolution with the purpose of modeling data are developed for the deceleration phase, where the self-similar BM76 solution for adiabatic \bw s is adopted {\cite{saripiran,PK00,granot_breaks}}. 
Since VHE emission can be detected at quite early times (a few tens of seconds), we are interested also in the description of the coasting phase and in a proper treatment of the transition between coasting and deceleration. 

In the following, to derive the evolution of the bulk Lorentz factor we adopt the approach proposed by \cite{nava13}.
This method allows to describe the hydrodynamics of a relativistic \bw\ expanding into a medium with an arbitrary density profile $\rho(R)$ and composition (i.e. enriched by pairs), and the transition from the free expansion of the ejecta to the deceleration phase, taking into account the role of radiative and adiabatic losses.
The internal structure is neglected (homogeneous shell approximation), and the \lf\ $\Gamma$ considered is the one of the fluid just behind the shock front. 
In the deceleration phase, the self-similar solutions derived in BM76 are recovered by this method, both for the adiabatic and the fully radiative cases, and for constant and wind-like density profiles of the external medium. 
The presented approach also allows to introduce time-varying radiative efficiency, either resulting from a change with time of \ee\ or a change in the radiative efficiency of the electrons.
Equations reported here are valid after the reverse shock has crossed the {ejecta. Corrections} to the hydrodynamics before the reverse-shock crossing time can be found in \cite{nava13}. 

\subsubsection{Equation describing the evolution of the bulk Lorentz factor}
The aim is to derive an equation describing the change $d\Gamma$ of the bulk Lorentz factor of the fluid just behind the shock in response to the collision with a mass $dm(R)=4\pi R^2\rho(R) dR$ encountered when the shock front moves from a distance $R$ to $R+dR${ and with $\rho$ being the mass density.}
The change in $\Gamma$ is determined by dissipation of the bulk kinetic energy,
conversion of internal energy back into bulk motion, and injection of energy into the \bw.
The latter is sometimes invoked to explain plateau phases in the X-ray early afterglow or to explain flux rebrightenings {\cite{zhang_meszaros_2001,flare_centralengine_zhang,laskar15}}.
The following treatment neglects energy injection, which however can be easily incorporated in this kind of approach.

To write the equation for energy conservation, from which $d\Gamma/dR$ can be derived, we first need to recall how the energy density transforms under Lorentz transformations.
In the following, {we denote quantities} measured in the frame comoving with the shocked fluid (comoving frame), {with a prime, }to distinguish them from quantities measured in the frame of the progenitor star (rest {frame, without a prime)}.

The energy density in the comoving frame is $u'=u'_{\rm int}+\rho'c^2$, where $u'_{\rm int}$ is the comoving internal energy and $\rho'$ is the comoving mass density. Applying Lorentz transformations, $u=(u'+p')\Gamma^2-p'$, where $p'$ is the pressure and is related to the internal energy density by the equation of state $p'=(\hat{\gamma}-1)\,u'_{\rm int}=(\hat{\gamma}-1)\,(u'-\rho'c^2)$, where $\hat{\gamma}$ is the adiabatic index of the shocked plasma.
The energy density is then given by: $u=u'_{\rm int}(\hat{\gamma}\Gamma^2-\hat{\gamma}+1) + \rho'c^2\Gamma^2$, which shows how the internal energy and rest mass density transform.
The total energy in the progenitor frame will be $E=u V=uV'/\Gamma$, where $V$ is the shell volume in the progenitor frame, and can be expressed as:
\begin{equation}
\label{eq:Etot}
E=\Gamma M c^2+\Gamma_{\rm eff} E'_{\rm int}~,
\end{equation}

where: 

\begin{equation}
\Gamma_{\it \it eff}\equiv\frac{\hat{\gamma}\Gamma^2-\hat{\gamma}+1}{\Gamma}~,
\end{equation}

which properly describes the Lorentz transformation of the internal energy. 
Here, $M=M_0+m=\rho' V'$ is the sum of the ejecta mass $M_0=E_0/\Gamma_0c^2$ and of the swept-up 
mass $m(R)$, and $E'_{int}=(u'-\rho'c^2) V'$ is the comoving internal energy. 
The adiabatic index can be parameterized
as $\hat{\gamma}=(4+\Gamma^{-1})/3$ to obtain the
expected limits $\hat{\gamma}\simeq4/3$ for $\Gamma\gg1$ and $\hat{\gamma}\simeq5/3$ for 
$\Gamma\rightarrow1$. 
The majority of analytical treatments use 
$\Gamma$ instead of $\Gamma_{\it eff}$, which implies an error up to a factor of $4/3$ in the ultra-relativistic limit \cite{nava13}.

The blast-wave energy $E$ in Eq.~\ref{eq:Etot} can change due to ({\it i}) the rest 
mass energy $dm\,c^2$ collected from the medium, ({\it ii}) radiative losses $dE_{rad}=\Gamma_{\it eff}\,dE'_{rad}$, and ({\it iii}) injection of energy. 
Ignoring possible episodes of energy injections into the \bw, the equation of energy conservation in the progenitor frame is:
\begin{equation}
d\left[\Gamma(M_0+m)c^2+\Gamma_{\it eff} E'_{int}\right]= dm\,c^2+\Gamma_{\it eff} dE'_{rad}~.
\label{eq:energy conservation}
\end{equation}
The overall change in the comoving internal energy $dE'_{int}$ results from the sum of 
three contributions:
\begin{equation}
dE'_{int} = dE'_{sh}+dE'_{ad}+dE'_{rad}~~.
\label{eq:int_ene_contrib}
\end{equation}
The first contribution, $dE'_{sh}=(\Gamma-1)\,dm\,c^2$, is the random kinetic energy 
produced at the shock as a result of the interaction with an element $dm$ of 
circum-burst material: as pointed out by BM76, in the post-shock frame,
the average kinetic energy per unit mass {$dE'_{sh}/dm$} is constant across the shock, and equal to 
$(\Gamma-1) c^2$. 
The second term in eq.~\ref{eq:int_ene_contrib}, $dE'_{ad}$, is the internal energy lost due to adiabatic expansion, that leads to a conversion of random energy back to bulk kinetic energy.
The third term, $dE'_{rad}$, accounts for radiative losses. 

From Eq.~\ref{eq:energy conservation}, it follows that the variation of the Lorentz factor is:
\begin{equation}
\label{eq:G solve}
\frac{d\Gamma}{dR}=-\frac{(\Gamma_{\it eff}+1)(\Gamma-1)\,c^2\frac{dm}{dR}
+\Gamma_{\it eff}\frac{dE'_{ad}}{dR}}{(M_0+m)\,c^2+E'_{int}\frac{d\Gamma_{\it eff}}{d\Gamma}}~,
\end{equation}
from which the evolution $\Gamma(R)$ of the bulk Lorentz factor of the fluid just behind the shock as a function of the shock front radius can be derived.

The term $\Gamma_{\rm eff}\,dE'_{ad}/dR$, accounting for adiabatic losses, allows to
describe the re-acceleration of the fireball: 
this contribution, usually neglected, becomes important only when the density decreases faster than $\rho\propto R^{-3}$.
To evaluate Eq.~\ref{eq:G solve} it is necessary to first specify $dE'_{ad}$ and $E'_{int}$.

\subsubsection{Internal energy and adiabatic losses}
In specific cases, the adiabatic losses and the internal energy content can be expressed in an analytic form.
The following treatment to estimate adiabatic losses and the internal energy content of the \bw\ assumes that, right behind the shock, the freshly shocked electrons instantaneously radiate a fraction \epsrad\ of their internal energy and then they cool only due to adiabatic losses{ \cite{nava13}}. 
By assuming that the accelerated electrons promptly radiate at the shock, and then they evolve adiabatically, one is implicitly considering either fast cooling regime or quasi-adiabatic regime, in which case the radiative losses do not affect the shell dynamics.

Defining \ee\ as the fraction of energy $dE'_{sh}$ dissipated by the shock and gained by the leptons, the mean random Lorentz factor of 
post--shock leptons becomes (for a more detailed discussion see section ~\ref{subsec:fermi_mechanism}): \begin{equation}
\gamma_{acc,e}-1= (\Gamma-1)\epsilon_{e}/\mu_e\,.
\end{equation}

Here, $\mu_{e}=\rho_e/\rho$ is the ratio between the mass density $\rho_e$ of shocked electrons and positrons 
(simply ``electrons'' from now on) and the total mass density of the shocked matter $\rho$.  
In the absence of electron-positron pairs $\mu_e=m_{e}/(m_e+m_{p})\simeq m_{e}/m_{p}$. 

Leptons then radiate a fraction  \epsrad\ of their internal energy, i.e., the energy lost to radiation is  $dE'_{rad}=-\epsilon_{rad}\epsilon_{e}\, dE'_{sh}=-\epsilon \,dE'_{sh}$, with 
$\epsilon\equiv\,$\epsrad\ee\ being the overall fraction of the shock-dissipated energy that goes into radiation.
After radiating a fraction \epsrad\ of their internal energy, the mean random \lf\ of the freshly shocked electrons decreases down to:
\begin{equation}\label{eq:gerad}
\gamma_{rad,e}-1=(1-\epsilon_{rad})(\gamma_{acc,e}-1)=
(1-\epsilon_{rad})(\Gamma-1)\frac{\epsilon_{e}}{\mu_e}~.
\end{equation}
The assumption of instantaneous radiative losses is verified in the fast cooling regime 
($\epsilon_{rad}\sim1$), which is required (but not sufficient) to have $\epsilon\sim1$ 
(i.e., a fully radiative blast-wave). 
In the opposite case $\epsilon_{rad}\ll1$, the evolution is nearly adiabatic ($\epsilon\ll1$), regardless of the value of \ee, 
and the details of the radiative cooling processes are likely to be unimportant for the shell dynamics. 
The case with intermediate values of \epsrad\ and $\epsilon$ is harder to treat analytically, 
since the electrons shocked at radius $R$ may continue to emit copiously also at larger distances, 
affecting the blast-wave dynamics. 

A similar treatment can be adopted for protons: if protons gain a fraction \epsp\ of the energy dissipated by the shock 
(with $\epsilon_p=1-\epsilon_e-\epsilon_B$), 
their mean post--shock Lorentz factor will be:
\begin{equation}\label{eq:gpacc}
\gamma_{acc,p}-1=(\Gamma-1)\frac{\epsilon_p}{\mu_p}~,
\end{equation}
where $\mu_p=\rho_p/\rho$ is the ratio between the mass density of shocked protons $\rho_p$ and the total shocked mass density $\rho$. 
In the standard case, when pairs are absent, $\mu_p\simeq1$. 
Since the proton radiative losses are negligible, the shocked protons will lose their energy only due to adiabatic cooling.

Adiabatic losses can be computed starting from 
$dE_{\rm int}^\prime=-p^\prime\,dV^\prime$, where $p^\prime$ is the pressure in comoving frame. 
For $N$ particles with Lorentz factor $\gamma$ the internal energy density is:
\begin{equation}
    u^\prime_{\rm int} = \frac{N (\gamma-1)m \,c^2}{V^\prime}~.
\end{equation}
The radial change of the 
Lorentz factor, as a result of expansion losses, is:
\begin{equation}
    \left(\frac{d(\gamma-1)}{dr}\right)_{ad}=-(\hat{\gamma}-1)(\gamma-1)\frac{d\ln V^\prime}{dr}\,.
    \label{eq:p_ad}
\end{equation}
To estimate the adiabatic losses, let us assume that the shell comoving volume scales 
as $V'\propto R^3/\Gamma$, corresponding to a shell thickness in the progenitor frame $\sim R/\Gamma^2$. 
This scaling is correct for both relativistic and non--relativistic shocks, in the decelerating phase (BM76). 
For re--accelerating relativistic shocks, \citet{shapiro80} showed that the thickness of the 
region containing most of the blast-wave energy is still $\sim R/\Gamma^2$. 
For the sake of simplicity, changes in the comoving volume due to a time-varying 
adiabatic index or radiative efficiency are neglected. 
If the scaling $V'\propto R^3/\Gamma$ is assumed, the equation can be further developed analytically, and reads:
\begin{equation}
   \left(\frac{d(\gamma-1)}{dr}\right)_{ad}=-(\hat{\gamma}-1)(\gamma-1)\left(\frac{3}{r}-\frac{d\ln \Gamma}{dr}\right)~.
\end{equation}
The comoving Lorentz factor at radius $R$, for a particle injected with $\gamma(r)$ when the shock radius was $r$, will be
\begin{equation}
\gamma_{\rm ad}(R,r)-1=\left(\frac{r}{R}\right)^{3(\hat{\gamma}-1)}\left[\frac{\Gamma(R)}{\Gamma(r)}\right]^{(\hat{\gamma}-1)}(\gamma(r)-1)~.
\label{pad}
\end{equation}
where $\gamma(r)$ is given by 
$\gamma_{rad,e}(r)$ (Eq.~\ref{eq:gerad}) for leptons, and by $\gamma_{acc,p}(r)$ (Eq.~\ref{eq:gpacc}) for protons.

Considering the proton and lepton energy densities separately,
the comoving internal energy at radius $R$ will be:
\begin{equation}
\label{eq:int energy}
E'_{int}(R)\!=\!4 \pi c^2\!\!\! \int_0^R\!\!\! 
dr r^2 \left\{\rho_{p}(r) [\gamma_{ad,p}(R,r)\!-\!1]\!+\!\rho_{e}(r) 
[\gamma_{ad,e}(R,r)\!-\!1] \right\}~ .
\end{equation}
With the help of Eq.~\ref{pad}, one can explicitly find $E'_{int}(R)$ and insert it
in Eq.~\ref{eq:G solve}.

\vskip 0.2 cm
\noindent
The other term needed in Eq.~\ref{eq:G solve} is $dE'_{ad}/dR$.
First, we have derived $(d\gamma/dR)_{ad}$ for a single particle. Now integrating over the total number of particles, again considering protons and leptons separately, one obtains:
\begin{equation}
\label{eq:ad losses}
\frac{dE'_{ad}(R)}{dR}=-4 \pi c^2 (\hat{\gamma}-1)\left(\frac{3}{R}-
\frac{d \log \Gamma}{dR}\right)\int_0^R dr r^{2} 
 \left\{\rho_{p}(r) (\gamma_{ad,p}-1)+\rho_{e}(r)(\gamma_{ad,e}-1)\right\}~.
\end{equation}
In Eqs. \ref{eq:int energy} and \ref{eq:ad losses}, it is assumed that only the swept-up 
matter is subject to adiabatic cooling, i.e., that the ejecta particles are cold. 

As long as the shocked particles remain relativistic, 
the equations for the comoving internal energy and for the adiabatic expansion losses assume 
simpler forms:
\begin{equation}
\begin{split}
 \label{eq:eint_rel}
E'_{int}(R)&= 4 \pi c^2 \int_0^R\,dr\,r^2 \frac{r}{R}\left[\frac{\Gamma(R)}{\Gamma(r)}\right]^{1/3} 
\Gamma(r)\left\{ \frac{\epsilon_p}{\mu_p}\rho_{p}+
(1-\epsilon_{rad})\frac{\epsilon_e}{\mu_e}\rho_{e}\right\}\\ &
= 4 \pi c^2 \int_0^R\,dr\,r^2 \frac{r}{R}
\left[\frac{\Gamma(R)}{\Gamma(r)}\right]^{1/3} \Gamma(r)\,\rho(r)\left(\epsilon_p+\epsilon_e-\epsilon \right)\\ &
\frac{dE'_{ad}(R)}{dR} = - E'_{int}(R)\left(\frac{1}{R}-\frac{1}{3}\frac{d \log \Gamma}{dR}\right)~.
\end{split}
\end{equation}

In the absence of significant magnetic field amplification, $\epsilon_p+\epsilon_e\simeq1$ so that $\epsilon_p+\epsilon_e-\epsilon\simeq1-\epsilon$, and the radiative processes of the blast-wave are entirely captured by the single efficiency parameter $\epsilon$.
In the fast cooling regime $\epsilon_{rad}\sim1$ and $\epsilon\simeq\epsilon_e$. 
In this case the term 
$\epsilon_p+\epsilon_e-\epsilon$  reduces to \epsp, meaning that, regardless of the 
amount of energy gained by the electrons, in the fast cooling regime the adiabatic losses are dominated by the protons, since the electrons lose all their energy to radiation.

Evaluating these expressions for adiabatic blast-waves in a power--law density profile 
$\rho\propto R^{-s}$, one obtains:
\begin{equation}
E'_{int}(R)=\frac{9-3s}{9-2s}\Gamma m c^2~~,~~~~~~  
\frac{dE'_{ad}(R)}{dR}=-\frac{(9-s)(3-s)}{2(9-2s)}\frac{\Gamma m c^2}{R} ~,
\end{equation}
where $\Gamma\propto R^{-(3-s)/2}$ as in the adiabatic BM76 solution has been used. 

In the fully radiative regime $\epsilon=1$, which implies $E'_{int}=0$ and $dE'_{ad}=0$, 
Eq.~\ref{eq:G solve} reduces to:
\begin{equation}
\label{eq:rad}
\frac{d\Gamma}{dm}=-\frac{(\Gamma_{\it eff}+1)(\Gamma-1)}{M_0+m}~~,
\end{equation}
which describes the evolution of a momentum--conserving (rather than pressure--driven) snowplow. 
Replacing $\Gamma_{\it eff}\rightarrow \Gamma$, the solution of this equation coincides with the result 
by BM76. 

Since the model is based on the homogeneous shell approximation, the adiabatic solution does not 
recover the correct normalization of the BM76 solution. 
In this treatment, the total energy of a relativistic decelerating adiabatic blast 
wave in a power--law density profile $\rho(R)\propto R^{-s}$ is
\begin{equation}
E_0\simeq \Gamma_{\it eff} E'_{int}\simeq\frac{4}{3}\Gamma E'_{int}\simeq\frac{12-4s}{9-2s}\Gamma^2 m c^2~~,
\end{equation}
so that the BM76 normalization can be recovered by multiplying the density of external 
matter in Eqs. \ref{eq:G solve}, \ref{eq:int energy} and \ref{eq:ad losses} 
by the factor $(9-2s)/(17-4s)$.
To smoothly interpolate between the adiabatic regime and the radiative regime, 
the following correction factor should be adopted:
\begin{equation}
\label{eq:corr}
C_{BM76,\epsilon}\equiv\epsilon+\frac{9-2s}{17-4s}(1-\epsilon)~~.
\end{equation}
No analytic model properly captures the transition between an adiabatic relativistic 
blast-wave and the momentum--conserving snowplow, as $\epsilon$ increases from zero to unity. 
The simple interpolation in Eq.~\ref{eq:corr} joins the fully adiabatic BM76 solution with the fully radiative momentum-conserving snowplow.

\vskip 0.2 cm
In summary, Eqs.~\ref{eq:G solve}, \ref{eq:int energy} and \ref{eq:ad losses}, 
complemented with the correction in Eq.~\ref{eq:corr} (which should by applied to every occurrence of  
external density and external matter) completely determine the evolution of the shell Lorentz factor $\Gamma$ 
as a function of the shock radius $R$.

%-----------------------------------

\subsection{Relativistic shock acceleration}
\label{subsec:fermi_mechanism}
The spectral shape of the afterglow emission is well described by power-laws over a wide energy range (from radio to GeV-TeV). This is the clear manifestation of the presence of an electron population that has been accelerated in a power-law energy distribution.
In GRB afterglows, the main candidate to explain the accelerated non-thermal particles is a Fermi-like mechanism that operates with similar general principles as the non-relativistic diffusive shock acceleration: particles are scattered back and forth across the shock front by magnetic turbulence and gain energy at each shock crossing. The particles themselves are thought to be responsible for triggering the magnetic instability that produces the turbulent field governing their acceleration. 
The outcome of this acceleration process is determined by the composition of the ambient medium (electron-proton plasma in the case of GRB forward shocks), the fluid Lorentz factor ($\Gamma_{GRB}>>1$, decreasing to non-relativistic velocity only after several weeks or months), and the magnetization $\sigma$ (i.e., the ratio between Poynting and kinetic flux in the pre-shocked fluid, $\sigma=B^2/(4\pi\,m_p\,n\,c^2)$){, with $B$ being the magnetic field strength.} For GRB forward shocks, the magnetization is low, around $10^{-9}$ in the interstellar medium and in any case below $10^{-5}$ even for a magnetized circumstellar wind. 

In this section we summarise the present understanding of particle acceleration and magnetic field generation in electron-proton, ultra-relativistic, weakly magnetized shocks. The statements and considerations reported in this section refer specifically to this case (which is the one relevant for forward external shocks in GRBs) and might not be valid for magnetized plasma and/or mildly-relativistic flows and/or electron-positron plasma.

In general, the information that one would extract from theoretical/numerical investigations and compare with observations are: i) the spectral shape of the emitting electrons (i.e., the minimum and maximum Lorentz factor $\gamma_{min/max}$ and the spectral index $p$), ii) the acceleration efficiency (i.e., the fraction of electrons $\xi_e$ and the fraction of energy $\epsilon_e$ in the non-thermal population), and iii) the strength of the self-generated magnetic field, usually quantified in terms of fraction $\epsilon_B$ of the shock-dissipated energy conveyed in the magnetic field. In particular, in order to compare with observations, the relevant $\epsilon_B$ is the one in the downstream, in the region where radiative cooling takes place and the emission is produced.

After revisiting the state-of-the-art of the theoretical understanding (for recent reviews, see~\cite{sironi15,vanthieghem20}), we discuss how particle acceleration and magnetic field amplification are incorporated in GRB afterglow modeling, and then we comment on the constraints on the above-mentioned parameters as inferred from the comparison between the model and the observations.

\subsubsection{Inputs from theoretical investigations}
Analytical approaches and Monte Carlo simulations generally rely on the assumption that electromagnetic waves, providing the scattering centers to regulate and govern the acceleration, are present on both sites of the shock, with a given strength and spectrum, so that the Fermi mechanism can operate. The particle distribution is then evolved under some assumption (such as diffusion in pitch angle) on the scattering process, and considering a test-particle approximation (i.e. the high-energy particles do not modify the properties of the waves).

\begin{figure}[!ht]
   \begin{center}
    \includegraphics[width=0.6\textwidth]{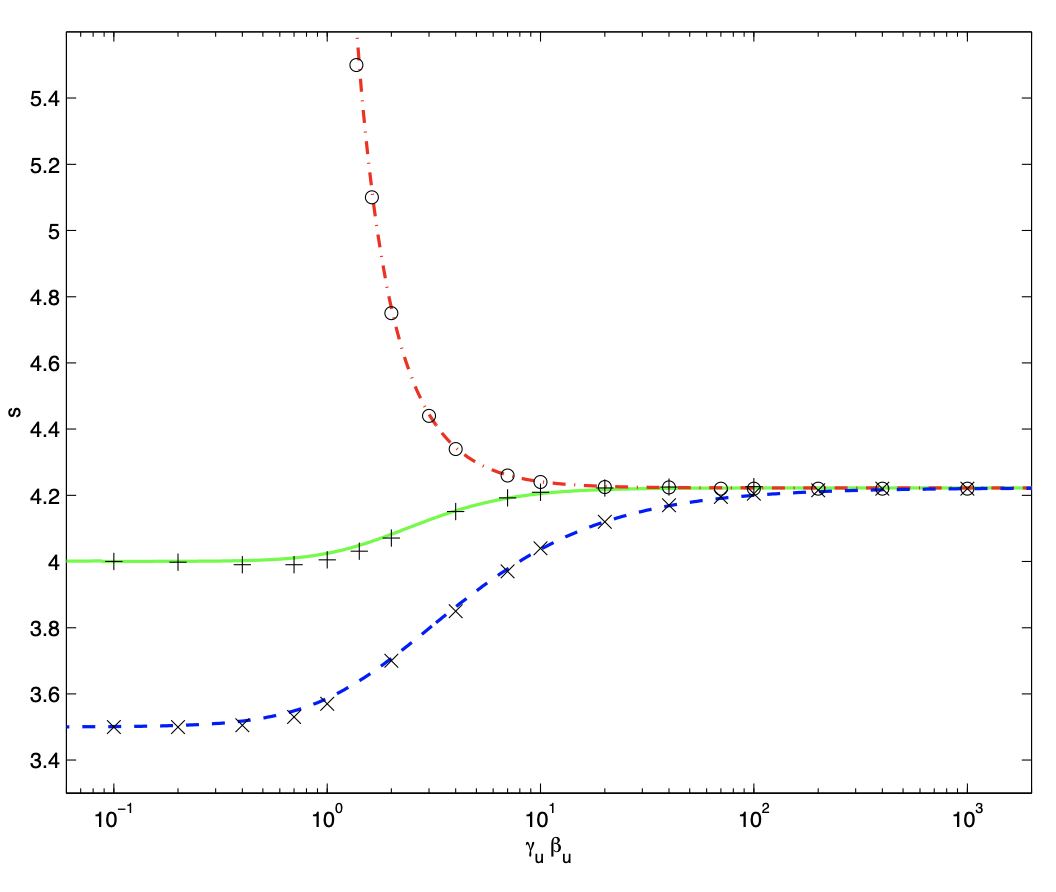}
    \caption{Spectral index ($s_p=p+2$) of the electrons accelerated in shocks as a function of the shock velocity. Curves refer to the equation derived by \cite{keshet_waxman05} under the hypothesis of isotropic, small-angle scattering and is a generalization of the non-relativistic formula. Symbols show the comparison with numerical studies. Different curves refer to different assumptions on the type of shock (see~\cite{keshet_waxman05} for details): in all cases, the value of the spectral index approaches the same value $s_p\sim 4.2$ (corresponding to $p\sim2.2$) in the ultra-relativistic limit.}
    \label{fig:universal_p}
    \end{center}
\end{figure}

The main success of these approaches is the verification that under these conditions power-law spectra are indeed produced and the predicted spectral index is in very good agreement with observations of afterglow radiation from GRBs {\cite{keshet_waxman05}.} 
The spectral index has been calculated for different assumptions on the equation of state, diffusion prescription and for a wide range of shock velocities {\cite{keshet_waxman05}}. A quasi-universal value $p\simeq2.2-2.3$  is found in the ultra-relativistic limit.
Figure~\ref{fig:universal_p} shows a comparison between analytical and numerical results as a function of the shock velocity for three different types of shocks (see~\cite{keshet_waxman05} for details). In the ultra-relativistic limit ($\gamma\beta\gg1$), the estimates of the spectral slopes converge to a universal value $p=s-2\sim2.2$.

%---------------------------

%---------------------------

The investigation of relativistic shocks is complemented by particle-in-cell (PIC) simulations, where the non-linear coupling between particles and self-generated magnetic turbulence is captured from first principles. 

The limitations of this technique are imposed by the
computation time: for accuracy and stability, PIC simulations need to resolve the electron plasma skin depth $c/\omega_{pe}$ of the upcoming electrons (where $\omega_{pe}=\sqrt{4\pi e^2 n_e / m_e}$ is the plasma oscillation frequency of the upstream plasma, $n_e$ is the proper density,{ and $m_e$ is the electron mass)}, which is orders of magnitudes smaller than the scales of astrophysical interest. It is then difficult to follow the evolution on time-scales and length-scales relevant for astrophysics. Low dimensionality (1D or 2D instead of 3D) and small ion-to-electron mass ratios are additional limitations imposed by the computation time. Results of PIC simulations need then to be extrapolated to bridge the gap between the micro-physical scales and the scales of interest.
With these caveats in mind, we summarize here the main achievements.

PIC simulations have shown that magnetic turbulence can be efficiently ($\epsilon_B\sim0.01-0.1$) generated by the accelerated particles streaming ahead of the shock (in the so-called precursor region), where they generate strong magnetic waves which in turn scatter the particles back and fourth across the shock. In particular, in the weakly magnetised shocks discussed in this section, the dominant plasma instability is thought to be the so-called Weibel (or current filamentation) instability {\cite{weibel_instabilities}}, generated by the counter-streaming of the accelerated particles against the background plasma in the precursor region \cite{spitkovsky08,sironi13}.
PIC simulations have shown that as long as the fluid is ultra-relativistic ($\Gamma>5$), the main parameter governing the acceleration is the magnetization $\sigma$, i.e. the efficiency of the process is insensitive to $\Gamma$, as the precursor decelerates the incoming background plasma. 

\begin{figure}[!ht]
   \begin{center}
    \includegraphics[width=0.7\textwidth]{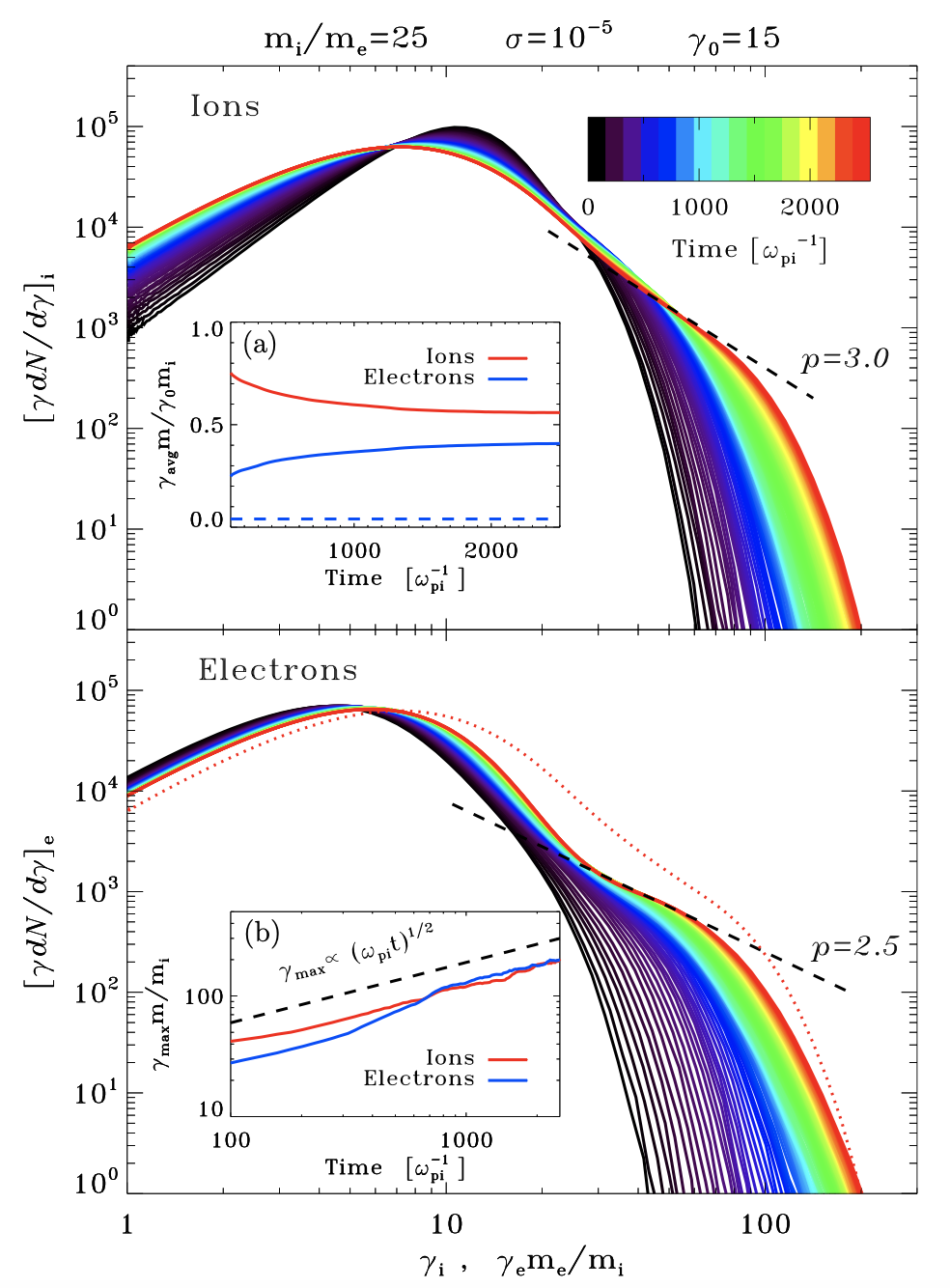}
    \caption{PIC simulations: temporal evolution of the downstream spectrum of ions (upper panel) and electrons (bottom) for mass ratio $m_i/m_e=25$, shock Lorentz factor $\Gamma=15$ and magnetization $\sigma=10^{-5}$. The evolution is followed until $t=2500\omega_{pi}^{-1}$. From~\cite{sironi13}{. \copyright AAS. Reproduced with permission.}}
    \label{fig:PIC}
    \end{center}
\end{figure}

An example of downstream particle spectra derived by PIC simulations is shown in Figure~\ref{fig:PIC} (\cite{sironi13}). The ion and electron spectra are shown for a 2D simulation with $\Gamma=15$, ion-to-electron mass {ratio} $m_i/m_e=25$, and $\sigma=10^{-5}$. The temporal evolution is followed up to {$t=2500\,\omega_{pi}^{-1}$}. The formation of a non-thermal tail is clearly visible.

The downstream non-thermal population is found to include around $\xi\simeq3\%$ of the electrons, carrying $\epsilon_e\simeq10\%$ of the energy. The spectral index is around $p\sim2.5$. The acceleration proceeds similarly for electrons and ions, since they enter the shock in equipartition (i.e. their relativistic inertia is comparable) as a result of efficient pre-heating in the self-excited turbulence in the precursor.

The maximum energy $\gamma_{max}$ increases proportionally to $t^{1/2}$ (see inset in Figure~\ref{fig:PIC}), slower than the commonly adopted Bhom rate {\cite{bohm1949characteristics}}, in which case  $\gamma_{max}\propto t$. Extrapolating the $\gamma_{max}$ behaviour to the relevant time-scales and considering that synchrotron cooling will limit the acceleration for high-energy particles, the electron maximal Lorentz factor is found to reach values $\gamma_{max}\sim10^7$ in the early phase of GRB afterglows, corresponding to synchrotron photon energies around 1\,GeV, roughly consistent with observations.
All these results on the particle spectrum are obtained on time-scales that are too short for the supra-thermal particles to reach a steady-state and their extrapolation to longer time-scales is not trivial.

A {still debated open question} (because computationally demanding) is how the magnetization evolves downstream. 
PIC simulations have {found }values of $\epsilon_B\sim0.1-0.01$ in the vicinity of the shock front. How this turbulence evolves on longer time-scales is still matter of debate.
The turbulence is expected to decay rapidly, on time-scales orders of magnitude shorter than the synchrotron cooling time. Magnetization is then predicted to be very different close to the shock and in the region where particle cooling takes place. Electrons would then cool in a region of weak magnetic field \cite{lemoine13,lemoine+13}.
These considerations suggest that {it might not be correct }to define a single magnetization $\epsilon_B$ {in GRB modeling, infer its value from observations }and compare {with} predictions
from PIC {simulations referring} to the magnetization near the shock {front.  Magnetization values }inferred from observations most likely probe a region downstream, far from the front shock (see Section~\ref{subsubsec:b} for a discussion). 

Theoretical efforts are fundamental to provide physically motivated inputs for the phenomenological parameters included in the afterglow {model. The} large number of unknown model parameters, coupled with a limited number of constraints provided by observations, implies that constraints from theory are of paramount importance for a correct interpretation of the emission in GRBs and for grasping the origin of their non thermal emission, from radio to TeV energies. On the other hand, despite the huge progresses in the theoretical understanding of relativistic acceleration, the theory is not quite yet to the point of providing robust inputs for modeling observations. It is then clear how the two approaches must be combined to gain knowledge on the micro-physics of acceleration and magnetic field generation on the one hand and on the origin of radiative processes and macro-physics of the emitting region (bulk Lorentz factor and energy content) of the sources on the other hand. 

\subsubsection{Description of shocks in GRB afterglow modeling}
\label{subsec:shocks_microphysics}
The theory of relativistic shock acceleration is applied to GRB afterglow by introducing several unknown parameters in the model. These are the fractions $\epsilon_e$ and $\epsilon_B$ of dissipated energy gained by the accelerated particles and amplified magnetic field, the spectral index $p$ of the accelerated particle spectrum, and the fraction $\xi_e$ of particles which efficiently enter the Fermi mechanism and populate the non-thermal distribution.

Recalling that the shock dissipated energy (in the comoving frame) is given by $dE'_{sh} = (\Gamma - 1) dm c^2 $ (see section~\ref{subsec:jet_dynamics}), the corresponding energy density is $u'_{\rm sh}=(\Gamma - 1) \rho' c^2$.
From shock jump conditions, the density in the comoving frame $\rho'$ is related to the density of the unshocked medium (measured in the rest frame) by the equation:
%---------------------------------------------
\begin{equation}
    \rho' = 4\Gamma\,\rho
\end{equation}
%---------------------------------------------
which is valid in both the ultra-relativistic and non-relativistic limits{ (see e. g., \cite{BM76})}.
In the GRB afterglow scenario it is usually assumed that pairs are unimportant and then the density of protons and electrons is the same: $n_p = n_e = n$. This implies that the mass is dominated by protons: $\rho=n m_p$. 
In this case, the available energy density that will be distributed to the accelerated particles (electrons and protons) and to the magnetic field can be expressed as:
%---------------------------------------------
\begin{equation}
    u'_{\rm sh}  = 4\Gamma(\Gamma - 1) n m_p c^2
\end{equation}
%---------------------------------------------
A fraction $\epsilon_B$ of this energy will be conveyed to the magnetic field:
%---------------------------------------------
\begin{equation}
    u'_B= \epsilon_B u'_{\rm sh} = \epsilon_B 4\Gamma(\Gamma - 1) n m_p  c^2 
\end{equation}
%---------------------------------------------
from which it follows that the magnetic field strength $B'$ is:
%---------------------------------------------
\begin{equation}
    B' = \sqrt{32\pi \epsilon_B m_p c^2 n (\Gamma - 1)\Gamma}
    \label{eq:magnetic_field_equation}
\end{equation}
%---------------------------------------------
Similarly, for the accelerated electrons:
%---------------------------------------------
\begin{equation}
    u'_e = \epsilon_e u'_{sh} = \epsilon_e 4\Gamma (\Gamma - 1)n m_p  c^2 = \langle\gamma\rangle m_e c^2 4\Gamma \xi_e n
\end{equation}
%---------------------------------------------
where $\langle\gamma\rangle$ is the average random Lorentz of the accelerated electrons: 
%---------------------------------------------
\begin{equation}
    \langle \gamma\rangle = \frac{\epsilon_e}{\xi_e} \frac{m_p}{m_e} (\Gamma - 1)
    \label{eq:gamma_medium_shock}
\end{equation}
%---------------------------------------------
The accelerated non-thermal electrons are assumed to have a power-law spectrum, as a result of shock acceleration. Their energy distribution can be described by a power-law $N(\gamma) d\gamma \propto \gamma^{-p} d\gamma$ for $ \gamma_{min} \leq \gamma \leq \gamma_{max}$ where $\gamma_{min}$ is the \textit{minimum Lorentz factor} of the injected electrons and $\gamma_{max}$ is the \textit{maximum Lorentz factor} at which electrons can be accelerated. 
To derive the relation between $\gamma_{\rm min}$, $\gamma_{\rm max}$ and the model parameters, we consider the definition of the average Lorentz factor $\langle\gamma\rangle$:
%---------------------------------------------
\begin{equation}
    \langle\gamma\rangle = \frac{\int_{\gamma_{min}}^{\gamma_{max}} N(\gamma)\gamma d\gamma }{\int_{\gamma_{min}}^{\gamma_{max}} N(\gamma)d\gamma} 
    \label{eq:gamma_medium}
\end{equation}
and solve the integrals. Equating \ref{eq:gamma_medium_shock} and \ref{eq:gamma_medium} leads to (for $p \neq 1$):
%---------------------------------------------
\begin{equation}
\begin{array}{cc}
    \Bigg[\frac{\ln\big({\frac{\gamma_{max}}{\gamma_{min}}}\big)}{\frac{1}{\gamma_{min}} - \frac{1}{\gamma_{max}}} \Bigg] = \frac{\epsilon_e}{\xi_e} \frac{m_p}{m_e} (\Gamma - 1) \hspace{3.1cm} \text{if $p = 2$} \\
    \Bigg[\frac{p - 1}{p - 2}\frac{\gamma_{min}^{-p+2} - \gamma_{max}^{-p + 2}}{\gamma_{min}^{-p+1} - \gamma_{max}^{-p + 1}} \Bigg] = \frac{\epsilon_e}{\xi_e} \frac{m_p}{m_e} (\Gamma - 1) \hspace{2cm} \text{if $p \neq 2$}
    \label{eq:gamma_min_implicit}
\end{array}
\end{equation}
%---------------------------------------------
A simplified equation for $\gamma_{min}$ can be obtained assuming that $\gamma_{max}^{-p+2} \ll \gamma_{min}^{-p+2}$:
%---------------------------------------------
\begin{equation}
    \gamma_{min} = \frac{\epsilon_e}{\xi_e} \frac{m_p}{m_e}\frac{p-2}{p-1} (\Gamma - 1)
    \label{eq:gamma_min}
\end{equation} 
%---------------------------------------------
Since $p$ is expected to be $2 < p < 3$, this condition is verified for $\gamma_{max} \gg \gamma_{min}$. 
The minimum Lorentz factor is then not treated as {a free parameter of the model}, as it is calculated from eq.~\ref{eq:gamma_min} as a function of the free parameters $\epsilon_e, \xi_e$ and $p$.
{Concerning the prescription for the value of $\gamma_{max}$ (for details see Section \ref{subsec:max_syn_energy}) usually} it relies on the condition that radiative losses between acceleration episodes are equal to the energy gains, where energy gains proceeds at the Bhom rate. As we mentioned in the previous section, PIC simulations however have shown that this might not be the case. 

A similar treatment can be adopted also for protons simply substituting $\epsilon_e$ with $\epsilon_p$, $m_e$ with $m_p$ and assuming a power-law energy distribution with spectral index $q$. As a result, the minimum Lorentz factor for protons can be derived as:
%-------------------------------
\begin{equation}
\begin{array}{cc}
\Bigg[\frac{\ln\big({\frac{\gamma_{max,p}}{\gamma_{min,p}}}\big)}{\frac{1}{\gamma_{min,p}} - \frac{1}{\gamma_{max,p}}} \Bigg] =\frac{\epsilon_p }{\xi_p} (\Gamma - 1) \hspace{2.8cm} \text{if $q = 2$} \\
    \Bigg[\frac{q - 1}{q - 2}\frac{\gamma_{min,p}^{-q+2} - \gamma_{max_p}^{-q + 2}}{\gamma_{min,p}^{-q+1} - \gamma_{max,p}^{-q + 1}}\Bigg] = \frac{\epsilon_p }{\xi_p} (\Gamma - 1) \hspace{2cm} \text{if $q \neq 2$}
    \label{eq:gamma_min_implicit_proton}
\end{array}
\end{equation}
%---------------------------------------------
{Solving} the equations assuming that $\gamma_{max,p} \gg \gamma_{min,p}$ leads to:
%---------------------------------------------
\begin{equation}
    \gamma_{min,p} = \frac{\epsilon_p}{\xi_p} \frac{q-2}{q-1} (\Gamma - 1)
    \label{eq:gamma_min_protons}
\end{equation}
%---------------------------------------------
The equations for $\gamma_{min}$ and{ Eq.~\ref{eq:magnetic_field_equation} for} $B$, coupled with the description of the blast-wave dynamics described in section~\ref{subsec:jet_dynamics}, provides all the necessary equations to derive the radiative output for a jet with energy $E$ and initial bulk Lorentz factor $\Gamma_0$ expanding in a medium with density $n(R)$. The derivation of the radiative output is detailed in section~\ref{subsec:radiation_processes}. To conclude the discussion about particle acceleration, in the next section we anticipate which constraints can be inferred on the physics of particle acceleration from multi-wavelength observations, once the afterglow model is adopted.

%----------------------------------------------------------------
\subsubsection{Constraints to the acceleration mechanism provided by observations}
Assuming that accelerated particles have a power-law spectrum ($dN^{\rm acc}/d\gamma\propto\gamma^{-p}$) and the cooling is dominated by synchrotron radiation, the spectral slope $p$ can be inferred from observations of the synchrotron spectrum and/or from the temporal decay of the lightcurves if observations are performed at frequencies higher than {the typical frequency $\nu_{\rm m}$ of photons emitted by electrons with Lorentz factor $\gamma_{min}$} (this is correct both in case of fast and slow cooling regime). 
The estimated value of $p$ from afterglow modeling are spread on a wide range, from $p\sim2$ to {$p\sim3$}, suggesting that the spectrum of injected particles does not seem to have a typical slope, at odds with theoretical predictions.
The determination of $p$ however, suffers from the uncertainties on the spectral index inferred from optical and X-ray observations, where the observed spectra are subject to unknown dust and metal absorption. A derivation of $p$ from the decay rate of the lightcurves is also subject to the correct identification of the spectra regime, and partially also to the assumption on
density profile of the external medium, which is often unconstrained (see section \ref{subsec:density_profile}).

The typical value of $\epsilon_{\rm e}$ inferred from afterglow modeling is around 0.1, meaning that 10\% of the shock-dissipated energy is gained by the electrons, spanning from 0.01 to large values, such as 0.8. Although this seems a large uncertainty, $\epsilon_e$ is perhaps the most well constrained parameter of the model, and is in good agreement with values predicted by numerical investigations {\cite{sironi13}}.
For the fraction $\epsilon_{\rm B}$, on the contrary, the inferred values varies in a very wide range, typically from $10^{-5}$ to $10^{-1}$ {\cite{santana14,wang15,zhang15}}. 
Recent studies that incorporate Fermi-LAT GeV observations \cite{beniamini15,beniamini16} have shown that {the typical values estimated for $\epsilon_B$ can be even smaller, in the range $\sim 10^{-7}$-$10^{-2}$. These values are needed } in order to model GeV radiation self-consistently with radiation detected at lower frequencies, with repercussions on the estimates of the other parameters, such as $n$ and $E$.
These small values of the $\epsilon_B$ needed to model the radiation have been tentatively interpreted as the sign of turbulence decay in the downstream {\cite{lemoine13,lemoine+13}. As a consequence,} even though the turbulence is strong ($\epsilon_B\simeq0.1$) in the vicinity of the shock, where particle are accelerated, it becomes weaker at larger distances, in the region where particles cool (see section~\ref{subsubsec:b}).
Small values of $\epsilon_B$ are confirmed by the modeling of recent TeV detections of afterglow radiation from GRBs  {(\cite{190114C_mwl_paper,190829A_salafia}, see }Section~\ref{Chap:TeVdiscovery}).

Another parameter that one would like to constrain from observations is the fraction of particles $\xi_e$ that are injected into the Fermi process. In the vast majority of the studies, this parameter  is not included (i.e. it is implicitly assumed that all the electrons are accelerated, $\xi_e=1$). This parameter is indeed difficult to constrain, as it is degenerate with all the other parameters \cite{eichler}.

Observations so far have not been able to identify the location of a high-energy cutoff in the synchrotron spectrum, that would reveal the maximal energy of the synchrotron photons and then the maximum energy $\gamma_{max}$ of the accelerated electrons. Observations by Fermi-LAT are in general consistent with a single power-law extending up to at least 1\,GeV. Photons with energies in excess of 1\,GeV have been detected from several GRBs, the record holder for Fermi-LAT being a $95$\,GeV photon {\cite{fermilat_130427A}}. These photons cannot be safely associated to synchrotron radiation on the basis of spectral analysis, as their paucity makes difficult to assess from spectral analysis whether they are consistent with the power-law extrapolation of the synchrotron spectrum or they are indicative of the rising of a distinct spectral component. 
In any case, the Fermi-LAT detections are suggesting that synchrotron photons should be produced at least up to a few GeV. This is consistent with the limit commonly invoked for particle acceleration: if the acceleration proceeds at the Bhom rate ($t_{acc}\simeq r_L/c$) {with $r_L=E/eB$ being the Larmor radius,} and is limited by synchrotron cooling ($t_{syn}\simeq6\,\pi\,m_e c/\sigma_TB^2\gamma$) then $\gamma_{max}\sim10^7-10^8$ can be reached.
Even though this does not necessarily imply that acceleration must proceed at the Bhom limit, the value of $\gamma_{max}$ inferred from the detection of GeV photons is quite large and barely consistent with what found by PIC simulations.
Whether or not the observations are in tension with the present derivation of $\gamma_{max}$ from PIC simulations and theoretical arguments, strongly depends on a clear identification of the origin of photons in the GeV-TeV energy range. Present and future observations with {Imaging Atmospheric Cherenkov Telescopes (IACTs)} are the main candidates to shed light on this issue.

%-----------------------------------

\subsection{Derivation of the radiative output}
\label{subsec:radiation_processes}
The expected radiative output can be estimated by means of analytical approximations, which provide prescriptions for the location of the synchrotron self-absorption frequency $\nu_{sa}$, the characteristic frequency $\nu_m$ emitted by electrons with Lorentz factor $\gamma_{min}$, the cooling frequency $\nu_c$ emitted by electrons with Lorentz factor $\gamma_c$, and the overall synchrotron flux {\cite{saripiran,PK00}}.
In these approaches, the synchrotron spectrum is in general approximated with power-laws connected by sharp breaks, but more sophisticated analytical approximations of numerically derived synchrotron spectra have also been proposed {\cite{granot_breaks}}.
The associated SSC component in Thomson regime \cite{sariesin} and corrections to be applied to the synchrotron and SSC spectra to account for the effects of Klein-Nishina \cite{nakar_kn_regime} {cross section (see Section \ref{subsubsec:sync_emission})} are also available in literature.
These prescriptions are usually developed for the deceleration phase, when the Blandford-McKee solution \cite{BM76} for the blast-wave dynamics applies, i.e., as long as the blast-wave is still relativistic. 
These models take as input parameters the kinetic energy content of the blast-wave $E_k$, the external density $n(R)=n_0\,R^{-s}$ (with $s=0$ or $s=2$), the fraction of shock-dissipated energy gained by electrons ($\epsilon_e$) and by the amplified magnetic field ($\epsilon_B$), and the spectral slope of the accelerated electrons $p$.
During the deceleration phase, the initial bulk Lorentz factor $\Gamma_0$ does not play any role, but its value determines the radius (or time) at which the deceleration begins.

An alternative approach to estimate the expected spectra and their evolution in time consists in solving numerically the differential equation describing the evolution of the particle spectra and estimate the associated emission \cite{petropoulou,pennanen,bosnjak09}.
In this section we describe a radiative code that solves simultaneously the time evolution of the electron and photon distribution. 
The code has been adopted, e. g., for the modeling of GRB\,190114C presented in \cite{190114C_mwl_paper}.  

The temporal evolution of the particle distribution is described by the differential equation:
%-------------------------------------
\begin{equation}
    \frac{\partial N(\gamma,t')}{\partial t'} = \frac{\partial}{\partial\gamma}\bigg[\dot{\gamma}N(\gamma,t')\bigg] + Q(\gamma)~,
    \label{eq:continuity_equation}
\end{equation}
%-------------------------------------
where $\dot{\gamma} = \frac{\partial\gamma}{\partial t'}$ is the rate of change of the Lorentz factor $\gamma$ of an electron caused by adiabatic, synchrotron and SSC losses and to energy gains by synchrotron self-absorption.  
In the SSC mechanism, the synchrotron photons produced by electrons in the emission region act as seed photons that are up-scattered at higher energies by the same population of electrons. Such scenario will generate a very high energy spectral component, which is the target of searches by IACTs such as {MAGIC\endnote{\url{https://magic.mpp.mpg.de/}} and H.E.S.S.\endnote{\url{https://www.mpi-hd.mpg.de/hfm/HESS/}}. 
In principle, also up-scattering of an external population of seed photons can be considered and included in the cooling term, but here we will ignore this mechanism (external Compton) and focus only on SSC.
The source term $Q(\gamma,t')=Q^{acc}(\gamma,t')+Q^{pp}(\gamma,t')$ describes the injection of freshly accelerated particles ($Q^{acc}(\gamma,t')=dN^{acc}/d\gamma\,dt'$) and the injection of pairs $Q^{pp}(\gamma,t')$ produced by photon-photon annihilation.

In the next sections we explicit each one of the terms included {in eq.~\ref{eq:continuity_equation}} and how to estimate the synchrotron and SSC emission.
To solve the equation, an implicit finite difference scheme based on the discretization method proposed by \cite{chang_cooper} can be adopted.

%----------------------------------------------------------------------
\subsubsection{Synchrotron and SSC cooling}\label{subsubsec:sync_emission}

%---------------------------------------------------
The synchrotron power emitted by an electron with Lorentz factor $\gamma$ depends on the pitch angle, i. e., the angle between the electron velocity and the magnetic field line.
{In the following, we} assume that the electrons have an isotropic pitch angle distribution and we use equations that are averaged over the pitch angle {(e. g., \cite{rybicki_lightman})}. The synchrotron cooling rate of an electron with Lorentz factor $\gamma$ is given by:
%-------------------
\begin{equation}
    \dot\gamma_{syn} \equiv \frac{d\gamma}{dt'}\bigg|_{syn} = -\frac{\sigma_T \gamma^2 B'^2}{6\,\pi\,m_e c}
\end{equation}
%-------------------

The cross section for the inverse Compton mechanism is constant and equal to the Thomson cross section ($\sigma_T$) as long as the photon energy in the frame of the electron is smaller than the rest mass electron energy $m_e\,c^2$. For higher photon energies, the cross section decreases as a function of the energy and is described by the Klein-Nishina (KN) cross section. To estimate SSC losses, we adopt the formulation proposed in \cite{jones}, which is valid for both regimes. 
Defining the SSC kernel as:
%-------------------------------------
\begin{equation}
    K(\gamma,\nu',\Tilde{\nu'}) = \left \{ \begin{array}{ll}
\frac{\varepsilon}{\Tilde\varepsilon} - \frac{1}{4\gamma^2}\hspace{6.1cm} \frac{\Tilde\varepsilon}{4\gamma^2} < \varepsilon < \Tilde\varepsilon\\
2q \ln{q} + (1 + 2q)(1 - q) + \frac{1}{2}(1 - q) \frac{(4\gamma \Tilde\varepsilon q)^2}{(1 + 4\gamma \Tilde\varepsilon q)} \hspace{0.2cm} \Tilde\varepsilon < \varepsilon < \frac{4\gamma^2 \Tilde\varepsilon}{1 + 4 \gamma \Tilde\varepsilon}~,
\label{eq:IC_kernel}
\end{array}
\right.
\end{equation}
%-------------------------------------
where:
%-------------------------------------
\begin{equation}
    \Tilde\varepsilon = \frac{h \Tilde{\nu'}}{m_e c^2} \hspace{0.5cm} \varepsilon = \frac{h \nu'}{m_e c^2} \hspace{0.5cm} q = \frac{\varepsilon}{4\gamma \Tilde\varepsilon (\gamma - \varepsilon)}~.
\end{equation}
%-------------------------------------
$\Tilde\varepsilon$ and $\varepsilon$ are the energies of the photons (normalized to the rest mass electron energy) before and after the scattering process, {respectively}. The two terms of Eq.~\ref{eq:IC_kernel} account respectively for the down-scattering (i.e. $\varepsilon < \Tilde\varepsilon$) and the up-scattering (i.e. $\varepsilon > \Tilde\varepsilon$) process. 
The energy loss term for the SSC can now be calculated with the equation:
%-------------------------------------
\begin{equation}
    \dot{\gamma}_{SSC} = \frac{d\gamma}{d t'}\bigg|_{SSC} = - \frac{3 h \sigma_t}{4 m_e c \gamma^2} \int d\nu' \nu' \int \frac{d\Tilde{\nu'}}{\Tilde{\nu'}} n_{\Tilde{\nu'}} (t') K(\gamma,\nu',\Tilde{\nu'})~.
    \label{eq:energy_losses_IC}
\end{equation}
%-------------------------------------

%---------------------------------------------------------------
\subsubsection{Adiabatic cooling}
As discussed in section \ref{subsec:jet_dynamics}, particles loose their energy adiabatically due to the spreading of the emission region. This energy loss term should be inserted in the kinetic equation governing the evolution of the particle distribution. 
To derive the adiabatic losses, we rewrite equation \ref{eq:p_ad} as a function of energy losses $d\gamma$ in a comoving time $dt'$:
%---------------------
\begin{equation}
    \dot{\gamma}_{ad} = \frac{d\gamma}{d t'}\bigg|_{ad}=-\frac{\gamma\beta^2}{3}\frac{d\ln V'}{dt'}~,
\end{equation}
%---------------------
{with $\beta$ being the random velocity of particles in unit of $c$.} The comoving volume $V'$ of the emission region can be estimated considering that the contact discontinuity is moving away from the shock at a velocity $c/3$. After a time $t'=\int dR/\Gamma(R)\,c$ the comoving volume is: 
%---------------------
\begin{equation}
    V' = 4 \pi R^2 \frac{ct'}{3}~,
\end{equation}
%---------------------
and:
%---------------------
\begin{equation}
    \dot{\gamma}_{ad} = \frac{d\gamma}{d t'}\bigg|_{ad} = - \frac{\gamma \beta^2}{3} \left(\frac{2\Gamma c}{R} + \frac{1}{t'}\right)~.
\end{equation}
%-----------------------------------------------------------------

\subsubsection{Synchrotron self-absorption{ (SSA)}}
Electrons can re-absorb low energy photons before they escape from the source region. The absorption coefficient $\alpha_\nu$ can be expressed as \cite{rybicki_lightman}:
%----------------------------
\begin{equation}
    \alpha_\nu = - \frac{1}{8\pi\nu'^2m_e} \int d\gamma P'(\gamma,\nu')\gamma^2\frac{\partial}{\partial\gamma}\bigg[\frac{N(\gamma)}{\gamma^2}\bigg]
    \label{eq:SSA_absorption}
\end{equation}
%----------------------------
valid for any radiation mechanism at the emission frequency $\nu'$, with $P'(\gamma,\nu')$ being the specific power of electrons with Lorentz factor $\gamma$ at frequency $\nu'$ and assuming $h\nu' \ll \gamma m_e c^2$. Thus, the SSA mechanism will affect mostly the low frequency range. This results in a modification of the lower frequency tail of the synchrotron spectrum as:
%----------------------------
\begin{equation}
    P'_{\nu} \propto \left \{ \begin{array}{ll}
\nu'^{5/2} \hspace{2.7cm} \nu'_i < \nu' < \nu'_{SSA}\\
\nu'^{2} \hspace{3cm} \nu' < \nu'_{SSA} < \nu'_i ~,
\end{array}
\right.
\end{equation}
%----------------------------
assuming a power-law distribution of electrons, with $\nu'_i = \min(\nu'_m,\nu'_{cool})$ and $\nu'_{SSA}$  the frequency below which the synchrotron flux is self-absorbed and the source becomes optically thick.

\subsubsection{Synchrotron and Inverse Compton emission}
Following \cite{ghisellini_svensson}, the synchrotron spectrum emitted by an electron with Lorentz factor $\gamma$, averaged over an isotropic pitch angle distribution is:
%---------------------
\begin{equation}
    P'^{syn}_{\nu'} (\nu',\gamma) = \frac{2\,\sqrt{3}\,e^3\,B'}{m_e\,c^2}\,x^2\,\left[K_{4/3}(x)\,K_{1/3}(x) - 0.6x(K^2_{4/3}(x)-K^2_{1/3}(x))\right]~,
    \label{eq:power_syn}
\end{equation}
%---------------------
where $x\equiv\nu'\,4\pi\,m_ec/(6\,e\,B'\gamma^2)$, and $K_{n}$ are the modified Bessel functions of order $n$.
The total power emitted at the frequency $\nu'$ is obtained integrating over the electron distribution:
%---------------------
\begin{equation}
    P'^{syn}_{\nu'} (\nu') = \int{P'^{syn}_{\nu'} (\nu',\gamma) \frac{dN}{d\gamma}d\gamma}  ~. 
\end{equation}
%---------------------

The SSC radiation emitted by an electron with Lorentz factor $\gamma$ can be calculated as:
%---------------------
\begin{equation}
    P'^{SSC}_{\nu'} (\nu',\gamma) = \frac{3}{4} h \sigma_T c \frac{\nu'}{\gamma^2} \int \frac{d\Tilde{\nu'}}{\Tilde{\nu'}} n_{\Tilde{\nu'}} K(\gamma,\nu',\Tilde{\nu'})~,
    \label{eq:power_IC}
\end{equation}
%---------------------
where $n_{\Tilde{\nu'}}$ is the photon density of synchrotron photons and the integration is performed over the entire synchrotron spectrum. Integration over the electron distribution provides the total SSC emitted power at frequency $\nu'$.

\subsubsection{Pair production}
Pair production by photon-photon annihilation is particularly important for a correct estimate of the radiation spectrum {in the GeV-TeV band. Indeed, some of the emitted VHE photons are lost due to their interaction} with photons at lower energies (typically X-ray photons). {As a result, the observed flux is attenuated and the resulting spectrum at VHE is modified}.
Here we follow the treatment presented in \cite{coppi_pp}. 
The cross section of the process $\sigma_{\gamma \gamma}$ as a function of $\beta'$, the centre-of-mass speed of the electron and positron is given by:
%---------------------------
\begin{equation}
    \sigma_{\gamma\gamma} (\beta') = \frac{3}{16}\sigma_T (1 - \beta'^2) \Bigg[(3 - \beta'^4)\ln{\bigg(\frac{1 + \beta'}{1 -\beta'}\bigg) - 2\beta'(2 - \beta'^2 ) \Bigg]}
\end{equation}
%---------------------------
where:
%---------------------------
\begin{equation}
    \beta'(\omega_t,\omega_s,\mu) = \Bigg[ 1 - \frac{2}{\omega_t \omega_s (1 - \mu)}\Bigg]^{\frac{1}{2}}
\end{equation}
%---------------------------
and $\omega_t = h \nu'_t /m_e c^2$ with $\nu'_t$ being the target photon frequency, $\omega_s = h \nu' /m_e c^2$ with $\nu'$ being the source photon frequency and $\mu = \cos{\phi}$, where $\phi$ is the scattering angle. Then, it is possible to derive the annihilation rate of photons into electron-positron pairs as:
%---------------------------
\begin{equation}
    R (\omega_t,\omega_s) = c \int_{-1}^{\mu_{max}} \frac{d\mu}{2} (1 - \mu) \sigma_{\gamma \gamma} (\omega_t,\omega_s,\mu)~,
    \label{eq:exact_pair_annihilation}
\end{equation}
%---------------------------
where $\mu_{max} = \max(-1, 1-2/\omega_s \omega_t)$ coming from the requirement $\beta'^2 > 0$. Considering $x = \omega_t \omega_s$ it is possible to derive asymptotic limits for $R(\omega_t, \omega_s) \equiv R(x)$ in two regimes. For $x \to 1$ (i.e. near the threshold condition) $R(x) \to c \sigma_T/2 (x-1)^{3/2}$, while for $x \gg 1$ (i.e. ultra-relativistic limit) $R \to \frac{3}{4}c \sigma_T \ln{x}/x $. An accurate and simple approximation which takes into account both regimes is given by:
%---------------------------
\begin{equation}
    R(x) \approx 0.652 c \sigma_T \frac{x^2 - 1}{x^3} \ln{(x)} \text{ H}(x-1)~,
    \label{eq:approx_pair_annihilation}
\end{equation}
%---------------------------
where $H(x-1)$ is the Heaviside function {\cite{coppi_pp}}. The approximation reproduces accurately the behaviour near the peak at $x_{peak} \sim 3.7$ and over the range $1.3 < x < 10^{4}$ which usually dominates during the calculations.{ A comparison between Eq.~\ref{eq:exact_pair_annihilation} and Eq.~\ref{eq:approx_pair_annihilation} is given in Figure~\ref{fig:pp_comparison} where the goodness of the approximation adopted in the mentioned $x$ range can be seen}. 
%---------------------------
\begin{figure}[!ht]
    \includegraphics[width=0.8\textwidth]{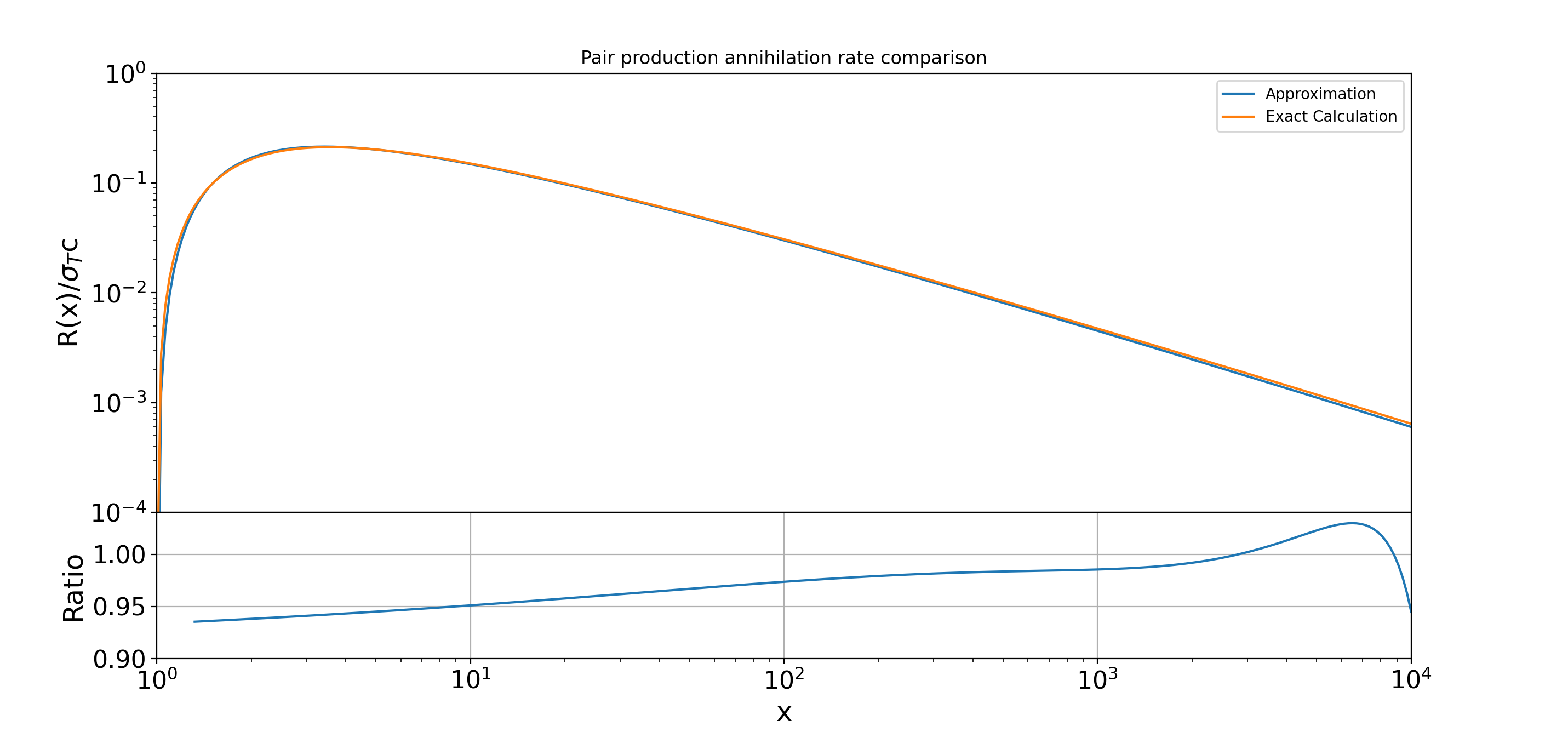}
    \caption{Comparison between the exact annihilation rate (eq.~\ref{eq:exact_pair_annihilation}) and the approximated formula (eq.~\ref{eq:approx_pair_annihilation}).{ The ratio between the two curves in the range ($1.3 < x < 10^{4}$) is shown in the bottom panel. In this range the ratio is always} $\lesssim 7\%$. }
    \label{fig:pp_comparison}
\end{figure}
%---------------------------
The impact of the flux attenuation due to pair production mechanism on the GRB spectra is estimated in terms of the optical depth value $\tau_{\gamma\gamma}$. From its definition:
%---------------------------
\begin{equation}
    \tau_{\gamma\gamma} (\nu') = \sigma_{\nu'\nu'_t} n' (\nu'_t) \Delta R'
    \label{eq:tau_pair_prod_standard}
\end{equation}
%---------------------------
where $n' (\nu'_t)$ is the number density of the target photons per unit of volume, $\sigma_{\nu'\nu'_t}$ is the cross section and $\Delta R'$ is the width of the emission region. Introducing the cross section in terms of the annihilation rate $R(x)$ in its approximated formula and integrating over all the possible target photon frequencies:
%---------------------------
\begin{equation}
    \tau_{\gamma\gamma} (\nu') = \frac{\Delta R'}{c} \int R(\nu',\nu'_t) n'_{\nu'} (\nu'_t) d\nu'_t
    \label{eq:tau_pair_production}
\end{equation}
%---------------------------
where $\nu'$ and $\nu'_t$ are the frequencies of the source and of the target interacting photons. The pair production attenuation factor can be then introduced simply multiplying the flux by a factor $(1 - e^{-\tau_{\gamma\gamma}})/\tau_{\gamma\gamma}$. This attenuation factor will modify the GRB spectrum giving a non-negligible contribution especially in the VHE domain. An example of the modification of a GRB spectrum due to pair production can be seen in Figure~\ref{fig:pp_attenuation}. Here the flux emitted in the afterglow external forward shock scenario by synchrotron and SSC radiation and the flux attenuation due to pair production have been calculated with a numerical code. For a set of quite standard afterglow parameters and assuming $\Delta R' = R/\Gamma$, the attenuation of the observed flux due to pair production become relevant above 0.2 TeV and it reduces the flux by $\sim$ 30$\%$ at 1 TeV and by $\sim$ 70$\%$ at 10 TeV.
%---------------------------
\begin{figure}[!ht]
    \centering
    \includegraphics[width=0.7\textwidth,trim={0 0.5cm 0 0.5cm},clip]{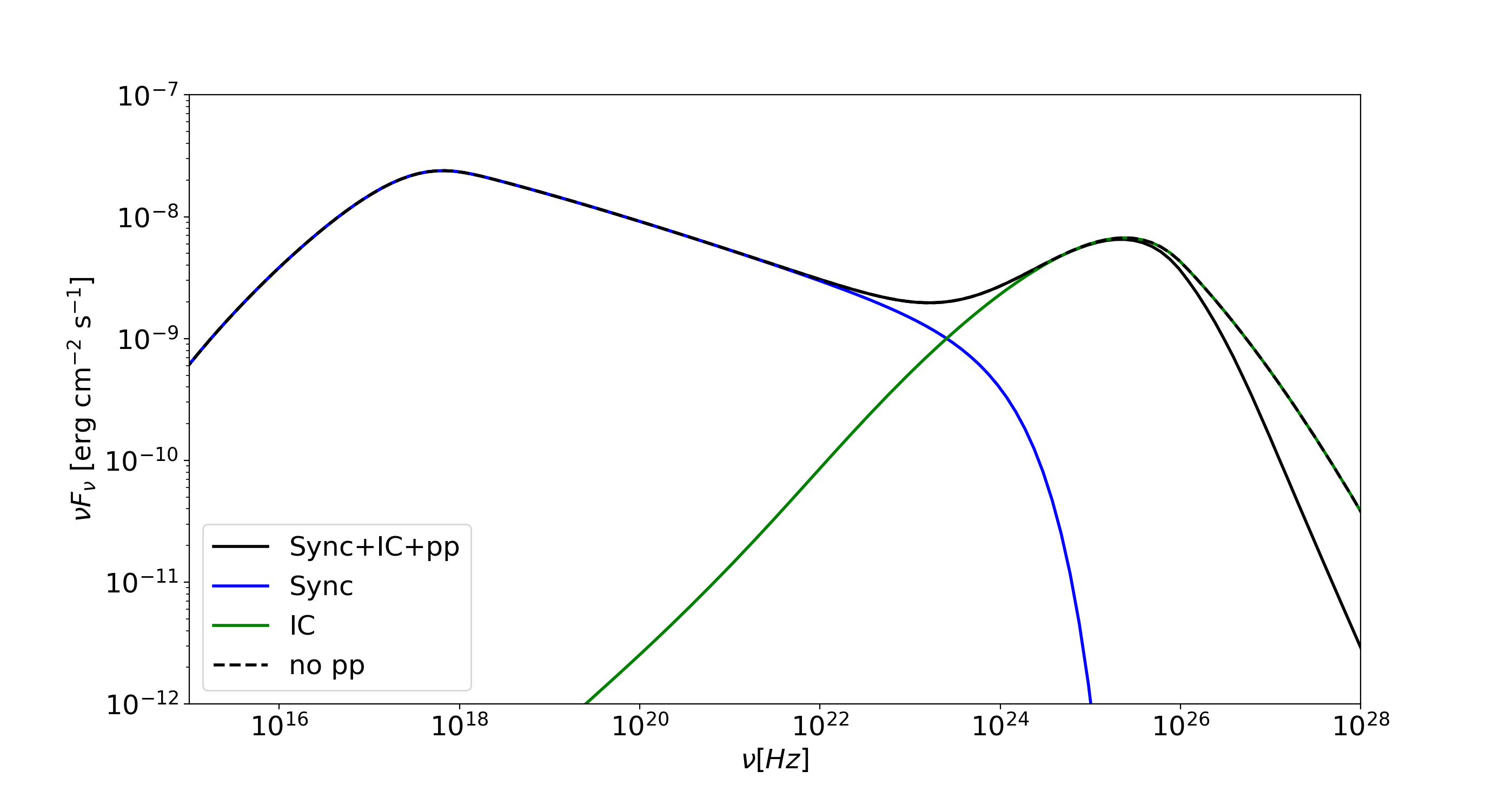}
    \caption{Spectral energy distribution in the GRB afterglow external forward shock scenario estimated with {the numerical code presented in this section for} $t_{obs} = 170 $ s. The effect of the pair production attenuation is clearly evident in the VHE tail. The set of afterglow parameters used are the following: $E_k = 1 \times 10^{53}$ erg, $s = 0$, $A_0 = 1$ cm$^{-3}$, $\epsilon_e = 0.2$, $\epsilon_B = 0.02$, $\Gamma_0 = 300$, $p = 2.5$ and $z = 0.5$.}
    \label{fig:pp_attenuation}
\end{figure}
%---------------------------
Similar considerations can be done also for the electron/positron production. Assuming that the electron and positron arises with equal Lorentz factor $\gamma$ and that $x_{peak} \sim 3.7$, a photon with energy $\omega_s \gg 1$ will mostly interact with a target photon of energy $\omega_t \approx 3/\omega_s$. Then, from the energy conservation condition:
%---------------------------
\begin{equation}
    2\gamma = \omega_s + \frac{3}{\omega_s} \approx \omega_s = \frac{h \nu'}{m_e c^2}
\end{equation}
%---------------------------
The $e^\pm$ production can be seen as an additional source term for the distribution of accelerated particles. As a result, an additional injection term $Q_e^{pp}$ to be inserted in the kinetic equation (equation \ref{eq:continuity_equation}) is calculated as:
%---------------------------
\begin{equation}
    Q_e^{pp} (\gamma,t') = 4 \frac{m_e c^2}{h} n_{\nu'} (\frac{2\gamma m_e c^2}{h},t') \int d\nu'_t n_{\nu'} (\nu'_t,t') R(\frac{2\gamma m_e c^2}{h},\nu'_t)
\end{equation}
%---------------------------

\subsubsection{Comparison with analytical approximations}
%---------------------------
\begin{figure}[!hb]
    \centering
    \includegraphics[width=0.8\textwidth]{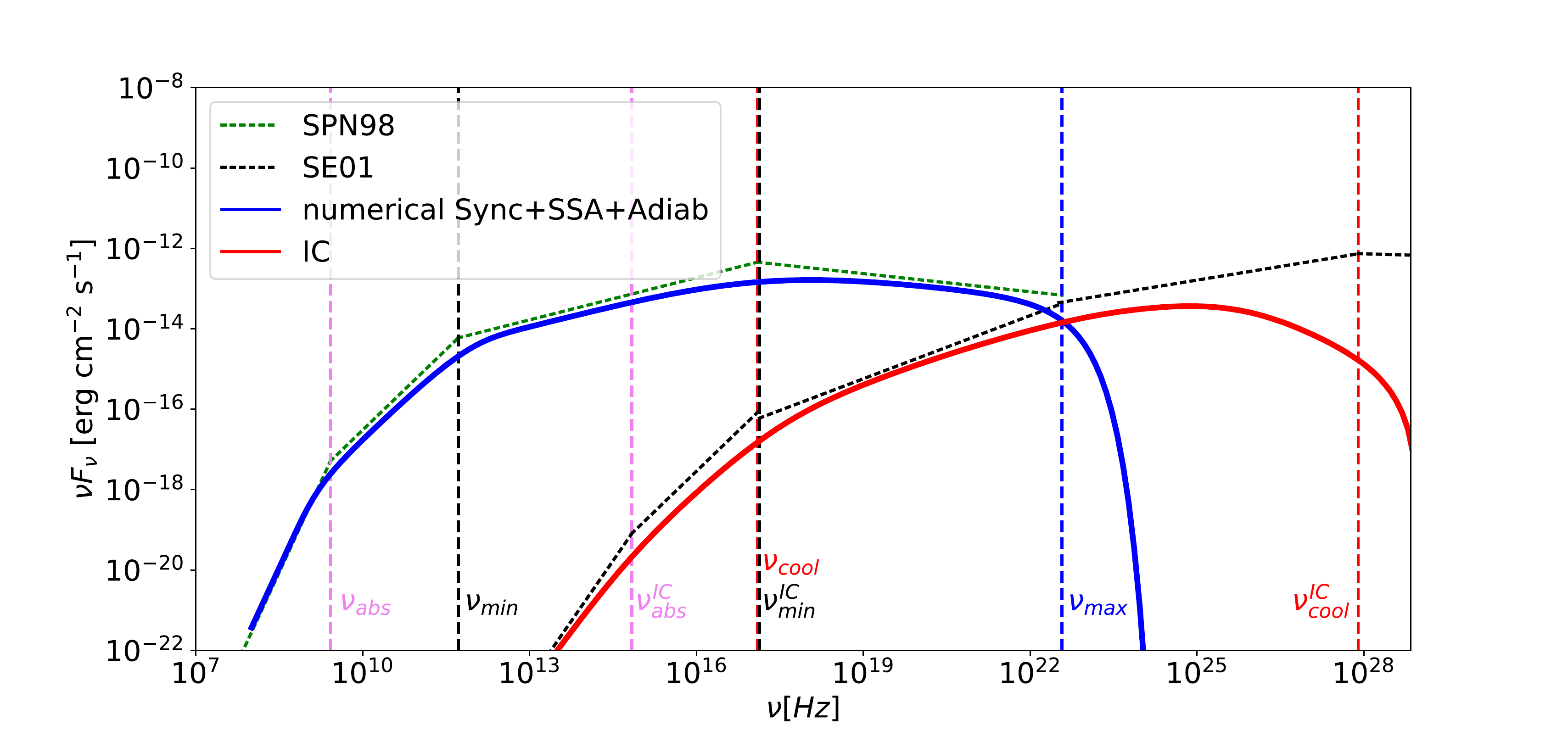}
    \caption{Comparison between spectra estimated with analytical approximations and numerical calculations. For the analytical method, the synchrotron emission (green dashed line) is estimated starting from \cite{saripiran} (SPN98, see legend) while the SSC (black dashed line) is taken from \cite{sariesin} (SE01). Vertical lines mark the break frequencies. The results from the numerical code described in section~\ref{subsec:radiation_processes} are shown with solid blue and red lines. 
    This example shows the spectrum calculated at $t = 10^4$\,s for $s = 0$, $p = 2.3$, $\epsilon_e = 0.05$, $\epsilon_B = 5\times 10^{-4}$, $E_k = 10^{52}$ erg, $n_0 = 1$ cm$^{-3}$, $\Gamma_0 = 400$, and $z = 1$.}
    \label{fig:comparison_num_an}
\end{figure}
%---------------------------
In order to compare results from the numerical method described in previous section and analytical prescriptions available in literature, we give an example in Figure~\ref{fig:comparison_num_an}. The analytical prescriptions for the synchrotron and the SSC component are calculated following \cite{saripiran} and \cite{sariesin}.
In \cite{saripiran} the synchrotron spectra and light-curves are derived assuming a power-law distribution of electrons in an expanding relativistic shock, cooling only by synchrotron emission. The dynamical evolution is described following BM76 equations for an adiabatic blast-wave expanding in a constant density medium. The resulting emission spectrum (green dashed lines in Figure~\ref{fig:comparison_num_an}) is described with a series of sharp broken power-laws.
The SSC component associated to the synchrotron emission was computed, as a function of the afterglow parameters, in \cite{sariesin}. In this work, calculations are performed assuming that the scatterings occur in Thomson regime. Modifications to the synchrotron spectrum caused by strong SSC electron cooling are also detailed. 

From the comparison proposed in Figure~\ref{fig:comparison_num_an}, it can be clearly seen that analytical and numerical results are in general in good agreement. Both curves follow the same behaviour except for the high-energy part of the SSC {component. Here the KN scattering regime, which is not taken into account in the analytical approximation, becomes relevant. As a result, the numerical calculations differ from the analytical ones showing a peak and a cutoff in the SSC spectrum due to the KN effects}.

Nevertheless, there are a few minor discrepancies between the two methods. 
The numerically-derived spectrum is very smooth around the break frequencies, with the result that the theoretically expected slope (e.g., the one predicted by the analytical approximations) is reached only in regions of the spectrum that lie far from the breaks, i.e. is reached only asymptotically. This puts into questions simple methods for discriminating among different regimes and different density profiles based on closure {relations, which are relations between the spectral and the temporal decay indices \cite{saripiran,chevalier99}.}
Regarding the flux normalization, there are minor discrepancies between the numerical and analytical results. This is due to the fact that in analytical prescriptions it is assumed that the radiation is entirely emitted at the characteristic synchrotron frequency. On the contrary, in the numerical derivation, the full synchrotron spectrum of a single electron is summed up over the whole electron distribution. Similar considerations apply to the SSC component when comparing with the analytical spectra. Moreover, the discrepancies observed between analytical and numerical SSC spectra are amplified by the differences observed in the target synchrotron spectra.

In general this comparison shows that the numerical treatment is a powerful tool able to predict the multi-wavelength GRB emission in a more accurate way than the analytical prescriptions. The latter ones, however, are still giving valid approximations of the overall spectral shape.
The main limitation of analytical estimates arises when TeV observations are involved. The importance of KN corrections is evident in this band and should be properly treated for a correct interpretation of the TeV spectra, as will be shown in section~\ref{Chap:TeVdiscovery}.

\section{Open Questions}\label{chap:openquestions}
As predicted by the basic standard model presented in the previous section, the afterglow emission is the result of particle acceleration and radiative cooling occurring in two different regions: the forward and the reverse shock. The temporal and spectral behaviour of the two emission components can be inferred after the jet/blastwave dynamics, acceleration mechanisms, and the radiation processes are modeled (section~\ref{chap:model}).
The general agreement between model predictions and observations convincingly proves that the long-lasting radio-to-GeV radiation is indeed produced in interactions between the ejecta and the external medium. Also, the radiative mechanisms involved and the nature of emitting particles are well established, with synchrotron (and possibly SSC) from the accelerated electrons (either at the forward or reverse shock) being the source of the detected radiation.

Despite the general success of the external shock scenario, there are several, longstanding open issues which represent a serious challenge for our present understanding of the afterglow emission and the GRB phenomenon in general. Moreover, even when observations seem to be in qualitative agreement with predictions, the extraction of the model parameters (which would give important feedbacks on our understanding of particle acceleration and GRB environments) is limited by the large degeneracy among parameters and lack of solid inputs from theoretical considerations.

Afterglow emission studies have not experienced relevant progresses in the last years, with observations and techniques which are the same since the launch of the Swift satellite.
The recent discovery of TeV radiation from GRBs is opening the possibility to renovate and boost afterglow studies, with major impacts on the general understanding of GRB sources.

In this section we list and comment on those aspects still lacking a clear explanation, and in particular we selected topics which might largely benefit from observations and detections in the VHE regime.

\subsection{X-ray flares}
Observations of the afterglow emission in X-ray and optical often display behaviours that are not predicted by the standard scenario, and require the inclusion of additional emission components contributing to the detected radiation. In the standard external forward shock scenario the afterglow light-curves in X-ray and optical band are expected to decay following a power-law or a broken power-law behaviour, where the breaks are interpreted as the cooling or injection frequency crossing the observed band {\cite{saripiran,PK00,granot_breaks}}. 
The advent of Swift-XRT and the increasing number of optical follow-up observations performed by ground based robotic telescopes have highlighted the presence, in a good fraction of cases, of unexpected features in the early time afterglow, such as flares and plateaus{ \cite{flare_centralengine_zhang,nousek06}}.

Flares are episodes of sudden rebrightenings characterised by a very fast rise of the flux, followed by an exponential decay profile.
Comprehensive studies of X-ray afterglows show that an X-ray flare is observed in $\sim33\%$ of the GRBs \cite{chincarini_flares,margutti_flares}. 
The times at which they are observed span a very wide range, from around $\sim 30$\,s up to $\sim 10^6$\,s after the trigger time. The time where the flare peaks is shown in Figure~\ref{fig:peakflux_flares} ($T_{pk}$, $x$-axis) for a large sample of 468 X-ray flares in long GRBs. Most of the flares occur within $10^3$ seconds, even though there are many cases of flares occurring several hours after the burst.
The width of the flare $\omega$ is found to evolve linearly with time to larger values following the trend $\omega \sim 0.2 T_{pk}$ \cite{chincarini_flares}. 
The average and peak luminosities $L$ of the flare also display a dependence from $T_{pk}$, with $L \propto T_{pk}^{-2.7}$ at least for early time ($T_{pk} < 10^3$\,s) flares  \cite{chincarini_flares,margutti_flares}.
When including also late time flares \cite{bernardini_latetime_flares,yi_flares_catalog} a {shallower} index is obtain, around $\sim -1.2$.
The energy emitted during flare episodes is quite large and, for early time flares, is around $\sim 10\%$ of the prompt emission or sometimes even comparable \cite{beniamini_xray_flares}.

Flares have been detected also in the optical, although the sample of optical flares is far smaller than the X-ray one {\cite{optical_flares}}.
A statistical study of optical flares detected by Swift/UVOT shows that most of them correlate with and share similar temporal properties to simultaneous X-ray flares. Nevertheless, there are a few dozen of GRBs for which no X-ray flaring activity is observed simultaneously with optical flares {\cite{optical_flares}}.

Flares are believed to have an inner origin and to be associated with a prolonged activity of the GRB central engine {\cite{flare_centralengine_burrows,willingale10,flare_centralengine_fan,flare_centralengine_lazzati,flare_centralengine_maxham,flare_centralengine_zhang}}.
However, the relatively long timescales on which they are detected represent a challenge for the model.
Many questions are still open, such as the location of the emitting region, what is powering the flares, and whether late time flares have a different origin than flares detected at early times.

Speculations about possible signatures of X-ray flares in the GeV-TeV range are present in literature \cite{wang_gev_tev_flare,lat_flares1,latflares2,latflares3}. Assuming that flares have an internal origin and are produced at $R<R_{dec}$, forward shock electrons will be exposed to the flare radiation, producing an IC emission component by up-scattering the flare photons. 
{Following these estimates, the} IC component peaks at $\sim100$\,GeV and has a flux comparable to the X-ray flux.
Alternatively, GeV-TeV radiation associated to flares can be produced by the SSC mechanism, where electrons responsible for X-ray synchrotron flares also upscatter these photons to higher energies. The process is considered less interesting for TeV radiation because the peak of this SSC component is expected to be around 1\,GeV \cite{wang_gev_tev_flare}, due to a relatively low minimum Lorentz factor $\gamma_{min}\sim60$. {Such value is estimated from theoretical considerations where $\gamma_{min} \simeq 60 \epsilon_{e,-1} (\Gamma_{IS} - 1)$ for $p = 2.5$, $\epsilon_e = 0.1$ and a relative shock Lorentz factor $\Gamma_{IS}$ of the order of unity.}
We notice that the recent estimates of the minimum electron Lorentz factor in the late prompt emission of GRBs \cite{oganesyan19} may modify these predictions, and place the expected SSC around $100$\,GeV.
The luminosity of this component will strongly depend on the size of the emitting region.
As a result, detection of flares in GeV-TeV band can provide relevant information to identify the properties at the emitting region and the production site of the flaring activity.

To understand what are the chances of current and future VHE ground based instruments to contribute to the study of flares, we perform some simplified estimates.
The MAGIC telescopes observed 138 GRBs in almost $\sim 16.5$ years, from 2005 up to June 2021 \cite{Berti_MG_MAGIC}. 
More than half of them (74 events) have been observed with delays from  shorter than $10^3$\,s, which means $\sim 4.5$ GRBs\,yr$^{-1}$, and 37 events observed with delays shorter than 100\,s (i.e. 2.2\,GRB\,yr$^{-1}$).
Considering that $\sim33\%$ of the long GRBs have an X-ray flare
and considering the distribution of their peak times (see Figure~\ref{fig:peakflux_flares}), we estimate that $\sim 1$ GRBs/yr is the rate of GRBs with an X-ray flare occurring during MAGIC observations. 

%---------------------------
\begin{figure}
   \begin{center}
    \includegraphics[width=0.8\textwidth]{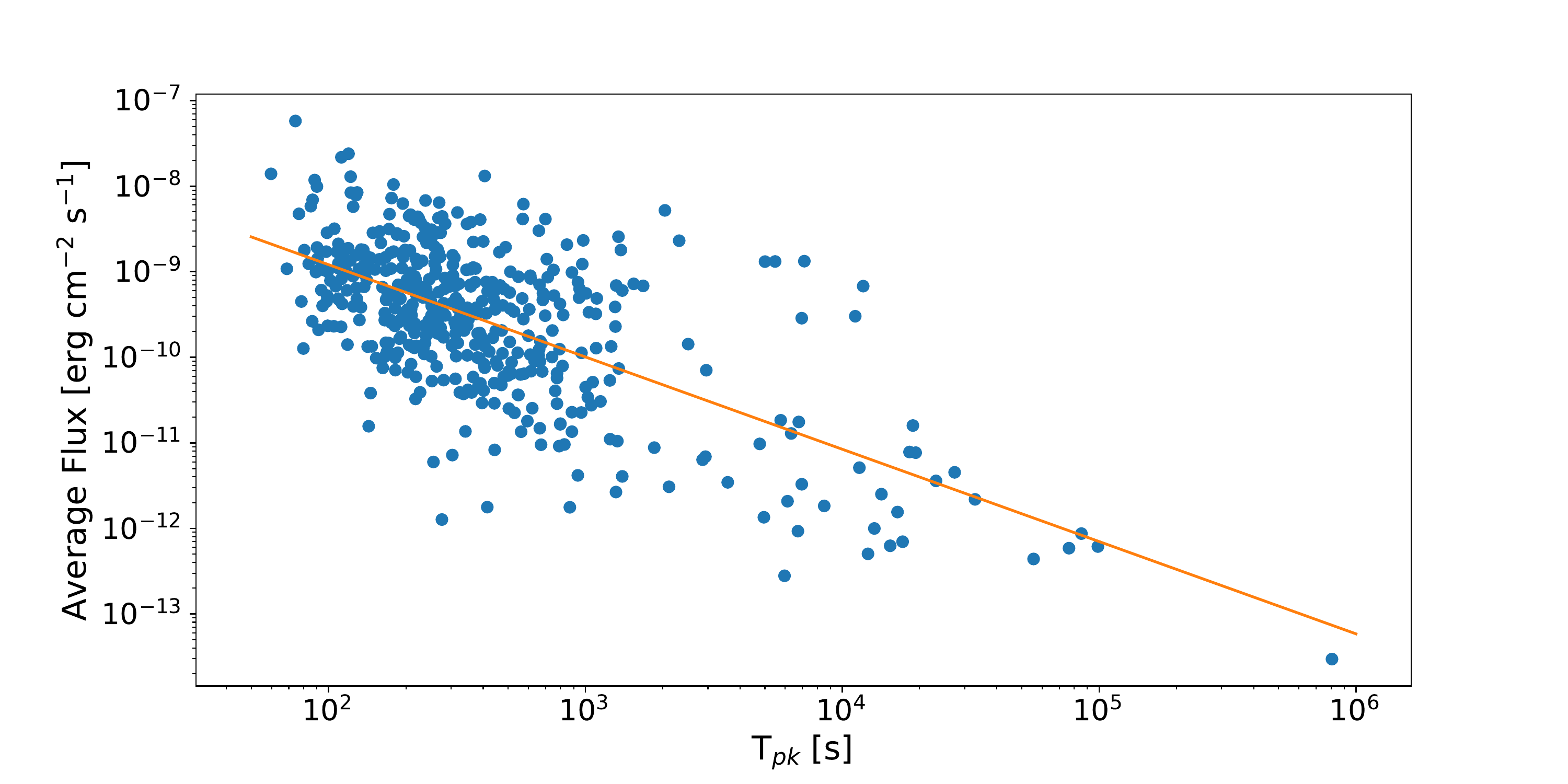}
    \caption{Average fluxes of Swift-XRT flares in the 0.3-10 keV energy range versus peak times $T_{pk}$. The blue points are the 468 flares observed by Swift from April 2005 to March 2015 (collected from \cite{yi_flares_catalog}). The orange line is the resulting linear regression which gives the expression: $\log{F_{av}} = -6.76 -1.08 \times \log{T_{pk}}$.}
    \label{fig:peakflux_flares}
    \end{center}
\end{figure}
%---------------------------

Let us go a bit further and estimate the detectability of a 
putative $\sim10^2$\,GeV counterpart of X-ray flares.
For the flux of the GeV-TeV flare, we consider as reference value the X-ray flux, and discuss what happens if a similar or ten times smaller flux is emitted at $\sim10^2$\,GeV.

We collect the X-ray flux of a large sample of flares from the catalog of X-ray flares presented in \cite{yi_flares_catalog}. 
Results are shown in Figure~\ref{fig:peakflux_flares}.
The {average flux of the flare and the flare peak time} correlate, and the orange line represents the best fit.
{To perform the estimates, we consider} two different flare peak times, $T_{pk} = 10^2$\,s and $T_{pk} = 10^3$\,s. The typical average fluxes at those times are $F=1\times 10^{-9}$\,erg cm$^{-2}$\,s$^{-1}$ and $F=1\times 10^{-10}$\,erg\,cm$^{-2}$\,s$^{-1}$, respectively.
Assuming that a similar amount of flux is emitter around $100$\,GeV we can compare these values with the differential sensitivity as a function of the observing time of IACT instruments.
Figure~\ref{fig:diffsensitivity_transient} (\cite{iact_sensitivity_transients}) shows the sensitivity for several telescopes to the detection of a point-like source at 5 standard deviations significance as a function of the exposure time and for four selected energies. 
Considering that the width of the flare is related to the peak time following the relation $\omega \sim 0.2 T_{pk}$, we can compare the  flare fluxes estimated at $T_{pk} = 10^2$\,s and $T_{pk} = 10^3$\,s with the differential sensitivity for observing time of $t_{obs} = 20$ s and  $t_{obs} = 200$\,s. 
The flare fluxes lie close to the differential sensitivity of the MAGIC telescopes (for 100 GeV at $t_{obs} = 20$\,s is $\sim$ $1.0 \times 10^{-9}$ erg cm$^{-2}$ s$^{-1}$ and at $t_{obs} = 200$\,s is $\sim$ $5.0 \times 10^{-10}$ erg cm$^{-2}$ s$^{-1}$). This indicates that MAGIC telescopes can barely detect such a flare. Moreover, {Extragalactic Background Light (EBL)} attenuation reduces the flux, that is why we are making the estimates at $100$\,GeV, where the attenuation is still small. We conclude that MAGIC would be able to detect (or place relevant constraints on) only the brightest X-ray flares (as it can be seen from Figure~\ref{fig:peakflux_flares}, the correlation has a large spread, and flares at $10^2$ or $10^3$\,s can easily have fluxes one order of magnitude larger than what assumed here).

Concerning future instruments, the {Cherenkov Telescope Array (CTA\endnote{\url{https://www.cta-observatory.org})} will have a sensitivity which is almost one order of magnitude lower than the MAGIC one and similar slewing capabilities.
The same estimates done for MAGIC can be applied to CTA, with the advantage that CTA will have a northern and southern sites, approximately doubling the possibility to follow GRBs within short time-scales.
This is a promising indication that the CTA array will be potentially able to detect possible counterpart at $E\sim30$\,GeV of X-ray flares, provided that this counterpart has a flux is which no less than ten times smaller than what detected in X-rays. As a result, it can play a major role in exploring and improving our knowledge of flares and their connection with prompt emission and with the prolonged activity of the central engine. 

%---------------------------
\begin{figure}
   \begin{center}
    \includegraphics[width=0.7\textwidth]{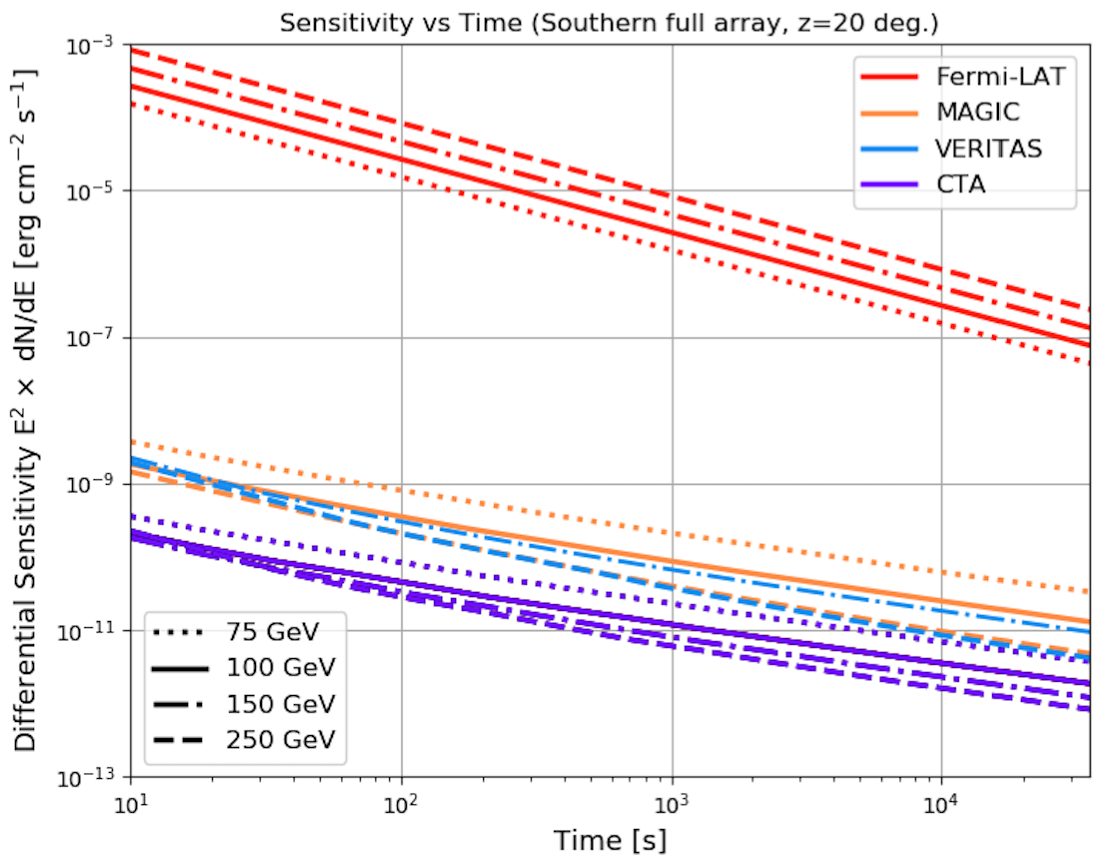}
    \caption{Differential sensitivity as a function of the observation time for several HE and VHE instruments (Fermi-LAT, MAGIC, VERITAS and CTA) at four selected energies (75, 100, 150, 250 GeV). From \cite{iact_sensitivity_transients}. }
    \label{fig:diffsensitivity_transient}
    \end{center}
\end{figure}
%---------------------------

%-----------------------------------------------------

\subsection{Density profile of the external medium}\label{subsec:density_profile}
Following the established connection between long GRBs and the core-collapse of massive stars, the jet is expected to produce the afterglow while propagating in the wind of the star in its free-streaming phase. Afterglow radiation of long GRBs should then be produced in the interaction with a medium with radial density profile $n\propto R^{-2}$. However, several investigations have shown that about half of the long GRBs are better explained if the blast-wave is assumed to run into a medium with constant density. We revise the evidence in support of the constant density medium and discuss the difficulties in reconciling these observations with expectations on the environment surrounding long GRB progenitors.

Long GRBs originate from the core-collapse of massive stars, most likely rapidly rotating, and with a possible evidence of a preference for low-metallicity.
The most convincing evidence in support of this paradigm is the association with type Ic supernovae and the proximity of GRBs to young star-forming regions.
{While} the connection of long GRBs (or at least with the bulk of the population) with the core-collapse of massive stars is solid, the role of metallicity and rotation in the launch of the GRB jets, and the identification of the progenitor star are still uncertain. 
The progenitor is usually identified with Wolf-Rayet stars, massive stars ($M>20\,M_\odot$) in the final stages of their evolution, characterised by powerful winds and a high mass loss rate {\cite{wolfrayet_progenitor_woosley}}.
The wind from the star is expected to interact and deeply modify the environment where the GRB explodes and leave imprints on its afterglow emission.

More in detail, from the interaction between the stellar wind and the ISM four concentric regions with different properties are expected to form. In the inner part (i.e., close to the star) the circumburst medium is permeated by the free-streaming wind, producing a density with radial profile $n\propto R^{-2}$. The density is related to the mass loss rate $\dot M$ and to the velocity $v_w$ of the free-streaming stellar wind by:
\begin{equation}
    n(R) = \frac{\dot M}{4\,\pi\,R^2\,m_p\,v_w}~.
    \label{eq:density_wind}
\end{equation}
A termination shock separates the unshocked from the shocked wind: the latter forms a hot bubble of thermalised wind material, with a nearly constant density profile, as the formation of pressure and density gradients is prevented by the high sound speed inside the bubble.
The hot bubble, in its outer part, is enclosed by a shell of shocked ISM, surrounded by the unshocked ISM.
The GRB jet is supposed to trill its way in this stratified medium {\cite{chevalier99}}. 

To understand where most of the afterglow evolution occurs, we have to estimate the deceleration radius $R_d$ and the non-relativistic radius $R_{NR}$ (i.e. the radius where the blast-wave has decelerated to non-relativistic velocity) and compare them to the termination shock radius.
For typical parameters ($\dot M=10^{-5}M_\odot$\,yr$^{-1}$ and $v_w=10^3$\,km\,s$^{-1}$), the fit to numerical models of Wolf-Rayet stars \cite{fryer06} give the following relation between the termination shock radius and the density of the unshocked ISM: $R_{T}=10\,n_{ISM}^{-1/2}$\,pc~,
where $n_{ISM}$ is the density of the unshocked ISM.
From the blast-wave dynamics, the deceleration and the non-relativistic radius are $R_d=6\times10^{-5}\,E_{52}\,v_{w,3}/(\dot M_{-5}\,\Gamma_{0,2})$\,pc and $R_d=0.6\,E_{52}\,v_{w,3}/\dot M_{-5}$\,pc, respectively.
It is evident how the complete evolution of the afterglow radiation occurs well inside the free-streaming region.

In afterglow modeling of long GRBs it is then customary to assume a density profile described by eq.~\ref{eq:density_wind}, where $\dot M$ and $v_w$ are treated as unknown parameters (normalised to the typical values of a Wolf-Rayet star) combined in one single free model parameter $A_\star$: $n(R)=3\times10^{35} A_\star R^{-2}$.
Despite this robust prediction, the modeling of afterglow observations shows that in a relevant fraction of cases, observations are better explained by adopting a circumburst medium with a constant density $n=n_0$.

The fraction of this cases varies depending on the method and on the selected sample, and is on average about 50\% \cite{panaitescu02,schulze11,li15,gompertz18}.

To place the termination shock at least inside the non-relativistic radius, one should invoke a very large density of the ISM, $n\gtrsim10^5$\,cm$^{-3}$, typical of dense cores of molecular clouds: $R_{T}=0.03\,(n_{ISM}/10^5$cm$^{-3})^{-1/2}$\,pc. Density profiles for different ISM densities are shown in Figure~\ref{fig:bubble}, upper panel.
Alternatively, one can try to variate the wind parameters. How the termination shock radius changes for different values of $\dot M$ and $v_w$ is shown in the bottom panel of Figure~\ref{fig:bubble}. A very low mass loss rate $\dot M=10^{-7}M_\odot$yr$^{-1}$ (which may find a justification in case of low-metallicity star) is needed to bring the termination shock radius below 1\,pc (for $n_{ISM}=10$\,cm$^{-3}$).
With this low mass{-loss} rate, the deceleration and non-relativistic radius increase ($R_{d}\sim6\times10^{-3}$\,pc and $R_{NT}\sim60$\,pc), placing the termination shock still after the deceleration radius but well within the non-relativistic radius, allowing for part of the observed emission to develop into a constant density environment. By increasing the blast-wave energy, the deceleration radius can further approach $R_T$. This suggests that it is more likely for a very energetic GRB to cross the termination shock at early times and then expand in a ISM-like medium, as compared to a faint GRB. An indication of an average larger $E_\gamma$ in GRBs with a wind-like medium as compared to GRBs with a ISM-like medium has been found in \cite{vanmarle06}, but is in contrast with results from the study performed by \cite{gompertz18} on a larger sample.

%---------------------------
\begin{figure}
   \begin{center}
    \includegraphics[width=0.6\textwidth]{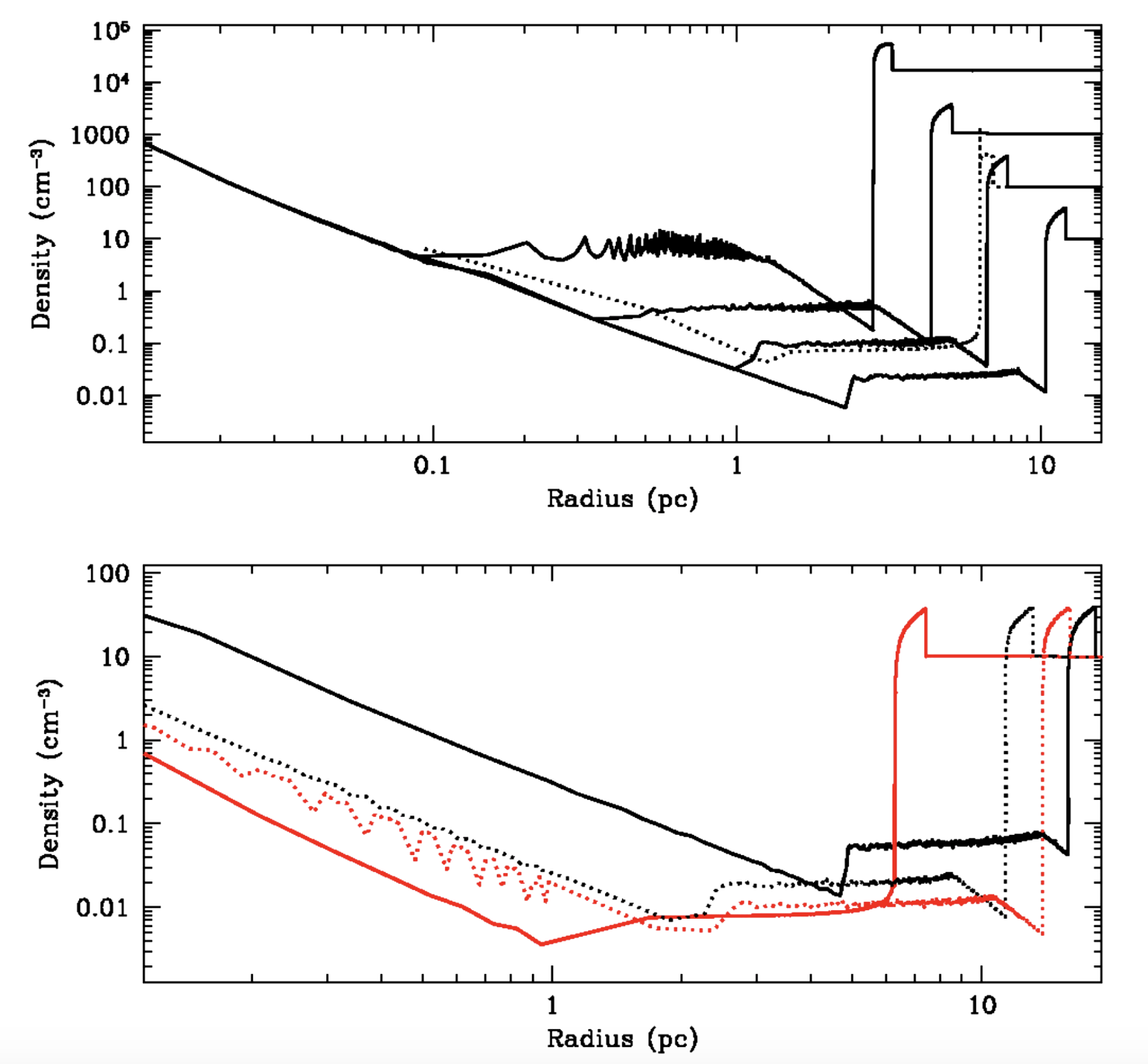}
    \caption{Density profile produced by a star wind as a function of the distance from the central object. The $R^{-2}$ profile characterises the region where the wind freely streams, which is separated from the shocked wind (with nearly constant density) by the termination shock. Top panel: the impact of different values of the ISM density on the termination shock is shown. For all the curves, it is assumed $\dot M=10^{-5} M_\odot$\,yr$^{-1}$ and $v_w=10^3$\,km\,s$^{-1}$, while the ISM density varies from 10 to $10^4$\,cm$^{-3}$, with the termination shock moving to smaller distances when the density increases. Bottom panel: for a fixed ISM density $n_{ISM}=10$\,cm$^{-3}$, the impact of different $\dot M$ and $v_w$ is shown. Black-solid: $\dot M=10^{-4} M_\odot$\,yr$^{-1}$ and $v_w=10^3$\,km\,s$^{-1}$; red-solid: $\dot M=10^{-6} M_\odot$\,yr$^{-1}$ and $v_w=10^3$\,km\,s$^{-1}$; black-dotted: $\dot M=10^{-5} M_\odot$\,yr$^{-1}$ and $v_w=500$\,km\,s$^{-1}$; red-dotted: $\dot M=10^{-5} M_\odot$\,yr$^{-1}$ and $v_w=2\times10^3$\,km\,s$^{-1}$. {From \cite{fryer06}. \copyright AAS. Reproduced with permission.}}
    \label{fig:bubble}
    \end{center}
\end{figure}
%---------------------------

The parameter space for which part of the afterglow emission can indeed be produced in the ISM-like density profile of the shocked wind is very limited, as it corresponds to the most energetic GRBs, low-metallicity progenitors and high-density ISM, or a combination of these factors \cite{vanmarle06}. 
These considerations on diversity of $E_k, \dot M, v_w$ and ISM density {may not be} sufficient to explain the results of the modeling {(i. e., the preference for a ISM-like environment). The} fraction of GRBs {which} might have these peculiar parameters can {hardly} account for the {large} fraction of GRBs for which a wind-like profile is excluded by observations. {The required conditions are too extreme to be verified in half of the population. However, it is not clear if this percentage has been overestimated by present studies.
To quantify the inconsistency, the first step would be} to perform a dedicated study of afterglow emission to assess the percentage of long GRB afterglows that are not consistent with a wind-like environment.

Methods based on closure relations may not be valid if the spectrum is modified by Compton scattering in Klein-Nishina regime {(see also \cite{closure_relations})}. Moreover these are based on simple approximation of the synchrotron spectrum into power-law segments, while the wide curvature of the real synchrotron spectra might lead to an incorrect estimates of {the value of $p$} if the observed frequency is in the vicinity of a synchrotron break frequency. A full modeling is then necessary to really assess the fraction of long GRBs for which an $R^{-2}$ density profile is excluded, and  ultimately understand if the paradigm for the environment of GRBs should be drastically modified.
Radio observations may be of great help, since the flux temporal behaviour does not depends on $p$ and is quite different in case of constant or wind-like density profile. Similarly, the detection of SSC radiation can help solving this ambiguity.

%----------------------

\subsection{Small values of $\epsilon_B$}\label{subsubsec:b}
For a long time the typical value of $\epsilon_B$ has been considered to variate between 0.01 and 0.1, both on the basis of theoretical considerations on particle acceleration and findings by numerical simulations. Indeed, the present understanding of the micro-physics at weakly magnetised shocks invoke the existence of self-generated micro-turbulence both behind and in front of the shock, at a level corresponding to $\epsilon_B\sim 0.01-0.1$. This layer of intense micro-turbulence is expected on theoretical grounds and recently corroborated by numerical PIC simulations.
Inferences of the value of $\epsilon_B$ from early modeling of afterglow radiation were broadly consistent with these numbers, confirming the presence of a large self-generated fields in ultra-relativistic weakly magnetized shocks.
More recently, several independent methods have provided evidence for significantly lower values. 

In particular, several studies on GRBs with GeV temporally extended emission detected by LAT arrive to the same conclusions that in order to explain GeV radiation as part of the synchrotron emission, multi-wavelength observations requires $\epsilon_B=10^{-6} - 10^{-3}$ \cite{kumar09,kumar10,barniolduran11,he11,beniamini16}.
Similar values have been inferred from studies that are based on radio, optical and/or X-ray emission and do not make use of high energy emission, such as \cite{barniolduran14,santana14,wang15,zhang15}.
A smaller magnetic field in the region where most of the particle cooling occurs might increase the expected relevance of the SSC component, as supported by recent detections of TeV radiation by IACTs.

Such a small values of $\epsilon_B$ may appear to be problematic \cite{pirannakar}, because strongly self-generated micro-turbulence must be present to ensure the scattering and acceleration of the particles, which otherwise would be simply advected away.

It was later pointed out that the inferred low values of magnetization might be indicative of a turbulence that is decaying on time-scales comparable with the electron cooling time \cite{lemoine13,lemoine+13}.
From a theoretical perspective, indeed, the micro-turbulence is expected to decay beyond some hundreds of skin depths. This picture has been validated by PIC simulations, which however are still far from probing time-scales comparable with the dynamical time-scale of the system. Dedicated simulations the magnetic field does decay behind the shock, on a time-scale much longer than $c\omega_{pi}$. Immediately behind the shock, the magnetic field carries a magnetization $\epsilon_B\sim 0.01$, which decays in time after $10^2-10^3$ plasma times. Eventually, the magnetic field will settle to the shock-compressed value $4\Gamma\,B_u$, where $B_u$ is the magnetization of the upstream unperturbed medium. 
In this scenario, high-energy particles, which produce MeV-GeV photons, feel only the region close to the shock, where the magnetization is large, due to their short cooling time. Particles that cool on longer time-scales (and produce radio, optical and X-ray photons) cool on longer time-scales, and then in a region where the magnetic field has decayed.

The application of cooling in a decaying magnetic turbulence to four GRBs detected by LAT has proved to be very successful and even able to give indications on how fast the turbulence decays, being consistent with a power-law decay $\epsilon_B\propto t^{-\alpha_t}$ with $\alpha_t\sim0.5$ \cite{lemoine+13}.

To understand and constrain the value of the magnetic field relevant for the particle cooling is of great importance, since an incorrect assumption or prior affects the estimates of all the other afterglow parameters, and in particular the density of the external medium {\cite{santana14,gompertz18}.} 

A low value of $\epsilon_B$ tends to increase the level of SSC luminosity for a given synchrotron luminosity. The recent detection of bright TeV emission from the afterglow of GRBs is an indication that this might indeed be the case.
Existing and future TeV observations will shed a light on this issue, fostering a revision of our prejudice on the value of the magnetic field in the region where particles cool.

%----------------

\subsection{Variation of micro-physical parameters with time} % 14C
Thanks to the increasing number of available observations on a wide range of frequencies (from radio to TeV) and times (from seconds to weeks), the basic assumption that micro-physical parameters (such as $\epsilon_e, \epsilon_e, p$ and $\xi_e$) are constant over the whole afterglow evolution can be testes. We comment on the hints (inferred from afterglow modeling) for temporal evolution of these parameters.

In case of well-sampled multi-wavelength light-curves, the modeling with synchrotron spectra is able not only to identify the location of the spectral breaks at a certain time but also evaluate their evolution in time. As a result, hints that micro-physical parameters $\epsilon_e$ and $\epsilon_B$ may vary with time have been found in some events with well detailed multi-wavelength follow-up campaigns. 

In \cite{filgas_timedependent} broad-band (from near infrared up to X-ray) afterglow data from GRB\,091127 were interpreted in the standard external forward shock scenario assuming a constant-density medium. The good quality of the data allows to identify the breaks in the light-curves and associate them with the synchrotron spectral breaks. As a result, the time evolution of the synchrotron breaks was estimated. In particular, it was calculated that the cooling break frequency $\nu_{cool}$ evolves as $\nu_{cool} \propto t^{-1.2}$ which is in contrast with synchrotron predictions for which a less steeper decay $\nu_{cool} \propto t^{-0.5}$ is expected. As a result, some assumptions of the standard model must be relaxed to remove the tension between  observations and theoretical predictions. 
The most likely option able to explain the cooling break observational behaviour without affecting the general interpretation of the data is to let the $\epsilon_B$ parameter variate with time. Assuming that $\epsilon_B \propto t^{0.49}$ the time evolution of $\nu_{cool}$ can be explained successfully. 

In \cite{vanderhorst_130427A} for GRB\,130427A modeling, in order to explain the observed fast evolution of the injection frequency $\nu_m \propto t^{-1.9}$ a temporal evolution of $\epsilon_e$ is claimed. Considering that $\nu_{m} \propto \epsilon_e^2$, a modest evolution of $\epsilon_e$ following the trend $\epsilon_e \propto t^{-0.2}$ is able to satisfactorily describe the observed light-curves. 

A time-dependent evolution of the micro-physical parameters has also been proposed in order to explain the features observed in the early afterglow phase which are not predicted by the external forward shock scenario such as  X-ray afterglow plateaus, chromatic breaks, and afterglow rebrightenings \cite{ioka_timedependent,kong_timedependent,huang_120729A,panaitescu_timedependent}.

Information from TeV observations can be certainly exploited in order to reduce the uncertainty on the values of the micro-physical parameters. The expansion of the broad band afterglow observations to a new spectral window will be a further test and a challenge for the multi-wavelength modeling based on the standard external forward shock scenario. In particular, the time evolution of the different energetic components including also TeV emission will give new insights useful to investigate the evolution of the micro-physical parameters. 
{A first proof is provided by the well-sampled multi-wavelength emission observed for GRB\,190114C, one of the few GRBs detected so far at TeV energies. The broadband emission can be explained only by invoking evolution of the micro-physical parameters with time \cite{misra_time_evolution_parameters}, as will be discussed in the next section.}

%-----------------------------------------------------

%------------------------------------------------------------
\subsection{Maximum synchrotron photon energy}
\label{subsec:max_syn_energy}
One of the expectations from Fermi-LAT observations of GRB afterglows was the identification of a spectral cutoff in the afterglow synchrotron spectrum{ marking the maximum energy of synchrotron photons \cite{guilbert83,dejager92}}. Such a cutoff has not been firmly identified. 
Its location is directly connected with the shock micro-physics conditions and the maximum energy at which electrons can be accelerated. 
This maximum energy is typically estimated equating the timescale for synchrotron cooling and the acceleration timescale, where acceleration is assumed to proceed at the Bohm level, considered as the fastest rate.
This estimate returns hence the maximum energy of the accelerated particles.
Assuming that the accelerated particles are electrons:
%-----------------------------
\begin{equation}
   t'_{L} = \frac{r_L}{c} = \frac{\gamma m_e c}{e B'} 
\end{equation}
%-----------------------------
where $r_L$ is the Larmor radius. For each crossing the electrons gain energy by a factor $\sim 2$. On the other hand, the energy losses for synchrotron radiation on this timescale are:
%-----------------------------
\begin{equation}
    \delta E' = t'_L P' = \frac{\gamma m_e c}{e B'} \frac{\sigma_T c \gamma^2 B'^2}{6 \pi} = \frac{1}{6\pi}\frac{\sigma_T m_e c^2 B' \gamma^3}{e}
\end{equation}
%-----------------------------
The particle stops to gain energy when:
%-----------------------------
\begin{equation}
    \delta E' = \gamma m_e c^2
\end{equation}
%-----------------------------
Therefore, the maximum Lorentz factor for electrons can be derived:
%-----------------------------
\begin{equation}
    \gamma_{max} = \sqrt{\frac{3\pi e}{\sigma_T B'}}
    \label{eq:gamma_max}
\end{equation}
%-----------------------------
The corresponding maximum synchrotron photon energy is:
\begin{equation}
    h\nu'_{max} = \frac{e B' \gamma^2_{max} h}{2 \pi m_e c}
    \label{eq:energy_max_sync}
\end{equation}
which for electrons is $\sim$ 50 MeV in their rest frame.
 
Similar considerations can be done also for protons. Following the same arguments presented below one obtains:
%-----------------------------
\begin{equation}
    \gamma_{cool,p} = \frac{6 \pi m^3_p c}{\sigma_T m^2_e B'^2 t'}
    \label{eq:gamma_cool_protons}
\end{equation}
%-----------------------------
for the cooling Lorentz factor and:
%-----------------------------
\begin{equation}
    \gamma_{max,p} = \sqrt{\frac{3\pi e m^2_p}{\sigma_T m^2_e B'}}
    \label{eq:gamma_max_proton}
\end{equation}
%-----------------------------
for the maximum Lorentz factor which sets a maximum photon energy of $\sim$ 100\,GeV. Synchrotron emission is less efficient for protons so they are less affected by cooling and they can reach higher maximum Lorentz factors than the electrons. 

Within this framework it is expected that observations in the GeV band can be exploited to identify the presence of a cut-off in the HE tail domain. At the current stage, Fermi-LAT observations indicate that the afterglow component of the HE energy emission is usually modelled with a single power-law component with index $\sim$ -2 and with no evidence of spectral evolution in time \cite{2nd_lat_catalog} and HE cut-offs. 

The absence of the cut-off in the observational data may be explained in several ways. The most likely interpretations are the limited sensitivity of the LAT instrument in the GeV range and the possible contamination due to the rising of the SSC spectral component. As a result, the synchrotron cut-off is hidden behind the VHE spectral component which can be detected in the GeV-TeV domain. This implies that TeV observations are fundamental in order to disentangle between the two spectral components and infer the cutoff of the synchrotron spectrum. 

Another possible interpretation is that the {lack of a} cutoff in the observational data is genuine. In this case, the synchrotron emission can exceed the limit assumed for the maximum photon energy and extend in the GeV-TeV domain. This interpretation can be tested with VHE observations. The extension of the HE power-law derived by LAT up to the TeV domain should be consistent with VHE data and no spectral hardening in the GeV-TeV band should be seen. Such scenario is in contrast with the standard particle acceleration model presented in Section~\ref{chap:model}. A deep revision of our understanding of acceleration mechanisms is required in order to make TeV emission from synchrotron radiation possible. In particular, a mechanism, able to accelerate electrons up to PeV energies is needed. 

Calculation performed so far assumes the presence of a uniform magnetic field $B'$ throughout shock heated plasma. If this assumption is rejected it is possible to consider a non-uniform magnetic field, stronger close to the shock front and decaying downstream. Following the calculation of \cite{kumar_maximum_sync} the magnetic field $B'$ can be expressed in terms of the distance from the shock front $x$ as:
\begin{equation}
    B'(x) = B'_{s} \bigg(\frac{x}{L_p}\bigg)^{-\eta} + B'_{w}~,
\end{equation}
where $B'_s$ and $B'_w$ are respectively the strongest and the weakest magnetic field strengths{, $\eta$ is the power-law decaying index,} and $L_p$ is the field decay length scale, which is estimated as \cite{medvedev_loeb}:
%---------------------------------
\begin{equation}
    L_p = \Bigg[\frac{m_p \Gamma_s c^2}{4 \pi n e^2}\Bigg]^{1/2} = 2.2 \times 10^{7} \frac{\Gamma_s}{n} \hspace{0.3cm} \text{cm}
\end{equation}
%---------------------------------
where $\Gamma_s$ is the shock front Lorentz factor and $n$ is the number density of the accelerated particles in the shocked fluid comoving frame. As a consequence, the Larmor radius $r_L$ increases with the distance from the shock front since $B'(x)$ becomes weaker and an electron travelling downstream will be likely sent back upstream when $r_L \leq x$. When considering the case $B'_s \gg B'_w$ and $x \gg L_p$ the particles will lose most of their energy in the region of low magnetic field. Therefore from the condition that losses in the low magnetic field region should be greater than losses in the high magnetic field region, after some algebra the following condition is obtained:
%---------------------------------
\begin{equation}
    \bigg(\frac{B'_s}{B'_w}\bigg)^2 \lesssim \frac{r_l}{L_p}
    \label{eq:mf_time_dependent}
\end{equation}
%---------------------------------
valid for $\eta > 1/2$ and $x_0/L_p \gg 1$ where $x_0$ is the width of the high magnetic field region. Considering that $x_0/L_p \equiv (B'_s/B'_w)^{1/\eta}$, eq. \ref{eq:mf_time_dependent} states that the Larmor radius in the high magnetic field region is larger than the actual width of the region and electrons will be barely deflected in such portion of the shocked plasma. As a result, it is possible to calculate the maximum Lorentz factor for electrons that loose most of their energy in the weak magnetic field region following the same conditions presented for the uniform magnetic field case:
%---------------------------------
\begin{equation}
    \gamma_{max} = \sqrt{\frac{3 \pi e}{\sigma_T B'_w}}
\end{equation}
%---------------------------------
As a result, the maximum synchrotron photon energy is given by:
%---------------------------------
\begin{equation}
    h\nu'_{max} = \frac{e \gamma^2_{max} h}{2 \pi m_e c} \bigg(\frac{B'_s}{B'_w}\bigg)
\end{equation}
%---------------------------------
which is greater than the one calculated in eq. \ref{eq:energy_max_sync} by a factor $B'_s/B'_w$. Numerical calculations \cite{sironi_spitkovsky} show that this ratio can be larger than $\sim 10^2$. As a result, photons of energies $\gtrsim 100$ GeV can be produced via synchrotron process when assuming a non-uniform magnetic field which decays downstream of the shock front and with particles loosing most of their energies in the weakest field region.

In both interpretations presented here, TeV observations are fundamental in order to investigate with unprecedented details the possible presence or absence of the synchrotron cutoff spectrum. This have also a direct impact on the study of the possible radiation mechanisms responsible for the VHE component in GRBs.

\subsection{Prompt emission efficiency}
The overall efficiency $\eta_\gamma$ of the prompt emission mechanism is the result of three processes: the efficiency $\eta_{diss}$ of the (still unidentified) mechanism responsible for dissipation of the jet energy, the efficiency $\epsilon_e$ of the acceleration mechanism in converting the dissipated energy into random energy of the electrons, and the radiative efficiency $\epsilon_{rad}$ of the electrons: $\eta_\gamma=\eta_{diss}\epsilon_e\epsilon_{rad}$.
Provided that it is reasonable to assume a fast cooling regime for the prompt emission ($\epsilon_{rad}=1$), the overall prompt efficiency is limited by the capability of the dissipation mechanism in extracting the kinetic or magnetic energy of the jet and the capability of the particle acceleration process to convey a fraction of this energy into the non-thermal electron population.
{The value of the efficiency }provides then a fundamental clue to place constraints on the origin of energy dissipation in GRBs, which is still quite uncertain, discriminating between matter and magnetic dominated jets.

From the definition of $\eta_\gamma=E_\gamma/E_0$ (where $E_0= E_\gamma + E_k $ is the initial explosion energy), we can write the relation $E_k = (1-\eta_\gamma)/\eta_\gamma E_{\gamma}$.
The parameter $\eta_\gamma$ can then be estimated from the comparison of the energy $E_{\gamma}$ emitted in the prompt phase and the energy $E_k$ left in the jet after the end of the prompt emission (i.e., at the beginning of the afterglow phase). While the {former} is directly estimated from observations, the latter one can be  inferred only indirectly, from the modeling of afterglow radiation.

One of the most adopted methods to infer $E_k$ for large samples of GRBs is to rely on the X-ray luminosity and use it as a proxy for the energy content of the blast-wave \cite{kumar00,berger03,lloyd04,berger07,davanzo12}. This method is solid as long as the X-ray band lies above $max(\nu_m, nu_c)$ and is not affected by inverse Compton cooling. If these two conditions are verified, then the electrons emitting X-ray photons are in fast cooling regime and their cooling is dominated by synchrotron losses. The luminosity produced is then proportional to the energy content of the accelerated electrons $E_k\epsilon_e$. Assuming a value (typically 0.1) for $\epsilon_e$, then $E_k$ can be estimated. 
Investigations based on the X-ray emission have inferred very large values of $\eta_\gamma$, between 0.5 and 0.9 \cite{lloyd04,granot06,nousek06,zhang07}. 

The very same approach can be applied also to 100\,MeV-GeV  photons detected by the LAT, under the assumption that these are synchrotron photons. 
A strong correlation between the GeV luminosity and $E_{\gamma,iso}$ has been indeed found, supporting the possibility that GeV photons lie in the high-energy part of the synchrotron spectrum, where the afterglow luminosity is proportional to $E_k\epsilon_e$ and can be use, similarly to the X-ray luminosity, to estimate $E_k$ \cite{nava14}.
A study by \cite{beniamini15} revealed that the energetics $E_k$ inferred independently from X-ray and GeV luminosities on a sample of 10 GRBs are inconsistent with each other. The {authors show that the }inconsistency is solved if $\nu_c>\nu_X$ {(where $\nu_X$ is the X-ray frequency),}  or if Compton losses are important in the X-ray band. Full modeling of the GeV, X-ray and optical data support this scenario. In both cases, $\epsilon_B$ is required to be much smaller than usually assumed, with values in the range $10^{-7}-10^{-3}$. This analysis shows that the GeV band is a much better proxy for $E_k$, since it is above $\nu_c$ and is not affected by inverse Compton cooling, due to Klein-Nishina suppression. Adopting GeV luminosities as a proxy for $E_k$, the estimated values of $E_k$ increase, affecting also the estimates of $\eta_\gamma$, which are around $5-10$\% \cite{beniamini16}.

A correct estimate of $\eta_\gamma$ is extremely important, since its value is related to the mechanism dissipating energy in the jet. Since internal shocks can hardly reach values{ of $\eta_\gamma$} larger than 10\%, values around 50-90\% have been used to argue that internal shocks are not a viable mechanism to explain prompt emission in GRBs, and more efficient mechanisms should be considered (e.g., magnetic reconnection). If the efficiency is however smaller than initially estimated, internal shocks may still be a viable solution. Moreover different estimates of $\eta_\gamma$ lead to different estimates on the total initial jet energy $E_0=E_{\gamma,iso}+E_k$, with repercussions on the energy budget of GRBs and finally on their progenitors and mechanisms for jet launching.
Small values of $\epsilon_B$ may then relax the problem with very large prompt efficiency, which is definitely unreasonable for internal shocks, but difficult to attain also for magnetic reconnection models {(for a discussion see e. g., \cite{zhang11,lazarian19,beniaminigranot}).} 

A scenario where the magnetic field strength is relatively low in the emitting region, implies a stronger SSC emission. Recent TeV detection of GRBs  support this scenario, and provide additional observations to constrain the magnetic field. Moreover, as shown by the first detections by IACTs (section~\ref{Chap:TeVdiscovery}), the energy in the TeV component is comparable to the energy in X-rays, providing better estimates for the energy budget in the afterglow phase. Future detections from a larger sample of GRBs can help in assessing more precisely the energy budget of the jet during the afterglow emission and add important information to constrain the efficiency of the prompt emission and favour or exclude some dissipation scenarios.

\subsection{Fraction of accelerated particles}

As described in section~\ref{chap:model}, the representation of the relativistic shock acceleration in GRB afterglow relies on some free parameters. These values describe the energy equipartition between particles and magnetic field and the non-thermal accelerated particle distribution. 
%----------------------------------
\begin{figure}
   \begin{center}
    \includegraphics[width=0.7\textwidth]{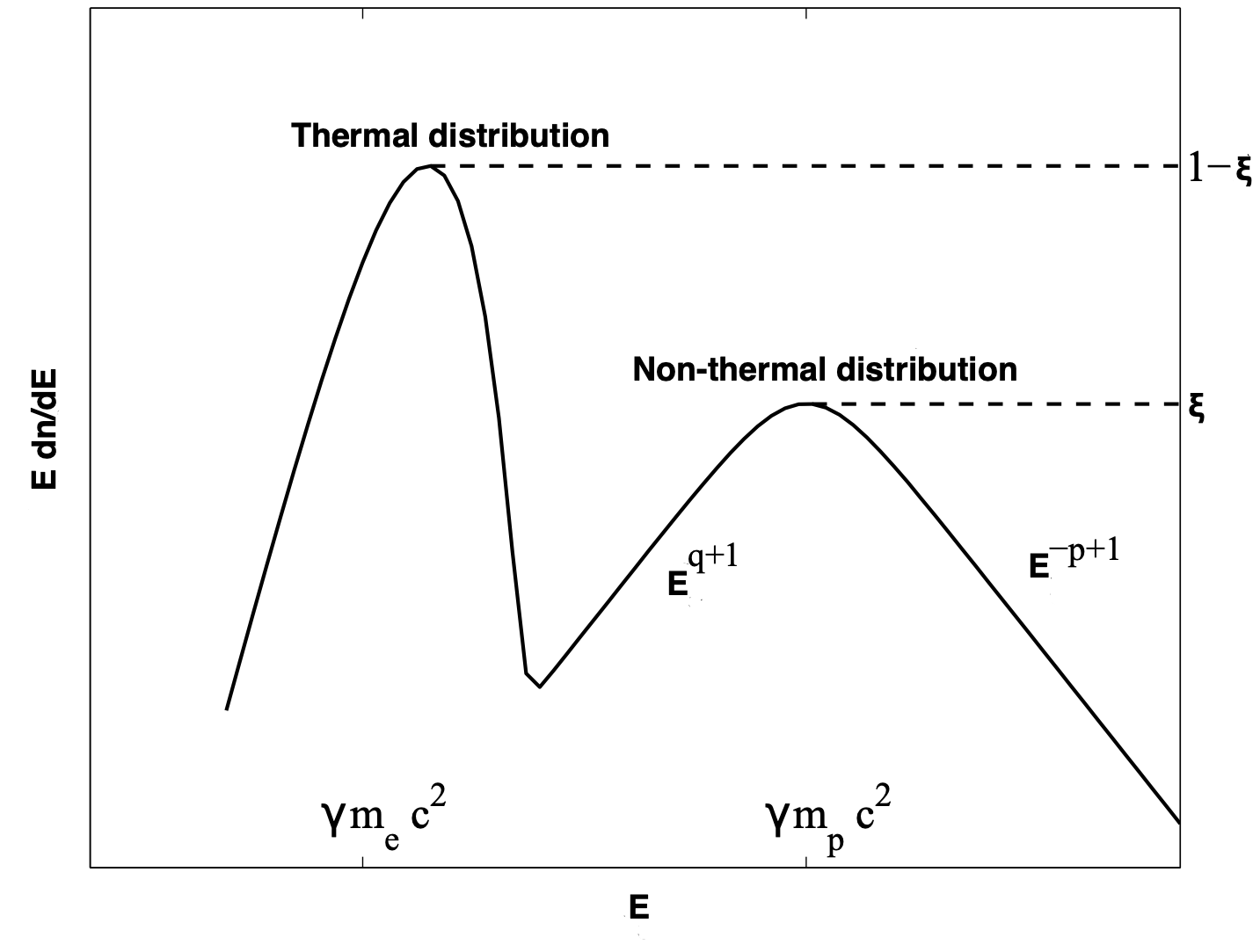}
    \caption{Distribution of accelerated electrons from a relativistic shock. The two bumps correspond respectively to the distribution {of the $(1 - \xi)$ fraction of particles heated }into a thermal component with post shock energy $\sim \gamma m_e c^2$ and {the fraction $\xi$ of particles accelerated }into a non-thermal component with post shock energy $\sim \gamma m_p c^2$. Adapted from \cite{eichler}. {\copyright AAS. Reproduced with permission.} }
    \label{fig:acc_distribution}
    \end{center}
\end{figure}
%---------------------------
In particular, the parameter $\xi_e$ is responsible to account for the fraction of particles (here we consider electrons but same considerations are valid also for protons) accelerated into a non-thermal distribution. This means that from relativistic shock theory it is expected that a fraction $1 - \xi_e$ of electrons is heated into a thermal distribution rather than a non-thermal one (see Figure~\ref{fig:acc_distribution}). For simplicity, usually in afterglow studies it is assumed that $\xi_e = 1$, i.e. all the particles are accelerated into a non-thermal distribution. Such assumption is used in order to avoid the large degeneracy which affects the GRB parameters when including this additional free value. In particular, afterglow modeling predictions obtained assuming $\xi_e = 1$ for parameters $E_k$, $n$, $\epsilon_e$ and $\epsilon_B$ cannot be distinguished from those one estimated for any $\xi_e$ in the range $m_e/m_p < \xi_e < 1$ and afterglow parameters $E'_k = E_k/\xi_e$, $n' = n/\xi_e$, $\epsilon'_e =\xi_e \epsilon_e$ and $\epsilon'_B = \xi_e \epsilon_B$ \cite{eichler}. This can be proven considering the dependencies of the jet dynamics and shock energy equipartition on the model parameters. As shown {in BM76 and} by previous calculations on the evolution of a relativistic blastwave, in the self-similar regime the bulk Lorentz factor evolve as $\Gamma \propto (E_k/n)^{1/2}$. As a result, the same flow evolution can be obtained with different values of $E_k$ and $n$ as long as their ratio is preserved. The fraction of energy given to the magnetic field is reduced by a factor $\xi_e$ but at the same time the energy density given to the shock is increased by a factor $1/\xi_e$ so that the magnetic field energy density is the same in the scenario when including or not $\xi_e$. Analogous considerations can be done also for the number and the energy density of the electrons so that their values are preserved. As a result, in principle it is not possible to distinguish between the two parameter sets obtained for $\xi_e$ or for any value $m_e/m_p < \xi_e < 1$. This imply that afterglow model parameters, when considering $\xi_e \neq 1$, are estimated with an uncertainty of factor $m_e/m_p$ and afterglow observations do not constrain directly their values (e.g. $E_k$ or $\epsilon_B$) but rather a fraction of their value multiplied or divided by $\xi_e$ (e.g. $E_k/\xi_e$ or $\epsilon_B \xi_e$).

It is possible to obtain information on the value of $\xi_e$ through PIC simulations or indirect features of the thermal component on the synchrotron afterglow spectra. As mentioned in the previous Section, PIC simulations of weakly magnetized shocks have found that the downstream population include around $\sim 3\%$ of the electrons which are accelerated into a non-thermal distribution. In case the efficiency is low (around $\sim 10\%$ or less) the presence of a large population of thermal electrons may affect the afterglow radiation spectra. The thermal electrons are distributed at lower energies than the non-thermal ones since $\eta \gamma m_e c^2$ $\ll$ $\gamma m_p c^2$ where $\eta \ll m_p/m_e$ is a factor describing the ignorance on the plasma physics governing electron heating beyond $\gamma m_e c^2$. As a result, the synchrotron radio component emission may be affected through the production of a new emission component from thermal electrons (for $\eta \gg 1$ and moderate $1/\xi_e$) or a large self-absorption optical depth (for $\xi_e \ll 1$) which may be visible in a time scale of few hours.
{Possible effects of the thermal component are discussed in \cite{laskar17,warren22,giannios_spitkovsky}.}

Information from the TeV component cannot completely solve the degeneracy between afterglow parameters and cannot provide additional clues on the non-thermal electron distribution. However, it can provide unique fundamental data useful to constrain the other afterglow parameters less constrained such as $\epsilon_B$ and the density. This will impact also the $\xi_e$ calculation since it can help to reduce the degeneracy between the sets of solutions available in the parameter space. 
Indeed, the multi-wavelength modeling of GRB\,190829A (detected at TeV energies by H.E.S.S.) showed that the only way to explain all the detected radiation, from radio to TeV, is to introduce the parameter $\xi_e$ in the modeling, which is constrained by data to be $\xi_e<0.13$ \cite{190829A_salafia}.

\section{Discovery of a TeV emission component in GRBs}\label{Chap:TeVdiscovery}

The robust theoretical framework developed throughout the years to explain the afterglow radiation predicts that, to some extent, GRBs should be TeV emitters (section~\ref{chap:model}). Observations in the HE band, and in particular the presence of $\gtrsim$ GeV photons with energies up to $\sim100$\,GeV, support this possibility. {On the other hand, }from the observational side the search for such emission is hampered by several drawbacks. Space-born telescopes, such as Fermi-LAT, sensitive up to few hundred GeV, have an hard time with GRBs due to their low $\gamma$-ray photon flux at the highest energies ($\sim10^2$\,GeV), caused by their cosmological distance and strong EBL absorption. These difficulties can be overcome by the much larger effective area of IACTs in the common energy range of sensitivity (50-300\,GeV). As a downside, IACTs have a small field of view (a few degrees wide), higher low-energy threshold ($\gtrsim50-100$\,GeV), and reduced duty cycle (less than $10$\%).

In the last decades, IACTs have performed a huge effort to become instruments suitable for GRB observations. In particular, the efforts have been focused in two directions: i) the development of fast repointing systems to promptly react to GRB alerts and start observations with delays of a few tens of seconds after the trigger time, and ii) the extension of the energy threshold below 100\,GeV, important to reduce the impact of the EBL attenuation on the detection probability of cosmological GRBs. 

After a decade of VHE observations resulting in non-detections, the first announcement of GRBs detected by IACTs arrived in 2019, thanks to the MAGIC and H.E.S.S. telescopes \cite{magic_gcn_190114c}. These detections have firmly established that GRBs can be bright sources of TeV radiation. Somewhat unexpectedly, VHE emission has been detected also several hours/a few days after the GRB onset, and up to energies of $\gtrsim 3$\,TeV. The timescales of the detections place the origin of the emission in the afterglow phase.
The TeV emission has been studied and interpreted in a multi-wavelength context, in order to evaluate the properties and the nature of the responsible radiation mechanisms. In particular, investigations have focused on SSC, external inverse Compton (EIC), and synchrotron radiation.

In this section, all GRBs for which a detection (significance $>5\sigma$) or a hint of detection (significance between 3 and 5$\sigma$) has been claimed by Cherenkov telescopes are presented. These are in total six events (one short and five long): GRB\,160821B (Section \ref{subsec:160821B}), GRB\,180720B (Section \ref{subsec:180720B}), GRB\,190114C (Section \ref{subsec:190114C}), GRB\,190829A (Section \ref{subsec:190829A}), GRB\,201015A (Section \ref{subsec:201015A}) and GRB\,201216C (Section \ref{subsec:201216C}). For each event we start with a brief description of the prompt and afterglow multi-wavelength observations. Then, we describe VHE observations and summarise the main results. Being these detections a novelty, and some of them laying close to the sensitivity detection threshold of the instrument, we describe in detail the VHE data analysis, the calculation of the significance excess at the GRB position (following the usual prescription used for VHE sources presented in \cite{li_ma_formula}), and the methods adopted for the derivation of the spectral energy distribution (SED) and of the light-curves. 
For each GRB, we also present the interpretations that have been put forward in the literature. 
A discussion on the main common properties and differences among this initial population of VHE GRBs and with respect to the whole GRB population is presented in section~\ref{chap:discussion}, where we address also the question of what we have been learning from these few detections.

In this section all quoted times refer to the time elapsed from the trigger time $T_0$ of the Swift-BAT or Fermi-GBM instrument, as will be specified. Photon indices are given in the notation $N_\nu\propto\nu^{\alpha}$, while temporal indices are defined by $F(t)\propto t^\beta_T$.

%--------------------------------------------------------------
\subsection{GRB\,160821B}\label{subsec:160821B}
GRB\,160821B is a short GRB at $z=0.162$ detected by the Swift-BAT \cite{160821B_swiftbat} on 21 August 2016 at $T_0=22:29:13$\,UT and by the Fermi-GBM \cite{160821b_fermiGBM}. The analysis of MAGIC observations shows a $\sim3\sigma$ excess at the GRB position.

\subsubsection{General properties and multi-wavelength observations}
The BAT prompt spectrum ($T_{90} = 0.48$\,s) is well described by a power-law with index  $\alpha=-0.11 \pm 0.88$ and an exponential high-energy cutoff, corresponding to a peak energy with $E_p = 46.3 \pm 6.4$ keV \cite{160821b_bat_refined}. The GBM prompt spectrum ($T_{90} = 1.088 \pm 0.977$ s) is fitted with a cutoff power-law function as well, with $E_p = 92 \pm 28$\,keV. Being located at redshift $z=0.162$, this is one of the nearest short GRBs detected up to date.{ Its} isotropic-equivalent energy $E_{\gamma,iso} \sim 1.2 \times 10^{49}$\,erg is toward the low energy edge of the known distribution for short GRBs  \cite{berger_review_short_grb}).

Afterglow observations are available in the radio, optical, X-ray and (V)HE band.
Fermi-LAT observations were performed from the trigger time up to $2315$\,s and from $5285$\,s to $8050$\,s and did not reveal any significant excess in the 0.3-3\,GeV band \cite{160821b_magic}.
Swift-XRT observations {\cite{xrt_detection_160821B}} started $57$\,s after the trigger time and revealed a complex behaviour of the X-ray afterglow light curve. An initial plateau is followed by a steep decay at around $10^2$\,s. Then, a power-law decay with index $\sim -0.8$ is observed after $\sim 10^3$\,s \cite{160821b_swift_xrt,afterglow_kilonova_160821b}. Optical observations were performed by several instruments \cite{gtc_160821b,hst_160821b,not_160821b,wht_160821b} revealing the presence of a fading source with a magnitude $r = 22.6 \pm 0.1$ mag
$0.95$ hrs after $T_0$. The identification of the host galaxy allowed the measurement of the spectroscopic redshift $z = 0.162$. 
The GRB is located in the outskirts of the host spiral galaxy, at $\sim 15$\,kpc projected distance from its center \cite{afterglow_kilonova_160821b,lamb_160821b}.
A fading radio afterglow counterpart was observed at 6\,GHz by VLA starting from 3.6\,h after the burst trigger \cite{vla_160821b}.
The multi-wavelength light-curves of GRB~160821B are shown in Figure~\ref{fig:160821B_lcs}.

\subsubsection{VHE observations and results}
The MAGIC telescopes started the follow-up of GRB\,160821B with a very short delay of $24$\,s after $T_0$ and continued observations for about $4$\,h \cite{160821b_magic}. The observations were performed with a relatively high Night Sky Background (NSB) (2-8 times brighter than in dark nights) due to the presence of the Moon, and in mid-high zenithal angle conditions (from 34$^{\circ}$ to $55^{\circ}$). Unfortunately, the first $\sim$ 1.7\,h of the data were strongly affected by clouds. 
As a result, dedicated and optimized software configurations were used. A more stringent image cleaning with respect to dark conditions was applied to take into account the spurious contribution coming from the high NSB. The analysis required cuts optimized on the Crab Nebula and on Mrk421 observed in similar conditions, and correction factors for the low atmospheric transmission, calculated thanks to the LIDAR facility {\cite{lidar}}. 

%-------------------------------------------------
\begin{figure}[!ht]
\begin{center}
    \includegraphics[width=0.5\textwidth]{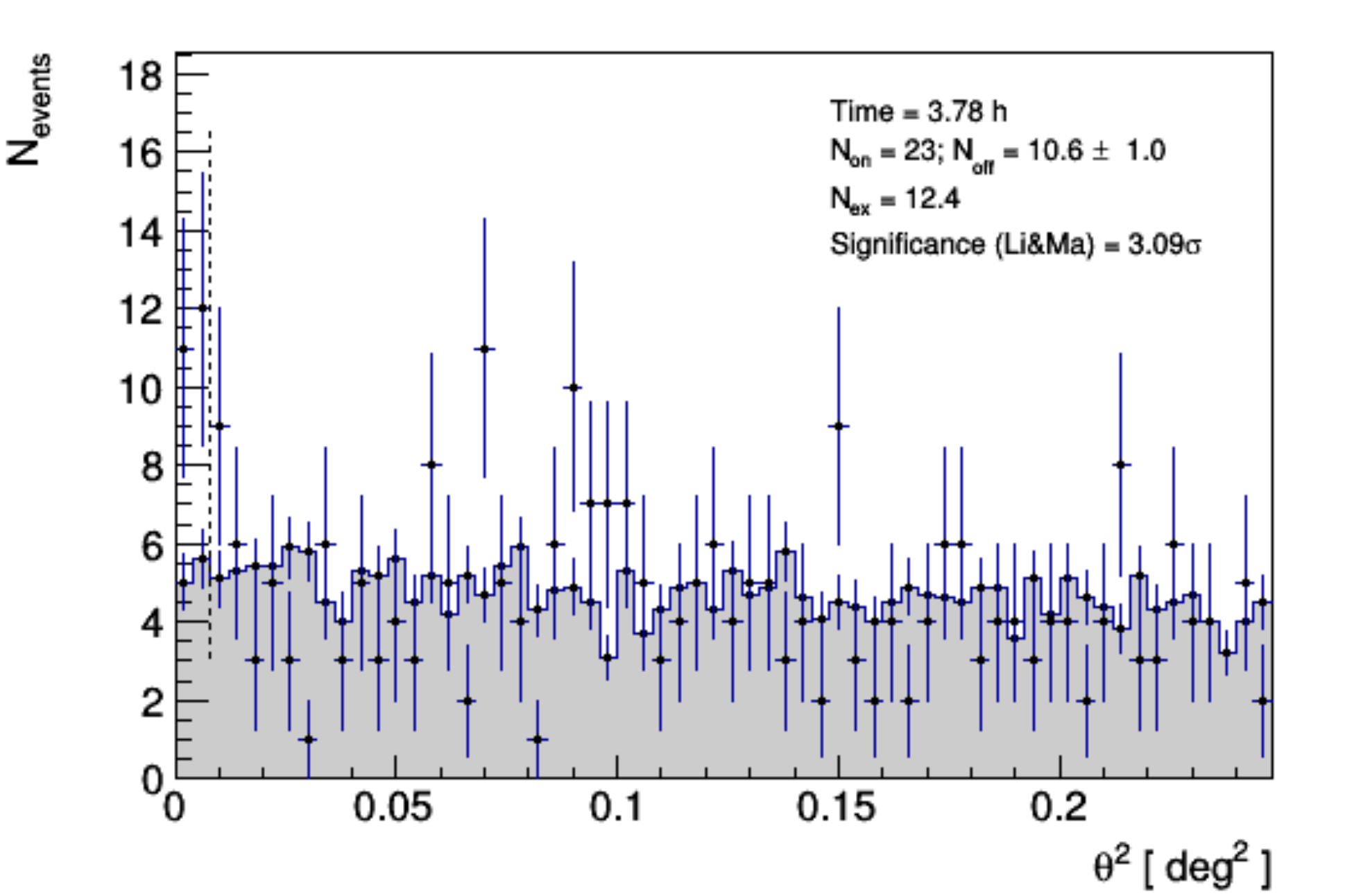}
    \caption{GRB~160821B: angular distance distribution $\theta^2$ between the nominal source position and the reconstructed event arrival directions. {The gray histogram represents the background events, while the black points with blue crosses are the $\gamma$-like events.} The vertical dashed line marks the $\theta^2$ cut value and defines the region in which excess events and signal significance are calculated. From \cite{160821b_magic}. {\copyright AAS. Reproduced with permission.}}
    \label{fig:160821B_theta2plot}
\end{center}
\end{figure}
%-------------------------------------------------

The pre-trial analysis showed the presence of a $3.1 \sigma$ ($2.9 \sigma$ post-trial\endnote{{A trial factor of 2 is considered due to the two sets of analysis cuts used for MAGIC data analysis.}}) significance excess at the GRB position provided by Swift-XRT (see Figure~\ref{fig:160821B_theta2plot}). 
The flux has been estimated for energies above $0.5$\,TeV assuming a power-law spectrum with photon index $\alpha=-2$. In the first 1.7\,h, where data taking was affected by bad atmospheric transmission, only flux upper limits could be derived. This time window has been divided into two time intervals ($24-1216$\,s and $1258-6098$\,s) and the derived upper limits are respectively $1.1 \times 10^{-11}$ cm$^{-2}$s$^{-1}$ and $5.4 \times 10^{-12}$ cm$^{-2}$ s$^{-1}$. 
In the subsequent 2.2\,h ($6134-14130$\,s) assuming that the signal is real, a flux value could  be calculated and is $9.9 \pm 4.8 \times 10^{-13}$ cm$^{-2}$ s$^{-1}$. For the same time interval, also a flux upper limit has been estimated and gives $3.0 \times 10^{-12}$ cm$^{-2}$ s$^{-1}$.
All the mentioned fluxes are shown in Figure~\ref{fig:160821B_lcs} (red symbols) and refer to the observed values, i.e., without correcting for EBL absorption.
Upper limits have been calculated at 95$\%$ confidence level following the prescriptions of \cite{rolke}. 
The low {($\sim 3\sigma$)} significance {estimated} did not allow to obtain an unfolded spectrum. As a result, in the SED (Figure~\ref{fig:160821B_sed}) the reconstructed flux in the third bin ($6134-14130$\,s) over the energy range $0.5 - 5$\,TeV is represented as an error box. The statistical error on the photon flux has been taken into account, while, for simplicity, the systematic error for the assumed spectral index was neglected.
The flux inferred by MAGIC observations in the $0.5 - 5$\,TeV energy range would imply a TeV luminosity at least 5 times larger (when de-absorbed by EBL) than the luminosity emitted in the X-ray band.

\subsubsection{Interpretation}
A modeling of the multi-wavelength observations, including MAGIC data, has been presented by the MAGIC Collaboration in \cite{160821b_magic}. 
The emission is interpreted as the sum of several components, dominating at different times and in different energy bands:
\begin{itemize}
\item synchrotron and SSC emission from electrons accelerated at the forward shock; this is in general the dominant emission component;
\item synchrotron emission from electrons accelerated by the reverse shock, which is found to dominate the radio band until $t \sim 4.8$\,h;
\item kilonova emission, powered by freshly synthesized $r$-process elements released in neutron star mergers; this component is found to dominate the optical/nIR from around 1 to 4 days \cite{afterglow_kilonova_160821b,lamb_160821b};
\item an X-ray extended emission component, widely attributed to long-lasting activity of the central engine, here dominating the X-ray band for $t < 10^3$\,s.
\end{itemize}
In performing this multi-component modeling, the synchrotron and SSC forward shock emission have been calculated with a one-zone numerical code (see \cite{190114C_mwl_paper} for details), while the reverse shock and kilonova emission contributions have been taken from \cite{afterglow_kilonova_160821b}. Only X-ray data at $t > 10^3$\,s have been included in the modeling, to exclude the extended emission component. The broad-band modeling is shown over-plotted to the light-curves in Figure~\ref{fig:160821B_lcs} (solid lines) and to the SED between 1.7 and\,4 h in Figure~\ref{fig:160821B_sed}.

%--------------------------------------
\begin{figure}[!ht]
\begin{center}
    \includegraphics[width=0.6\textwidth]{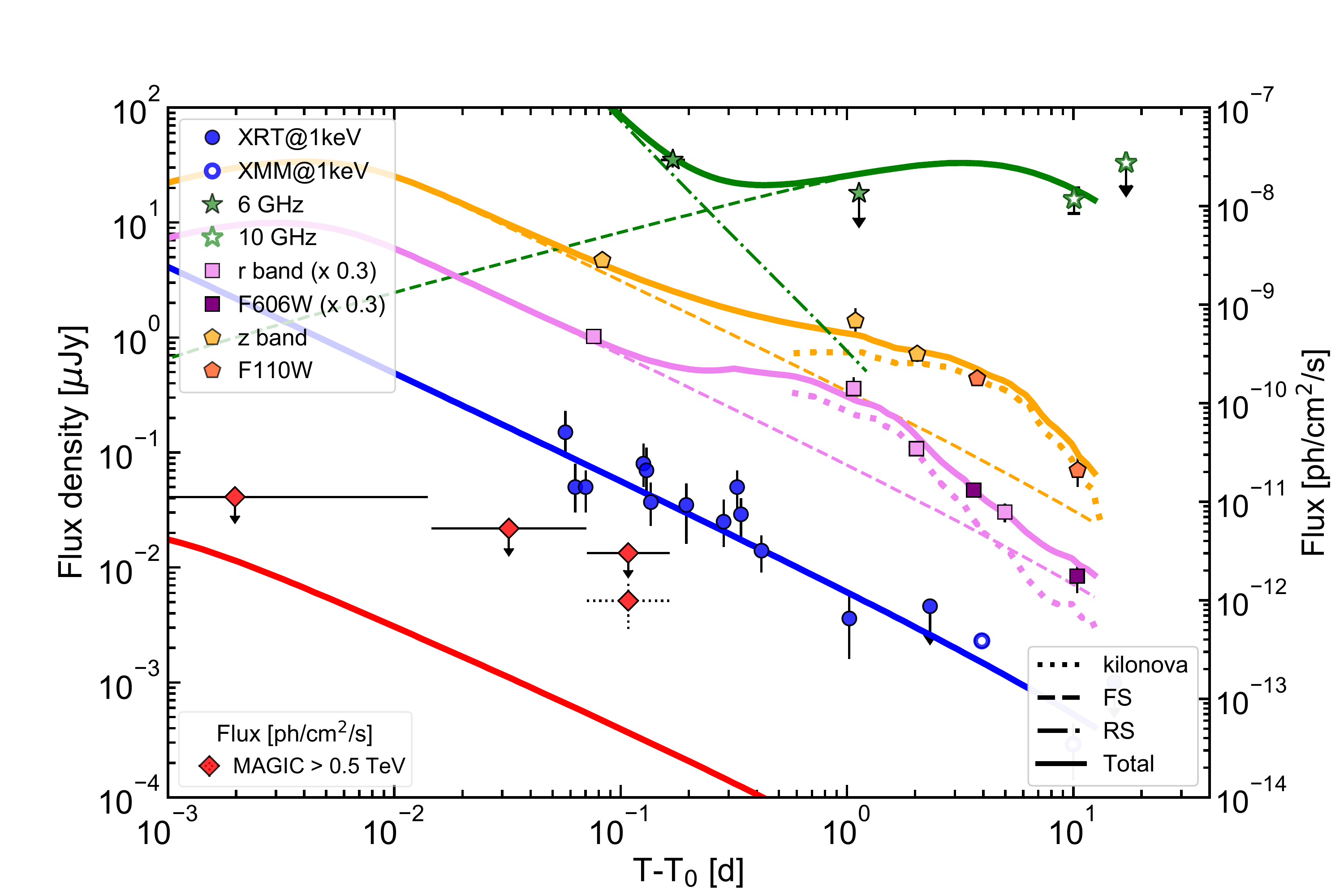} 
    \caption{GRB~160821B: multi-wavelength light curves (from radio to TeV) and their modeling according to \cite{160821b_magic}. {\copyright AAS. Reproduced with permission.} The different contributions from the forward shock {(FS)}, reverse shock {(RS)}, and kilonova are shown (see legend).}
    \label{fig:160821B_lcs}
\end{center}
\end{figure}
%--------------------------------------
%--------------------------------------
\begin{figure}[!ht]
\begin{center}
  \includegraphics[width=0.6\textwidth]{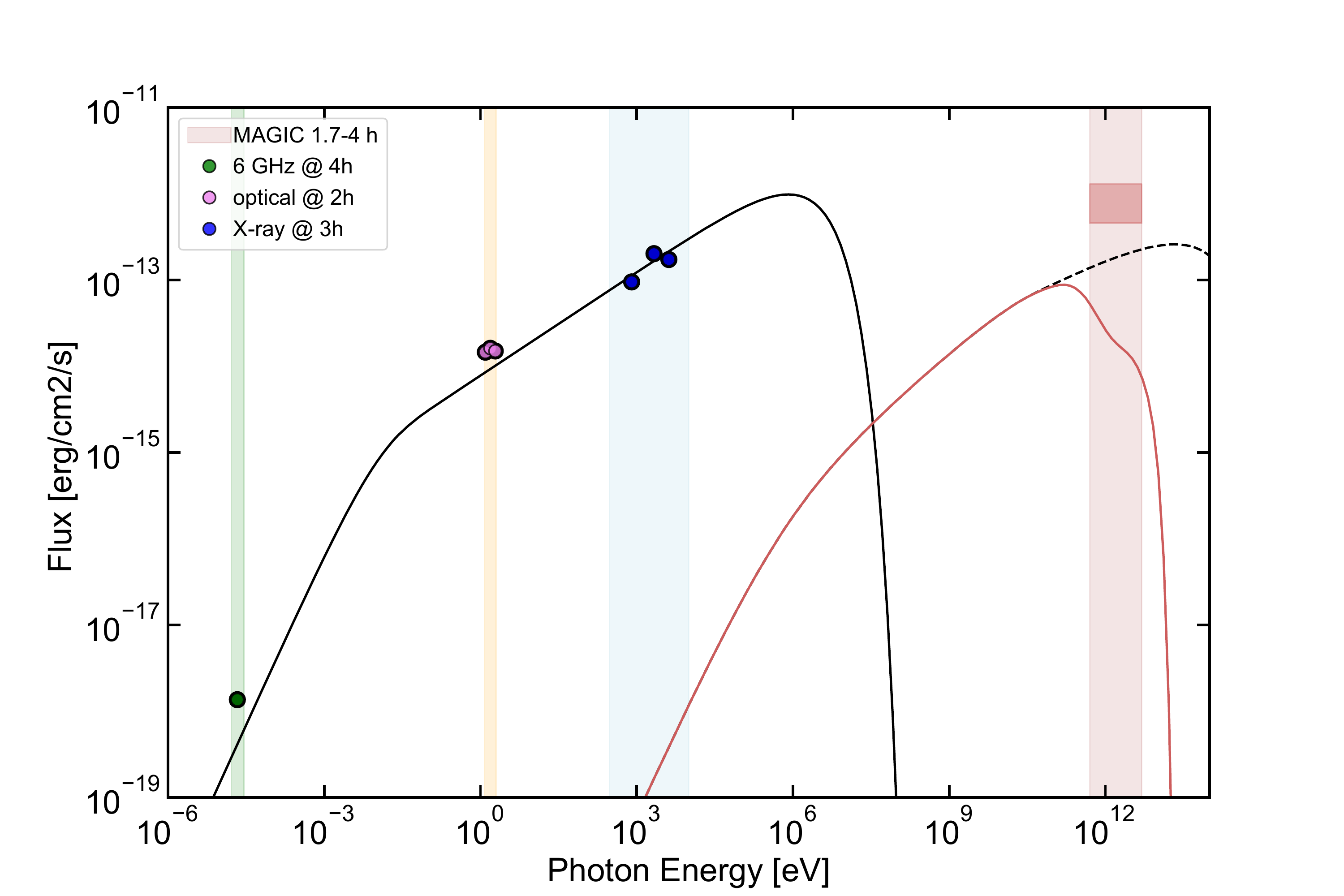}
    \caption{GRB~160821B: modeling of the simultaneous multi-wavelength SED at approximately $\sim 3$\,h according to \cite{160821b_magic} {(\copyright AAS. Reproduced with permission.)}, for the same parameters used to model the lightcurves in Figure~\ref{fig:160821B_lcs}. The shaded areas show the sensitivity energy range of the different instruments. The MAGIC error box on the reconstructed flux is also shown. Synchrotron (solid black line), intrinsic SSC (before EBL absorption, dashed black line) and SSC emission with EBL attenuation (solid red line) estimated from the numerical modeling are shown.}
    \label{fig:160821B_sed}
\end{center}
\end{figure}
%--------------------------------------

A very good agreement between data and modeling is found {in radio (green lines and points), optical (yellow and pink lines and points) and X-rays (blue lines and points).} 
A large degeneracy is present in the parameters, and the data modeling only allows to identify the ranges for the permitted values of each parameter. These are reported in Table~\ref{tab:160821B_modeling} and we note that they are very similar to those estimated in \cite{afterglow_kilonova_160821b} and in a later work by \cite{160821B_murase_eic}. In the allowed parameter space defined by radio, optical and X-ray observations, different combinations of the parameters predicts different SSC fluxes at 1\,TeV are found, reaching at most $F_{SSC}^{(1 TeV)} \sim 2 \times 10^{-13}$ erg cm$^{-2}$ s$^{-1}$. This value, when attenuated by EBL, is at least one order of magnitude fainter than the one inferred from data analysis of the MAGIC observations.
In other words, the parameter space constrained by the observations at lower frequencies is unable to account for such energetic TeV emission, and the SSC forward shock scenario fails to reproduce the observations, provided that the hint of excess found by MAGIC is a real signal from the source.

An alternative scenario that has been explored is the external inverse Compton (EIC) scenario, investigated by \cite{160821B_murase_eic}. These authors first consider a one-zone SSC model, and reach similar conclusions to those presented by the MAGIC Collaboration \cite{160821b_magic}: the SSC mechanism predicts a TeV flux around 1-2 orders of magnitude lower than the MAGIC observations (see Figure~\ref{fig:160821B_modeling_eic}, orange curves).
The alternative EIC scenario is then considered by the authors, where the seed photons are provided by the extended X-ray emission and the X-ray plateau. 
The extended emission and the plateau are fitted using two phenomenological functions. The energy spectrum of the late-prompt emission is described by a broken power-law (see Figure~\ref{fig:160821B_modeling_eic}, top and bottom panels). 
For the EIC model, the VHE spectrum is inferred for three different observed times ($t = 1.1, 1.8, 2$\,h) and compared to the MAGIC flux averaged between 1.7 and 4\,h. As it can be seen in Figure~\ref{fig:160821B_modeling_eic} {(bottom panel)}, the model flux at 2\,h under-predicts the MAGIC flux (the {MAGIC observed flux,} green shaded area, should be compared with the EBL-absorbed model flux). We conclude that also the EIC model is unable to explain the large TeV flux suggested by MAGIC observations.

%-----------------------------------------
\begin{figure}[!ht]
\begin{center}
    \includegraphics[width=0.55\textwidth,trim={0 7.3cm 0 0},clip]{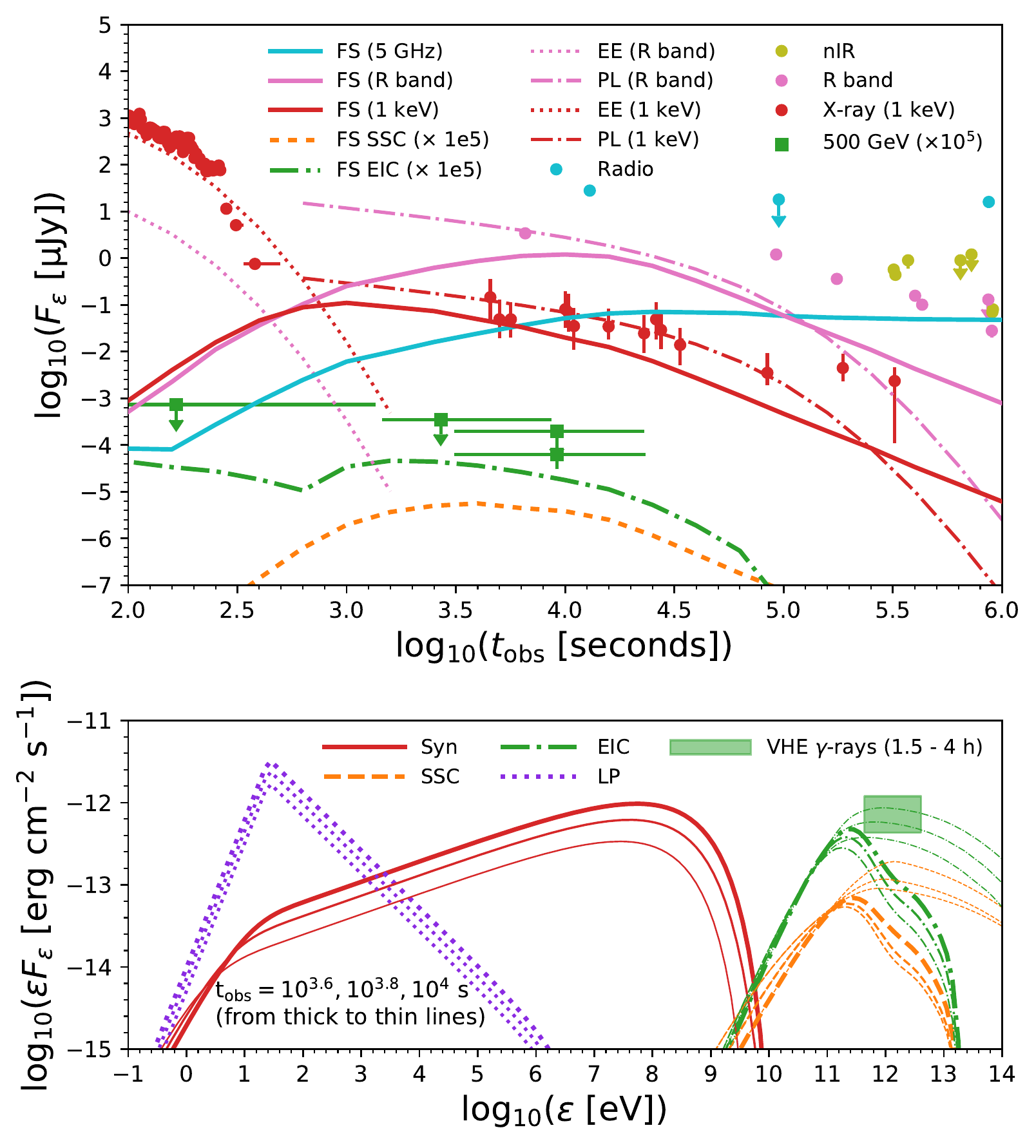} 
    \caption{GRB\,160821B: modeling of multi-wavelength light curves (top panel) and simultaneous SED (bottom panel) {in radio, optical, X-ray and VHE band} presented by \cite{160821B_murase_eic}.{\copyright AAS. Reproduced with permission.} Two scenarios are considered: SSC (dashed orange lines) and EIC (dash-dotted green lines). {In the top panel the flux contribution in the different energy bands due to the forward shock (FS) both in the SSC and in the EIC scenario, the extended X-ray emission (EE) and the X-ray plateau emission (PL) are shown. In the bottom panel for the VHE band both} the intrinsic (thin lines) and the absorbed (thick lines) VHE curves are shown {for the SSC and EIC scenarios for three different observed times. For the lower energies the broken power-law of the late prompt emission (dashed violet line) and the expected synchrotron SEDs (solid red line) are displayed.} The MAGIC observations (green symbols and green shaded area) are absorbed by EBL.}
    \label{fig:160821B_modeling_eic}
\end{center}
\end{figure}
%-----------------------------------------

%-----------------------------------------
\begin{table}[!ht]
\begin{center}
\begin{adjustbox}{width=0.72\textwidth}
\begin{tabular}{lccccccc}
\toprule
    & $E_k$ & log($\epsilon_e$) & log($\epsilon_B$) & log(n)  & $p$ & $\xi_e$ & $\theta_j$ \\
    & erg & & & cm$^{-3}$ & & & rad \\
\midrule
 \small{MAGIC Coll.}   & $10^{51}-10^{52}$ & [-1 ; -0.1] & [-5.5 ; -0.8] & [-4.85 ; -0.24] & 2.2-2.35 & 1 & / \\
  \small{Troja + 2019}  & $10^{50}-10^{51}$ & [-0.39 ; -0.05] & [-3.1 ; -1.1] & [-4.2 ; -1.7] & 2.26-2.39 & 1 & 0.08-0.50 \\
  \small{Zhang + 2021 (SSC)}  & $3 \times 10^{51}$ & -0.52 & -5 & -1.3 & 2.3 & 0.5 & 0.15 \\
  \small{Zhang + 2021 (EIC)}  & $2 \times 10^{51}$ & -0.3 & -6 & -1 & 2.5 & 0.1 & 0.1 \\
  \bottomrule
\end{tabular}
\end{adjustbox}
\end{center}
\caption{GRB\,160821B. List of the best fit parameters inferred from multi-wavelength modeling of the afterglow radiation by different authors.}
\label{tab:160821B_modeling}
\end{table}
%-----------------------------------------

%------------------------------------------------------

%-----------------------------------------------------------------
\subsection{GRB\,180720B}\label{subsec:180720B}
GRB\,180720B is a long GRB at $z=0.654$ triggered on 20 July 2018 by the Fermi-GBM \cite{gbm_180720B} at $T_0=14:21:39$\,UT and 5\,s later by the Swift-BAT instrument \cite{swift_180720B}. 
The H.E.S.S. telescopes observed and detected GRB\,180720B about 11\,h after the prompt emission, with a $\sim5\sigma$ statistical significance.

\subsubsection{General properties and multi-wavelength observations}
The extremely bright prompt emission of this event, which is the seventh in brightness among the GRBs detected by the Fermi-GBM until then, lasted for $T_{90} = 48.9 \pm 0.4 $\,s and released an isotropic-equivalent energy $E_{\gamma,iso} = 6.0 \pm 0.1 \times 10^{53}$\,erg in the 50-300\,keV range. 

%---------------------------------
\begin{figure}[!ht]
\begin{center}
  \includegraphics[width=0.75\textwidth,trim={1.8cm 0.9cm 0 2.3cm},clip]{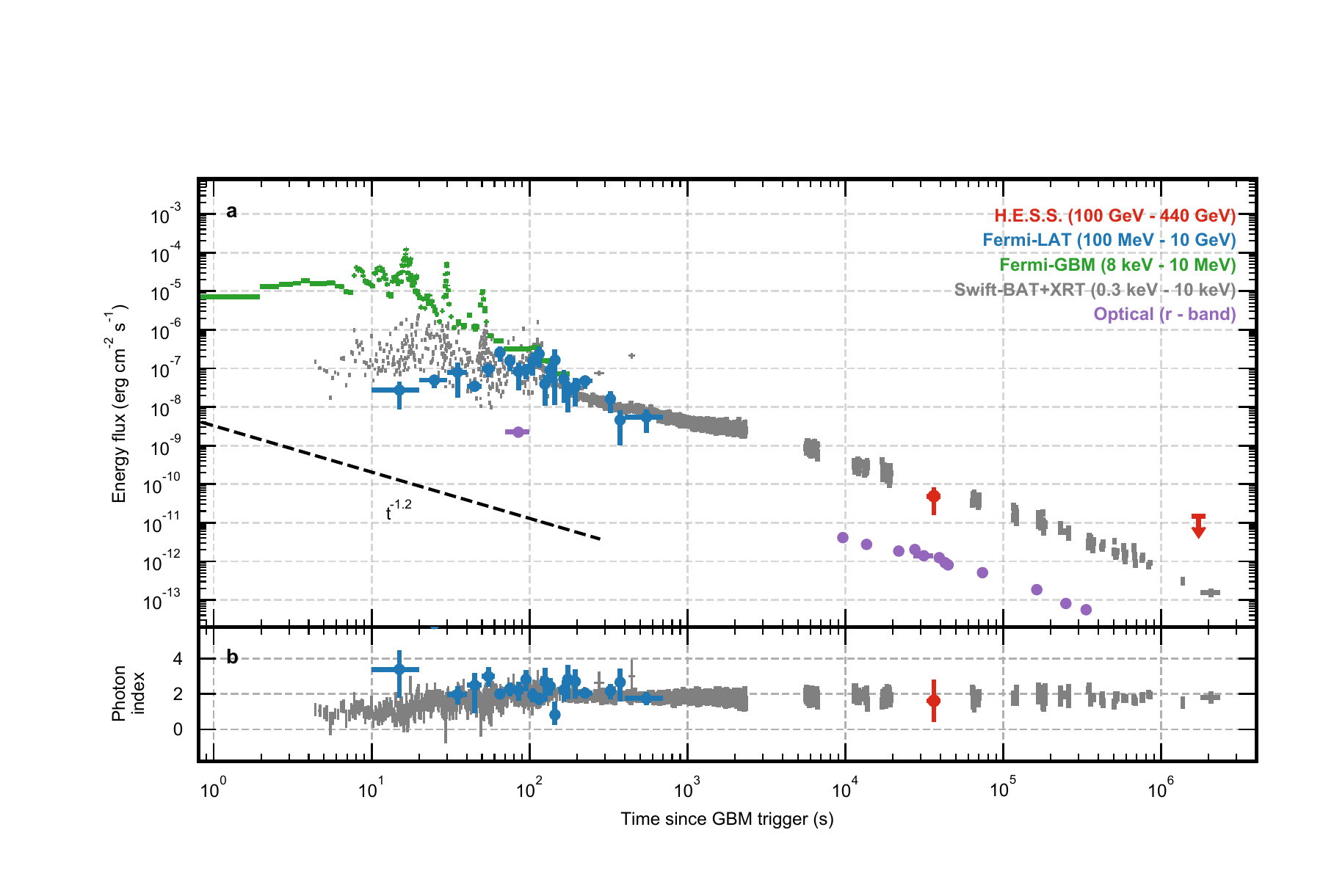}
    \caption{GRB\,180720B: broad-band light curves and photon index evolution including optical, X-ray and VHE data. For the late time H.E.S.S. observation (18 days after the trigger) an UL has been derived. A flux temporal decay with index $1.2$ (similar to what observed in optical and X-ray) is also shown in black dashed line. From \cite{hess_180720B}.}
    \label{fig:180720B_lcs}
\end{center}
\end{figure}
%---------------------------------

Multi-wavelength afterglow observations covered the entire electromagnetic spectrum (see Figure~\ref{fig:160821B_lcs}). 
Significant signal was detected by Fermi-LAT from the trigger time up to 700\,s, with the highest photon energy of 5\,GeV detected 137 s after the burst trigger \cite{lat_180720B}.
The Swift-XRT telescope observed and identified a bright afterglow starting from $90$\,s. This was still visible almost 30 days after the trigger time. The late-time light curve (from $2\times10^3$\,s to $4\times10^6$\,s) can be modelled with an initial power-law decay with an index $-1.19^{+0.01}_{-0.02}$ followed by a break at $t_{break}=8\times10^4$\,s to an index of $-1.55^{+0.04}_{-0.05}$ \endnote{\url{https://www.swift.ac.uk/xrt\_live\_cat/00848890/}}. Several optical observations \cite{optical_kanata_180720B,optical_lco_180720B,optical_tshao_180720B,optical_coatli_180720B,optical_ison_180720B,optical_kait_180720B,optical_master_net_180720B,optical_mitsume_180720B,optical_oaj_180720B,optical_rem_180720B} revealed the presence of a counterpart and allowed to estimate the redshift value of $z = 0.654$. The optical afterglow was seen to be slowly fading at an almost constant rate from around 10-11\,h after the trigger time \cite{optical_d50_180720B,optical_osn_180720B}{ as discussed in \cite{radio_180720B}}. Radio observations (not shown in the figure) were also performed starting from $\sim 1.7$ days after the burst showing a steep power-law decaying emission \cite{radio_180720B}.  

\subsubsection{VHE observations and results}
The H.E.S.S. telescopes followed-up the event for $\sim 2$\,h starting from 10.1\,h, revealing the presence of a source with a $ 5.3 \sigma$ pre-trial significance ($5.0 \sigma$ post-trial{\endnote{{post-trial significance is estimated by accounting for the previously well-localized GRBs observed by H.E.S.S. in the same array configuration as of GRB\,180720B}}). The observation was performed in standard dark and good weather conditions with a mean zenith angle of 31.5$^{\circ}$. Another observation was performed under similar conditions 18 days after the previous one with results consistent with background events. The inferred flux at $\sim11$\,h and the flux upper limit at 18\,d are shown in Figure~\ref{fig:180720B_lcs} (red symbols).

The observed spectrum in the $0.1 - 0.44$\,TeV energy band has been fitted both with a power-law (panel on the left in Figure~\ref{fig:180720B_sed}) and with a power-law with a cutoff $F_{int} = F_{0,int} \big(\frac{E}{E_0}\big)^{-\gamma_{int}}$, to describe an intrinsic power-law spectrum affected by the EBL attenuation $e^{-\tau(E,z)}$ (see Figure~\ref{fig:180720B_sed}, panel on the right). With reference to the second fit, the analysis returns a photon index $\gamma_{int} = 1.6 \pm 1.2$ (statistical) $\pm 0.4$ (systematic) and a flux normalization $F_{0,int} = (7.52 \pm 2.03$ (statistical) $^{+4.53}_{-3.84})$ (systematic) $ \times 10^{-10}$ TeV$^{-1}$ cm$^{-2}$ s$^{-1}$, evaluated at an energy $E_{0,int} = 0.154$\,TeV. 

%-------------------------------
\begin{figure}[!ht]
\begin{center}
  \includegraphics[width=0.7\textwidth]{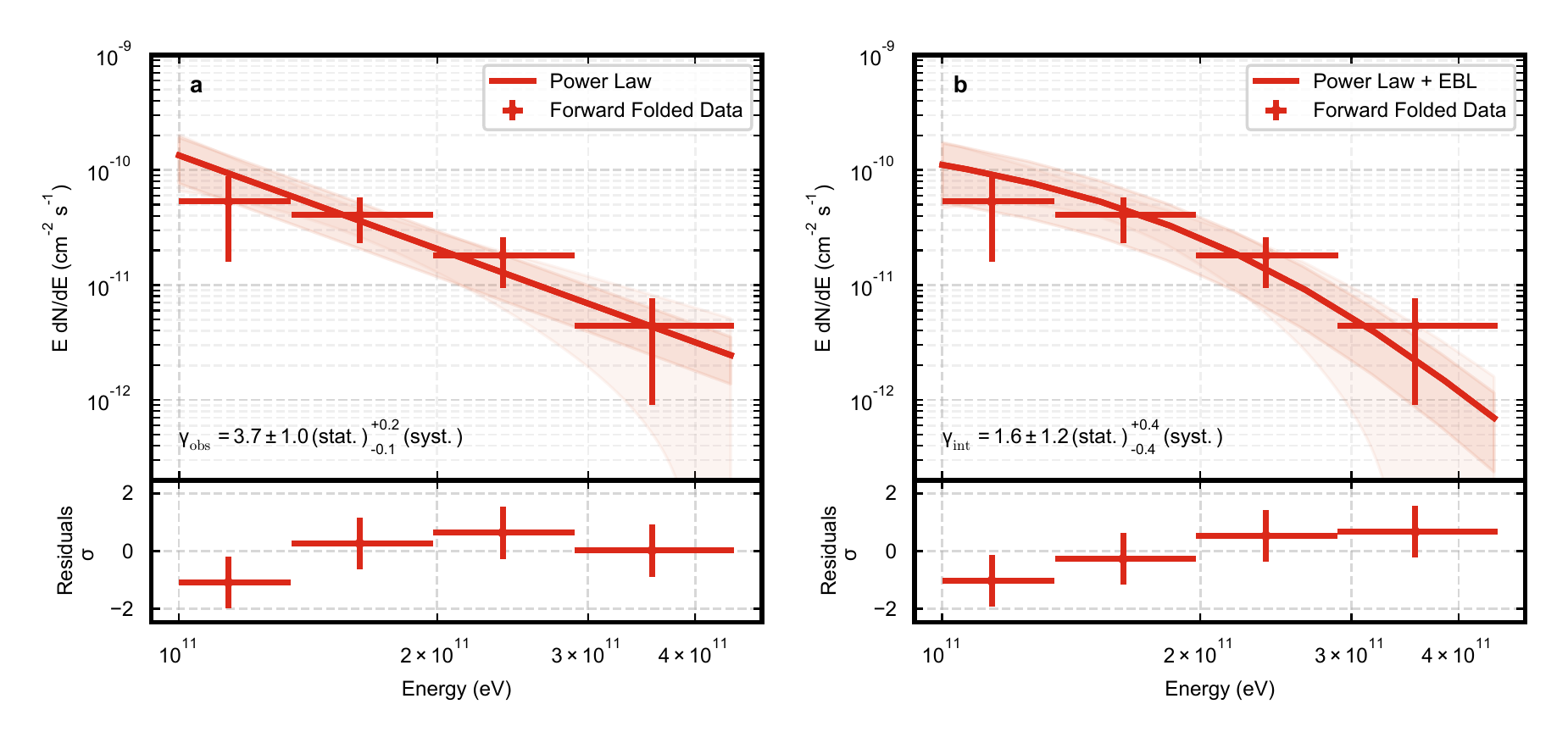}
    \caption{GRB\,180720B: spectral fit to the observed emission in the $0.1-0.44$\,TeV energy range. In the left panel the observed spectrum is assumed to be described by a power-law model, while in the panel on the right, the spectrum is an power-law attenuated by EBL. Statistical and systematic uncertainties at $1 \sigma$ confidence level are shown with shaded areas. From \cite{hess_180720B}. }
    \label{fig:180720B_sed}
\end{center}
\end{figure}
%-------------------------------

\subsubsection{Interpretation}
The H.E.S.S. Collaboration explored two possible radiation mechanisms to explain the VHE emission from GRB\,180720B  \cite{hess_180720B}: synchrotron emission and SSC radiation. A full modeling is not performed, and the discussion and comparison among the two different scenarios is based on estimates of the typical and maximal electron energy necessary in the two cases and on the comparison between spectral and temporal indices in different energy ranges.
A synchrotron spectrum with a flat ($\alpha\sim-2$) slope extending from X-ray to VHE could model the emission with one single broad component and explain the similarity between the H.E.S.S., Fermi-LAT, and Swift-XRT luminosities, and the consistency among their photon index values. The large error on the VHE photon index however is not placing strong constraints, leaving open both the possibility of a consistency with the extrapolation of the synchrotron spectrum but also the possibility of a spectral hardening, indicative of a second component.
A synchrotron origin of $10^2$\,GeV photons would require to find a process able to accelerate electrons up to PeV energies, which is in excess of the maximum electron energy achievable in external shocks (for a discussion, see {Section}~\ref{subsubsec:sync_emission}). 
Adopting the standard Bhom limit, $> 100$\,GeV emission 10\,h after the burst would require a huge bulk Lorentz factor $\Gamma \sim 1000$ which at these late times is really unlikely. As a result, these strong requirements disfavour the synchrotron emission as responsible of the VHE component in GRB\,180720B.

%---------------------------------
\begin{figure}[!ht]

  \includegraphics[width=0.35\textwidth,trim={0 14.3cm 0 0},clip]{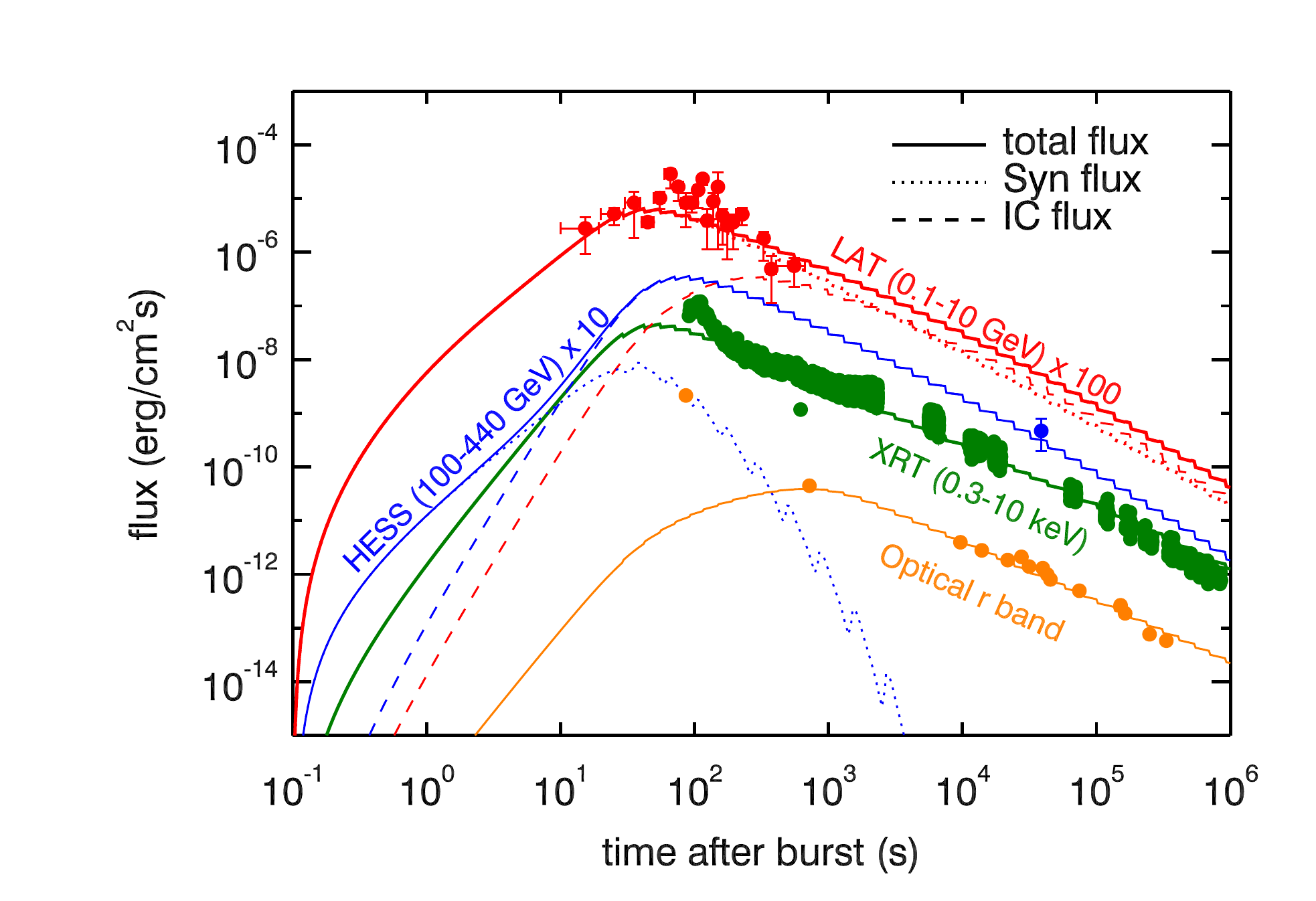} \quad
   \includegraphics[width=0.35\textwidth,trim={0 0 0 14.3cm},clip]{Images/180720B-new_2.pdf}
    \caption{GRB\,180720B: modeling of the broad-band light curves (top panel) and SED at the time of the H.E.S.S. detection (bottom panel) proposed by \cite{zhang_180720B_190114C}. Both the synchrotron and the SSC contribution to the total flux are shown (see legend). In the SED, X-ray and H.E.S.S. data are shown respectively with the green and the blue boxes.}
    \label{fig:180720B_modeling}
\end{figure}
%---------------------------------
The SSC scattering, on the contrary, arises as a natural candidate. 
A full broad-band modeling of GRB\,180720B data in this scenario is presented in \cite{zhang_180720B_190114C}. A numerical code reproducing the synchrotron and SSC emission in the afterglow shocks has been used (see \cite{modeling_180720B}). The resulting light-curves and SED in the H.E.S.S. observational time window are shown in Figure~\ref{fig:180720B_modeling}. The full emission is explained as afterglow forward shock radiation (except for the initial peak in the optical and X-ray curves at $t \sim 10^{2}$\,s which is attributed to reverse shock emission). 
In the case of a constant-density ISM environment, the parameters that best reproduce the data are: $E_k = 10^{54}$\,erg, $n = 0.1$\,cm$^{-3}$, $\epsilon_e = 0.1$, $\epsilon_B = 10^{-4}$, $\Gamma_0 = 300$ and $p = 2.4$. 
As it can be noticed, the equipartition factor $\epsilon_B$ needs to assume a quite low value in order to explain the observations.
A stellar wind-like environment is discarded by the authors on the basis of the comparison between the expected flux at $\sim 1-10$\,GeV, following the prescriptions of \cite{sariesin} ($\gtrsim 2 \times 10^{-6}$ erg cm$^{-2}$ s$^{-1}$ at $t \approx 100$\,s) and the one observed by Fermi-LAT ($\sim 10^{-8}$\,erg cm$^{-2}$ s$^{-1}$ at $t \approx 100$\,s). 
A low magnetic field equipartition factor is derived from the condition $E_{KN} \gtrsim 0.44$\,TeV at $t \approx 10$\,h where $E_{KN}$ is the energy at which the KN scattering becomes relevant. A transition energy between synchrotron and SSC component of $\sim 1$\,GeV is derived. Such value falls into the Fermi-LAT energy range and is compatible with a hardening of the spectrum in the VHE band. However, since Fermi-LAT sensitivity is above the predicted flux of GRB 180720B at $10$\,h in the GeV band, the data cannot firmly confirm the presence of this transition.  

%------------------------------------------------

%------------------------------------------------------------------
\subsection{GRB\,190114C}\label{subsec:190114C}
GRB\,190114C is a long GRB at redshift $z=0.42$ triggered by the Swift-BAT \cite{swift_gcn_190114C} on 14 January 2019 at $T_0=20:57:03$\,UT, and by the Fermi-GBM \cite{fermigbm_gcn_190114C}. The event was detected also by several other space $\gamma$-ray instruments such as AGILE, INTEGRAL/SPI-ACS, Insight/HXMT, and Konus-Wind \cite{190114C_mwl_paper}.
MAGIC detected GRB\,190114C starting $\sim60$\,s after $T_0$ with a significance above $50\sigma$.

\subsubsection{General properties and multi-wavelength observations}
The duration of the prompt emission is $T_{90} \approx 116$\,s as measured by Fermi-GBM and $T_{90} \approx 362$\,s by Swift-BAT. 
However, the prompt light curve showed a multi-peak structure only for about 25\,s, suggesting that the remaining activity, which is characterised by a smooth power-law decay, recorded by these instruments may be already the afterglow emission. Support to such interpretation is also obtained from a joint spectral and temporal analysis of the Fermi-GBM and Fermi-LAT data \cite{ravasio_190114C}.
The total radiated prompt energy is $E_{\gamma,iso} = (2.5 \pm 0.1) \times 10^{53}$\,erg in the energy range 1–10$^4$\,keV \cite{fermi_swift_190114C}.

%-----------------------------------
\begin{figure}[ht]
\centering
\includegraphics[width=0.7\textwidth]{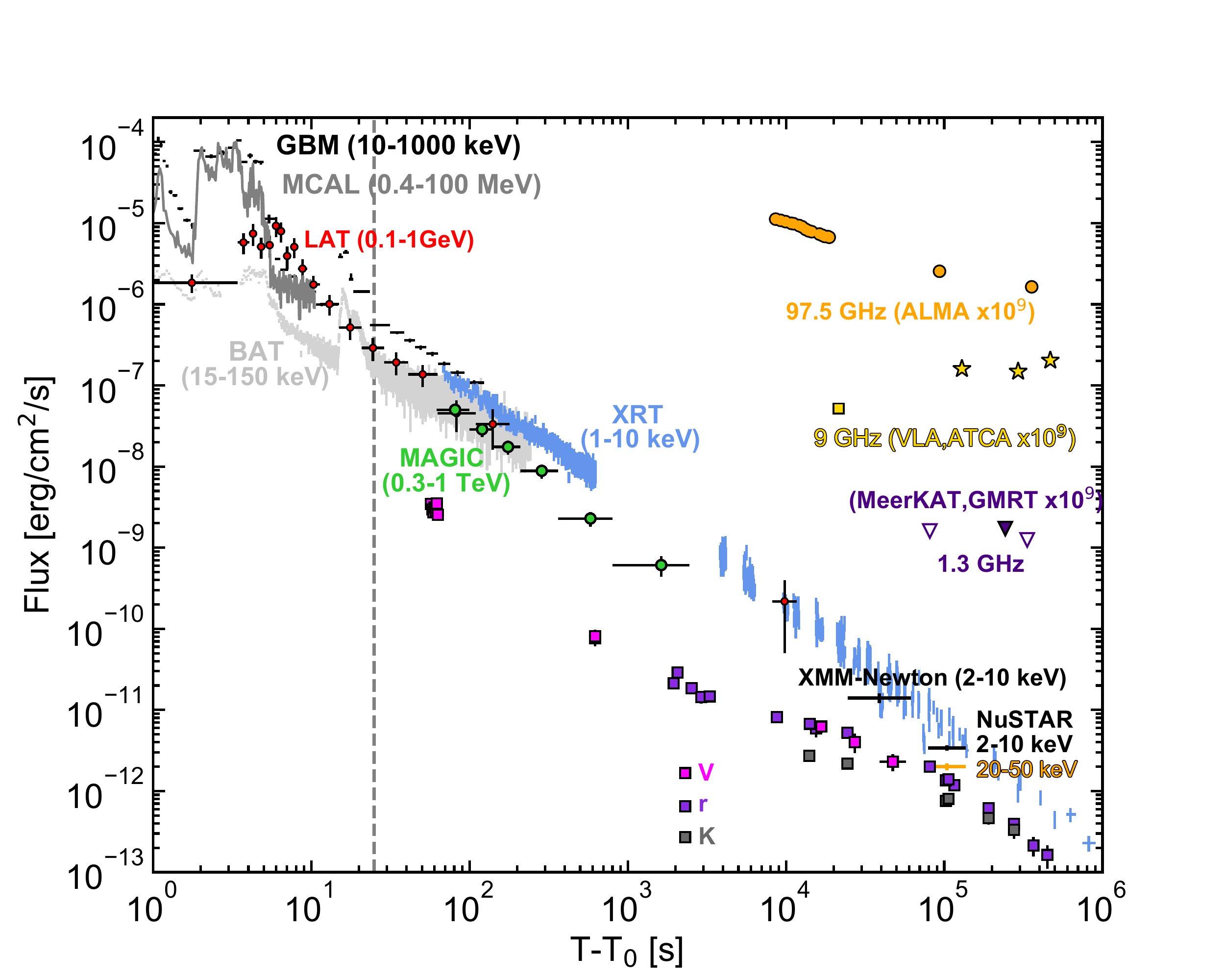}
\caption{GRB\,190114C: light-curves at different frequencies. From \cite{190114C_mwl_paper}}
\label{fig:all_lc}
\end{figure} 
%-----------------------------------

Extensive follow-up observations from several different instruments from GeV to radio are available. Light-curves are shown in Figure~\ref{fig:all_lc}.
Fermi-LAT observations started since the beginning of the prompt phase. A GeV counterpart was detected from $T_0$ to $150$\,s, when the burst left the LAT field of view and remained outside it until $8600$\,s. When LAT resumed observations significant signal was still detected at a flux level $\sim2\times10^{-10}$\,erg\,cm$^{-2}$\,s$^{-1}$ (0.1-1\,GeV).
After $\sim 60$\,s from the burst trigger, Swift-XRT started follow-up observations, which covered in total $\sim 10^{6}$\,s. The light-curve in the 1-10\,keV energy band is consistent with a power-law decay $F \propto t^\alpha$ with $\alpha =  -1.36 \pm 0.02$ \cite{190114C_mwl_paper}. NuSTAR and XMM-Newton observations are also available around 1-2\,days.
The NIR, optical and UV data were taken from around $\sim 100$\,s. The early emission is particularly bright and is interpreted as dominated by the reverse shock component \cite{laskar_rs_190114C}. Afterwards, the decay rate flattens and then steepens again after $\sim3\times10^4$\,s (see Figure~\ref{fig:all_lc}). The Nordic Optical Telescope measured a redshift of $z = 0.4245 \pm 0.0005$ \cite{not_190114C} which was then confirmed by Gran Telescopio Canarias \cite{gtc_190114C}.
Radio and sub-mm data were taken from $\sim 10^4$\,s and exhibit an achromatic behavior, possibly dominated by the reverse shock in the sub-mm range, followed by emission at late times with nearly constant flux.

\subsubsection{VHE observations and results}
%-----------------------------------
\begin{figure}[ht]
\centering
\includegraphics[width=0.5\textwidth]{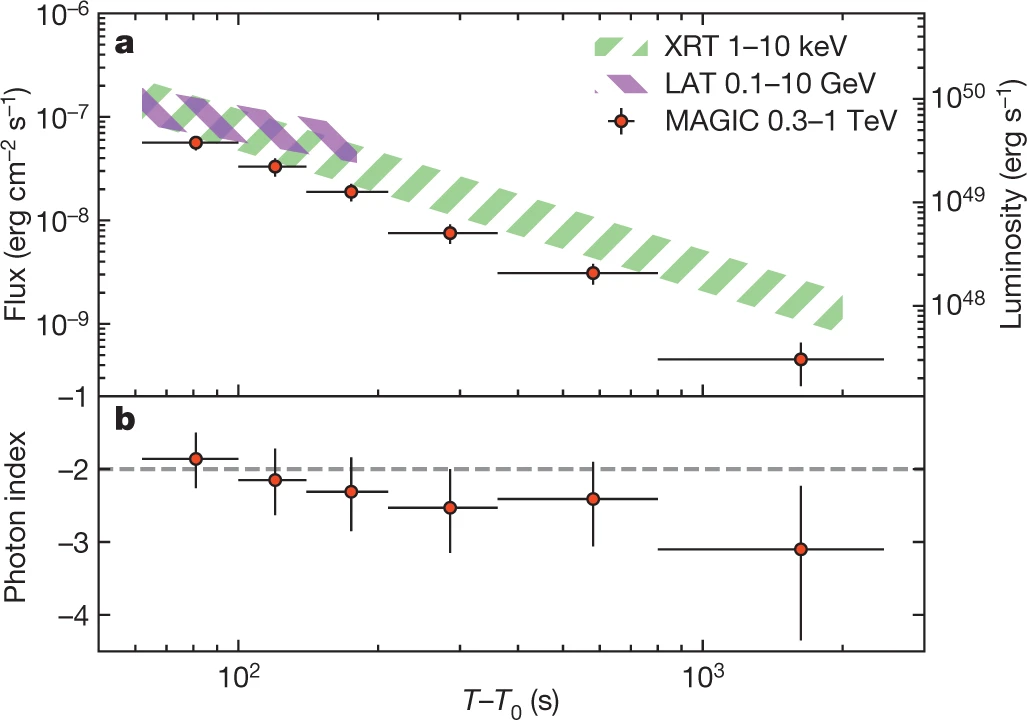}
\caption{GRB\,190114C. Upper panel: MAGIC light-curve (red circles), compared with the XRT (green band) and LAT (red band) emission. Bottom panel: temporal evolution of the intrinsic spectral photon index in the MAGIC data analysis time bins. From \cite{190114C_discovery_paper}.}
\label{fig:lc}
\end{figure} 
%-----------------------------------

After receiving (at 22\,s after the BAT trigger time) and validating (at 50\,s) the GRB alert, the MAGIC telescopes started observing GRB\,190114C at 57\,s and operated stably from 62\,s, starting from a zenith angle of 55.8$^\circ$. Observations lasted until 15912\,s when a zenith angle of 81.14$^\circ$ was reached. The observation was performed in good weather conditions but in presence of the Moon, resulting in a night sky background approximately 6 times higher than the standard dark night conditions.
The results of the offline analysis show a clear detection above the $50 \sigma$ level in the first 20 minutes of observation {\cite{190114C_discovery_paper}}. 

The light-curve (see Figure~\ref{fig:lc}, upper panel) for the intrinsic flux (i.e., corrected for the EBL absorption) in the 0.3–1\,TeV range was derived starting from 62\,s and up to 2454\,s. The TeV light curve is well described by a power-law with temporal decay index $\beta_T = - 1.60 \pm 0.07$, steeper than the one exhibited by the X-ray flux. The temporal evolution of the intrinsic spectral photon index $\alpha_{int}$ of the TeV differential photon spectrum is shown in the bottom panel. A constant value of $\alpha_{int}$ $\approx -2$ is consistent with the data, considering the statistical and systematic errors, but there is evidence for a softening of the spectrum with time.
The spectral fit in the 0.2-1\,TeV energy range for the time-integrated emission ($62-2454$\,s) returns $\alpha_{obs} = -5.34 \pm 0.22$ and $\alpha_\textup{int}=-2.22^{+0.23}_{-0.25}$ for the observed and EBL-corrected spectrum, respectively.

\subsubsection{Interpretation}
%-------------------------------
\begin{figure}[!ht]
\centering
\includegraphics[width=0.6\textwidth]{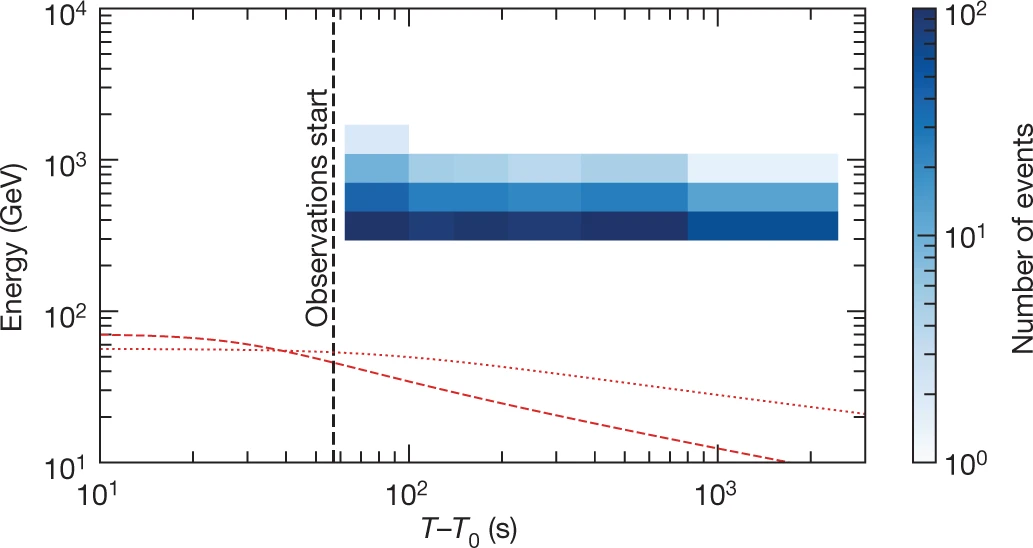}
\caption{GRB\,190114C: comparison between the distribution of the observed $\gamma$-ray events binned in time and energy (blue shaded areas) and the limiting curves for synchrotron maximum energy. Two different density radial profiles are considered for the derivation of the theoretical curves: constant ({red} dotted curve) and wind-like ({red} dashed curve). From \cite{190114C_discovery_paper}.}
\label{fig:burnoff}
\end{figure}
%-------------------------------
\label{subsubsec:interpretation_190114C}

The properties of the VHE light curve and spectrum of GRB\,190114C were studied by the MAGIC Collaboration in \cite{190114C_discovery_paper}. The PL behaviour, the absence of variability, and the relatively long timescale of the emission support the evidence that the VHE component belongs to the afterglow phase. 
An estimate of the amount of radiated power in the TeV range can be derived, assuming that the afterglow onset is at {$\sim6$\,s \cite{ravasio_190114C}. In} this case the energy radiated in the TeV band is $\sim 10\%$  of the isotropic-equivalent energy of the prompt emission $E_{\gamma,iso}$ considering the temporal evolution estimated from the MAGIC light-curve.

The energy of the photons observed by the MAGIC telescopes was compared with the maximum energy of synchrotron photons assuming two possible scenarios for the radial profile of the external density, namely constant and wind-like (Figure~\ref{fig:burnoff}).
These estimates of the maximum energy are based on the widely adopted limit on the maximum electron Lorentz factor set by equating the acceleration at Bhom rate with the synchrotron cooling rate {(see Eq.~\ref{eq:energy_max_sync})}.
Adopting this limit, synchrotron emission can not account for the TeV photons detected by MAGIC, and a different radiation mechanism must be invoked. 

An additional, model independent indication for the presence of a spectral component other than synchrotron is evident after multi-wavelength simultaneous SEDs are built. In \cite{190114C_mwl_paper} the VHE data were rebinned into five time intervals and XRT, BAT, GBM and LAT data were added when available, i.e., in the first two time intervals (Figure~\ref{fig:sed_magic_all}). 
The spectrum shows a double-peaked behaviour with a first peak in the X-ray band and the second one in the VHE band. 
The Fermi-LAT data play a particularly important role in revealing the shape of the SED, as they show a dip in the flux, strongly supporting an interpretation of the whole SED as superposition of two distinct components.

%-------------------------------
\begin{figure}[!ht]
    \centering
    \includegraphics[width=0.55\textwidth]{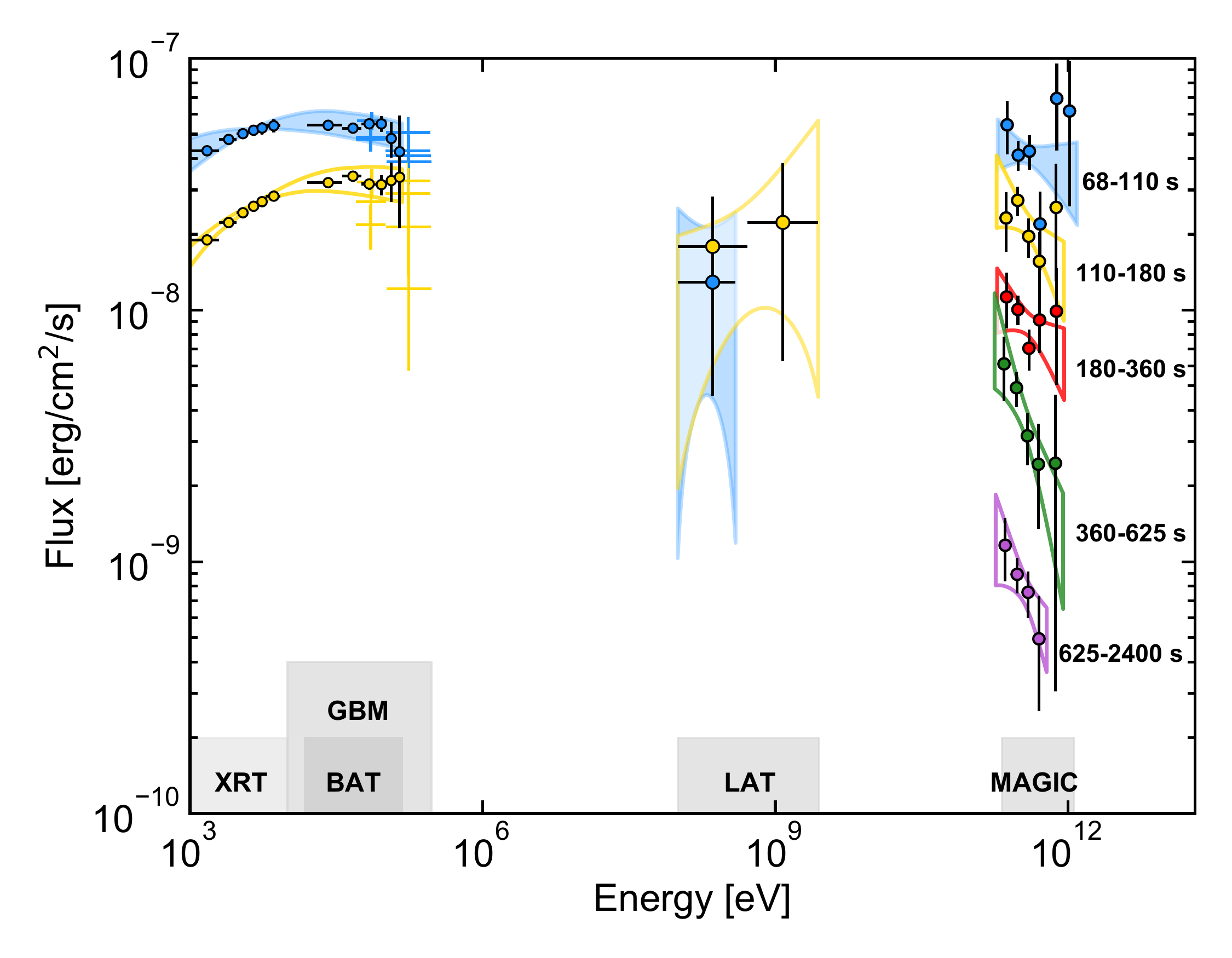}
    \caption{GRB\,190114C SEDs from soft X-rays to 1\,TeV in five different time intervals. The contour regions for flux uncertainties are also shown. For MAGIC and LAT the 1$\sigma$ error of their best-fit power-law functions are displayed. For Swift data, the 90\% confidence contours for the XRT-BAT joint fit is shown. MAGIC spectra are EBL-corrected assuming the model by \cite{dominguez_ebl}. From \cite{190114C_mwl_paper}.}
    \label{fig:sed_magic_all}
\end{figure}
%-------------------------------

Following these considerations, in \cite{190114C_mwl_paper} the SSC mechanism is explored. The broad-band emission is modeled with a numerical code reproducing the synchrotron and SSC radiation in the external forward shock scenario, including the proper KN cross section and the effects of $\gamma-\gamma$ annihilation. 

{The predicted spectra and lightcurves are compared with the data in  Figure~\ref{fig:modelling} and Figure~\ref{fig:grb_lc_190114c}.}
Acceptable modeling of the multi-wavelength afterglow spectra have been found for a constant medium with $E_k \gtrsim 3 \times 10^{53}$\,erg, $\varepsilon_e \approx 0.05-0.15$, $\varepsilon_B \approx (0.05-1) \times 10^{-3}$, $n \approx 0.5-5$ cm$^{-3}$ and $p \approx 2.4-2.6$. 
It is found that, being the peak of the SSC component below 200\,GeV, the KN suppression and the $\gamma-\gamma$ internal absorption have a non-negligible role in shaping the peak of the VHE spectrum. 
The modeling reproduces very well the XRT, LAT and TeV emission ({solid blue curve in Figure~\ref{fig:modelling} and solid blue, green and red curves in Figure~\ref{fig:grb_lc_190114c})}, while it overproduces both the optical and radio flux at late times {(solid violet, yellow and cyan curves in Figure~\ref{fig:grb_lc_190114c}).
According to \cite{190114C_mwl_paper}, a} similar fit is found also assuming a wind-like profile for the external density. In this case the parameters are $E_k = 4 \times 10^{53}$\,erg, $\varepsilon_e = 0.6$, $\varepsilon_B = 1 \times 10^{-4}$, $A_{\ast} = 0.1$ and $p = 2.4$.
Very interestingly, the modeling shows that the late LAT observation (around $10^4$\,s) is completely dominated by SSC emission (red dashed curve in Figure~\ref{fig:grb_lc_190114c}).
A different type of modelization is also investigated by \cite{190114C_mwl_paper}, under the requirement to model optical data. In this case (dotted curves in Figure~\ref{fig:grb_lc_190114c}), the fit is very good for optical, X-ray and LAT observations, but fails in reproducing the MAGIC light-curve. 

%-------------------------------
\begin{figure}[ht]
\centering
\includegraphics[width=0.55\textwidth]{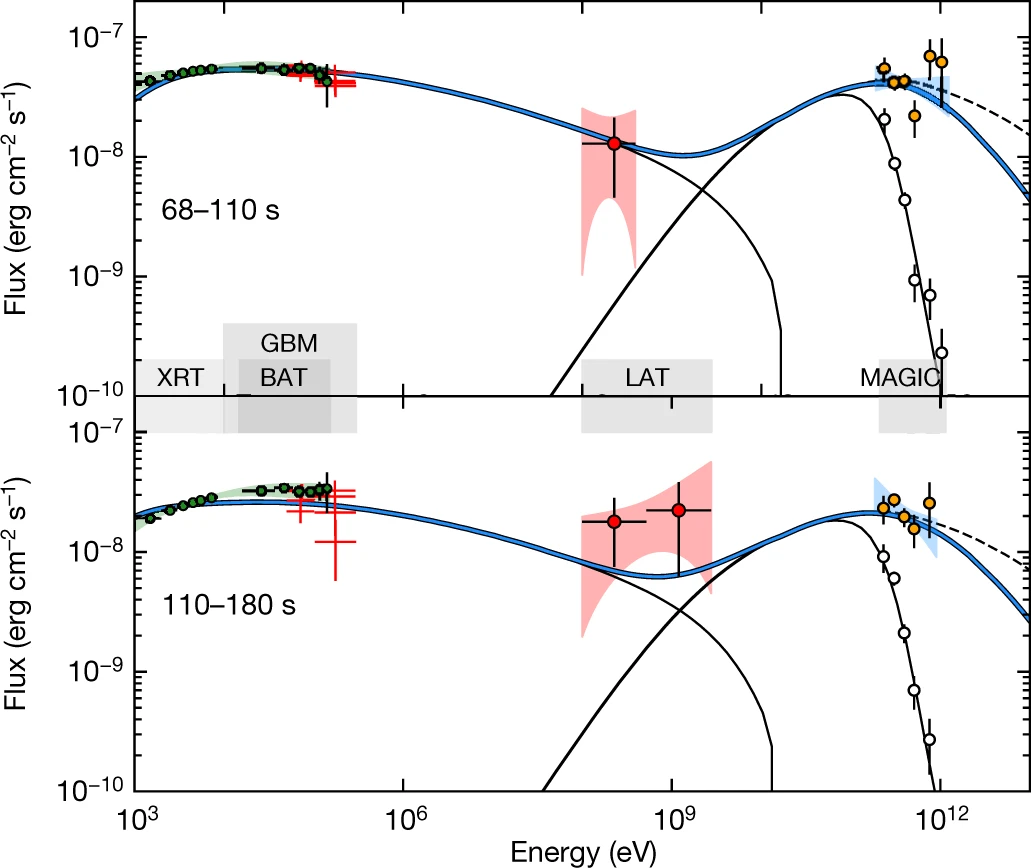}
\caption{GRB\,190114C: modeling of two SEDs in consecutive time intervals. The different curves refer to: the observed spectrum (thin solid line), the EBL deabsorbed spectrum (thick blue line) and the SSC component neglecting the effects of internal $\gamma$-$\gamma$ opacity (dashed line). From \cite{190114C_mwl_paper}.}
\label{fig:modelling}
\end{figure}
%-------------------------------

The values inferred for the GRB afterglow parameters are similar to those used for past GRB afterglow studies at lower frequencies. This is an indication that the SSC component can be a relatively common process for GRB afterglows, since it does not require peculiar values of the parameters to be explained. 

%-----------------------------------------
\begin{figure}[!ht]
\begin{center}
    \includegraphics[width=0.7\textwidth]{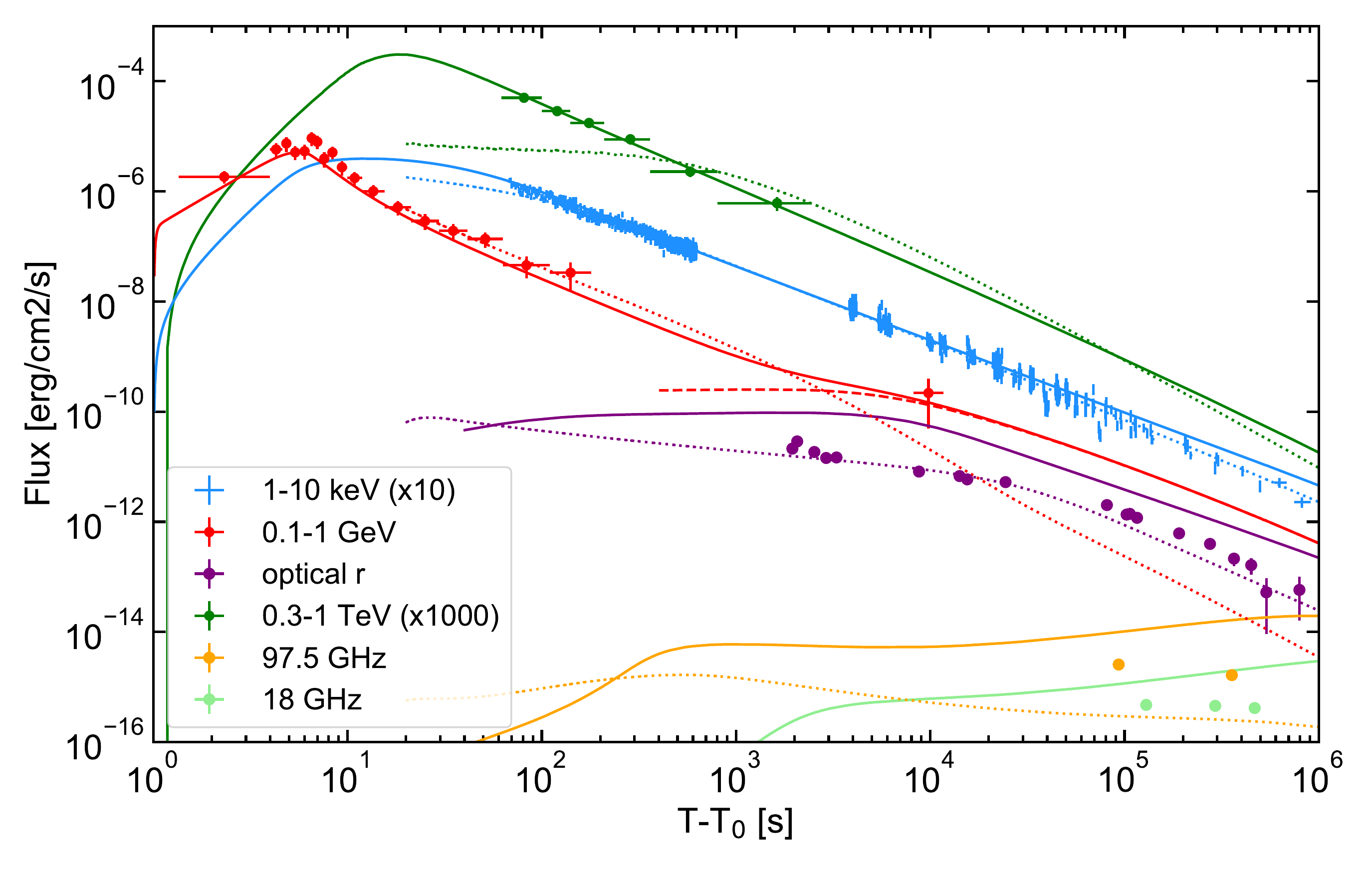}
    \caption{Modeling of broad band GRB\,190114C light curves. Description of the different modelings used is given in Section~\ref{subsubsec:interpretation_190114C}. From \cite{190114C_mwl_paper}.}
    \label{fig:grb_lc_190114c}
\end{center}
\end{figure}
%-----------------------------------------

Several other successful modelings of GRB\,190114C data within the synchrotron and SSC external forward shock scenario have been published in literature \cite{zhang_180720B_190114C,derishev_190114C,asano_190114C,razzaque_190114C}. A summary of the parameters inferred by different works can be found in Table~\ref{tab:190114C_modeling}.

In \cite{zhang_180720B_190114C} the X-ray, optical and LAT data before 100\,s are attributed to reverse shock emission or prompt contribution. A constant-density environment for the circumburst material is assumed. A time-averaged SED ($50-150$\,s) is estimated showing that at GeV energies a transition between the synchrotron and SSC component can be identified. From re-analysis of LAT data, a hard photon index ($1.76 \pm 0.21$) is derived, in agreement with the hardening of the spectrum caused by the rising of the SSC component. Differently from what seen by \cite{190114C_mwl_paper,derishev_190114C} $\gamma-\gamma$ absorption does not contribute significantly in shaping the VHE spectrum.

{A similar interpretation is given in \cite{asano_190114C}, although the inferred value of $\epsilon_B$ is larger ($\epsilon_B\sim10^{-3}$). A consistent modeling of the multi-wavelength observations as synchrotron and SSC radiation is found, both} the ISM and wind-like scenarios. The SED at 80\,s {(see figure~2 and figure~5 in \cite{asano_190114C})} and the broad-band light curves {(figure~3 and figure~6)} are reproduced, despite at 10$^{3}$\,s, the
model predictions in the $0.3-1$ TeV band and X-rays are slightly brighter than the observed data.

In \cite{razzaque_190114C}, analytical approximations are adopted for the description of the synchrotron and SSC components. In addition, the KN cut-off energy and the $\gamma-\gamma$ absorption contribution are calculated and compared with the data. A wind-like environment was used for the circumburst medium. The SEDs in the time intervals $68 - 110$\,s and $110 - 180$\,s are {modelled, and the} two values of the KN cut-off energies calculated at these times are $\sim 3.7$\,TeV and $2.1$\,TeV. This implies that the KN effect is relevant only at TeV levels and the VHE data can be modelled assuming that the SSC scattering is in Thompson regime. The $\gamma-\gamma$ absorption is also considered negligible since the estimated attenuation factor is way lower than the one due to EBL attenuation and it reaches values around unity only for energies $\gtrsim 1$\,TeV.

In \cite{derishev_190114C} the multi-wavelength data were fitted with a single-zone numerical code with an exact calculation of KN cross-sections as well as the attenuation due to the pair production mechanism. A smoothed analytical approximations for the electron injection function was used. A systematic scan over a 4-dimensional parameter space was performed to search for the best-fit solution at early and later times. The SED calculated at 90\,s and 150\,s {(figure~3 and figure~9, respectively)} are found to be well described by a fast cooling regime. The KN effect and the pair production mechanism shape the VHE spectrum significantly. It is estimated that $\simeq 10\%$ of the total emitted power, i.e., $ \simeq 25\%$ of initially produced IC power, is absorbed.

%----------------------------------------
\begin{table}[!ht]
\begin{center}
\begin{adjustbox}{width=0.72\textwidth}
\begin{tabular}{lcccccc}
\toprule
    & $E_k$ & $\epsilon_e$ & $\epsilon_B$ & n         & $p$ & $\xi_e$ \\
    & erg   &              &              & cm$^{-3}$ &     &         \\
\midrule
 \small{MAGIC Coll.}   & $ \gtrsim 3 \times 10^{53}$ & 0.05-0.15 & 0.05-1 $\times 10^{-3}$ & 0.5-5 & 2.4-2.6 & 1  \\
  \small{Wang + 2019}  & $6 \times 10^{53}$ & 0.07 & $4 \times 10^{-5}$ & 0.3 & 2.5 & 1  \\
  \small{Asano + 2020}  & $ 10^{54}$ & 0.06 & $9 \times 10^{-4}$ & 1 & 2.3 & 0.3  \\
  \small{Asano + 2020}  & $ 10^{54}$ & 0.08 & $1.2 \times 10^{-3}$ & 0.1 (wind) & 2.35 & 0.3  \\
  \small{Joshi + 2021}  & $4 \times 10^{54}$ & 0.03 & 0.012 & $2 \times 10^{-2}$ (wind) & 2.2 & 1  \\
  \small{Derishev + 2021}  & $ 3 \times 10^{53}$ & 0.1 & $2-6 \times 10^{-3}$ & 2 & 2.5 & 1  \\
  \bottomrule
\end{tabular}
\end{adjustbox}
\end{center}
\caption{GRB\,190114C: parameters inferred by different authors from the modeling of observations with a synchrotron-SSC scenario.}
\label{tab:190114C_modeling}
\end{table}
%----------------------------------------

\subsection{GRB\,190829A} \label{subsec:190829A}
GRB\,190829A is a nearby ($z=0.078$) long GRB triggered by Swift-BAT \cite{swift_gcn_190829A} and Fermi-GBM \cite{fermigbm_gcn_190829A}. The Fermi-GBM trigger time is $T_0=19:55:53.13$\,UTC.
H.E.S.S. detected this GRB over three consecutive nights, with significance of $21.7\sigma$ during the first night ($\sim4$\,h after the GRB trigger).

\subsubsection{General properties and multi-wavelength observations}
The prompt emission detected by the two instruments consists of two episodes with the first one seen in the time interval from the trigger time to 4\,s and the second brighter episode from 47\,s to 61\,s. The two episodes have very different spectral properties: the first one is described by a power-law with index $-1.41 \pm 0.08$ and an exponential high-energy cutoff function with $E_p = 130 \pm 20$\,keV, the second one can be described with a Band function  with $E_p = 11 \pm 1$ keV , $\alpha = -0.92\pm 0.62$ and $\beta_T = -2.51 \pm 0.01$ \cite{chand_190829A}.  The (isotropic equivalent) prompt emission energy inferred from spectral analysis of Fermi-GBM data is $E_{\gamma,iso} \sim 2 \times 10^{50}$\,erg.

A multi-wavelength observational campaign of the event was performed covering the entire electromagnetic spectrum (see Figure~\ref{fig:hess_lc_190829A} and \ref{fig:om_modeling_190829A}). 
The event was not detected in the HE range by Fermi-LAT. Nevertheless, ULs have been reported in the MeV-GeV band up to $3 \times 10^4$\,s \cite{lat_gcn_190829A}.
The Swift-XRT started observations at 97.3\,s and detected a bright X-ray afterglow, which was monitored until $\sim 7.8 \times 10^6$\,s \cite{swift_xrt_190829A}. The X-ray light curve in the 0.3-10\,keV energy range (observer frame) shows a peculiar behaviour with an initial steep decay phase followed by a plateau and a strong flare episode (Figure~\ref{fig:om_modeling_190829A}, upper panel, blue points). After the flare the standard afterglow phase starts with a decay following a power-law with a possible steepening around 10 days.
In the UV/optical/NIR band the event was followed by several instruments. 
The redshift was estimated to be $z = 0.0785 \pm 0.005$ \cite{gtc_190829A}, which makes this event one of the closest GRBs ever detected. Starting from 4.5 - 5.5 days after the GRB trigger an associated supernova has been reported \cite{supernova_gcn_190829A}. Also in the optical data a flare is seen simultaneously with the one in X-rays.  In the radio band the detection was reported by several instruments starting from $\sim 1$ day after the trigger \cite{190829A_radio_atca,190829A_radio_meerkat,190829A_radio_noema,190829A_radio_uGMRT}. The radio flux initially slowly increases and then starts to decay after 20-30 days.

\subsubsection{VHE observations and results}
%--------------------------------
\begin{figure}[!ht]
    \centering
    \includegraphics[width=0.6\textwidth]{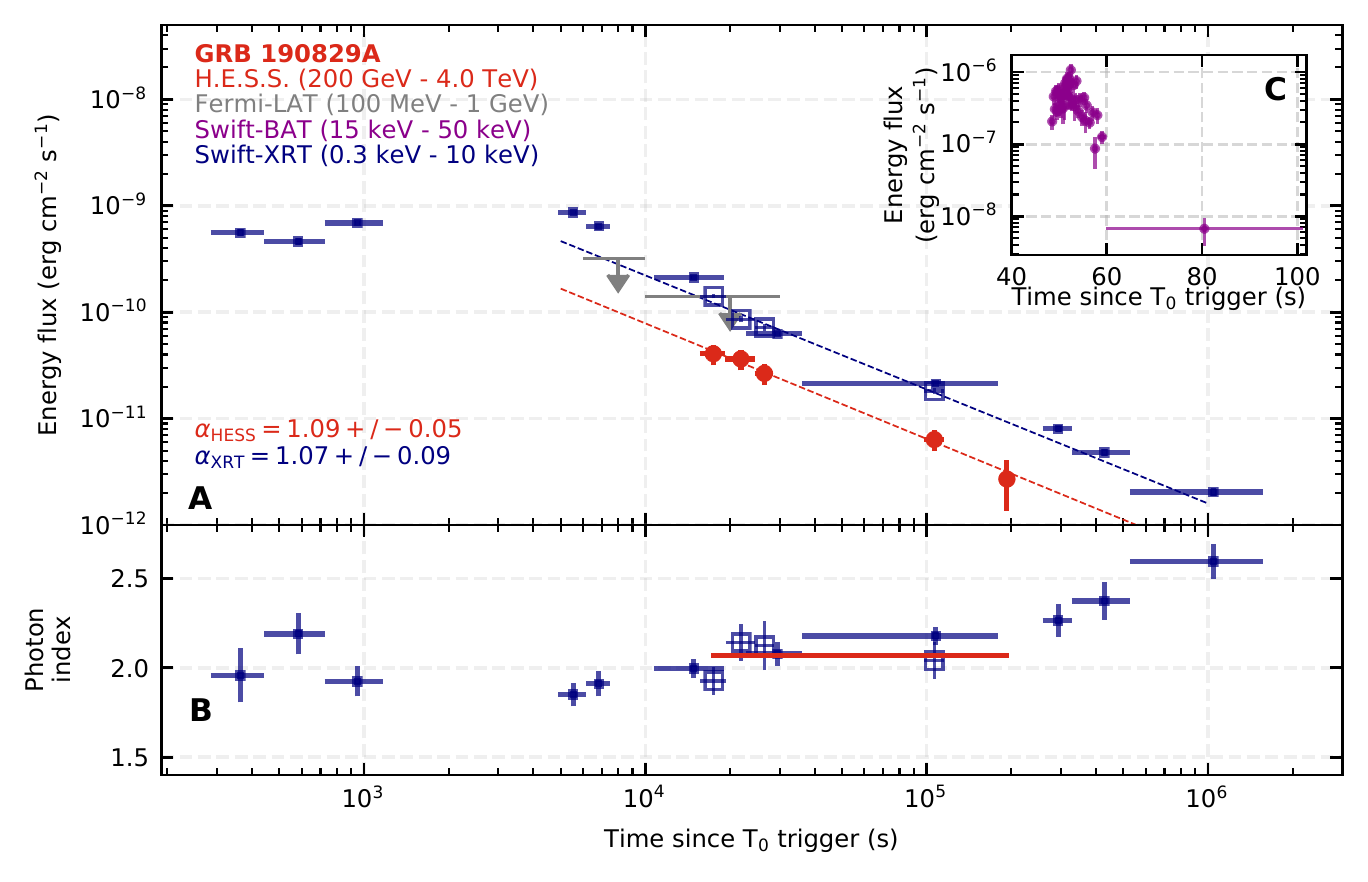}
    \caption{GRB\,190829A: multi-wavelength light curves {(A)} (upper panel) and photon index evolution (bottom panel) in the X-ray, HE and VHE band {(B)}. The BAT prompt light curve is shown in the inset {(C)}. From \cite{190829A_hess}.}
    \label{fig:hess_lc_190829A}
\end{figure}
%--------------------------------

The afterglow emission of GRB\,190829A was followed-up by the H.E.S.S. telescopes starting at 4.3\,h and continued for three consecutive nights (up to 55.9\,h). Observations were performed using the four medium-size telescopes of H.E.S.S. for a total amount of 13\,h divided respectively in 3.6\,h (starting at 4.3\,h), 4.7\,h (starting at 27.2\,h) and 4.7\,h (starting at 51.2\,h). The statistical significance at the GRB position found in the three nights are respectively $21.7 \sigma$, $5.5 \sigma$ and $2.4 \sigma$. 

Spectral analysis was performed for the first two nights fitting the observed photon spectrum with a power-law model. The following values are found: $\alpha_{obs} = - 2.59 \pm 0.09$ (stat.) $\pm 0.23$ (syst.) in the $0.18 - 3.3$\,TeV (first night) and $ \alpha_{obs} = - 2.46 \pm 0.22$ (stat.) $\pm 0.14$ (syst.) in the 0.18 - 1.4\,TeV energy range (second night). 
Fitting a power-law attenuated by EBL, the photon indices inferred for the intrinsic spectrum are: $\alpha_{int} = - 2.06 \pm 0.10$ (stat.) $\pm 0.26$ (syst.) in the $0.18 - 3.3$\,TeV energy range (first night) and $ \alpha_{int} = - 1.86 \pm 0.26$ (stat.) $\pm 0.17$ (syst.) in the 0.18 - 1.4\,TeV energy range (second night).  The photon indices in each night are consistent, within systematical uncertainties, with those of the simultaneous X-ray emission.
Combining all three nights, the photon index is $\alpha_{int} = - 2.07 \pm 0.09$ (stat.) $\pm 0.23$ (syst.) in the 0.18 - 3.3\,TeV energy range.

The light-curve in the 0.2 - 4.0\,TeV energy range derived up to 56\,h is compared in Figure~\ref{fig:hess_lc_190829A} with the XRT light-curve and the LAT upper limits. The time-evolving flux was satisfactorily modeled with a power-law decay $F(t) \propto t^{\alpha}$ with $\alpha = - 1.09 \pm 0.05$. Such decay index is similar to the X-ray one derived in the same time interval ($\alpha_X = - 1.07 \pm 0.09$).

\subsubsection{Interpretation}
The interpretation of the VHE emission from GRB\,190829A is debated and different radiation mechanisms including synchrotron, SSC or EIC emission have been proposed so far to explain the origin of the TeV emission.

%--------------------------------
\begin{figure}[!ht]
    \centering
    \includegraphics[width=0.6\textwidth]{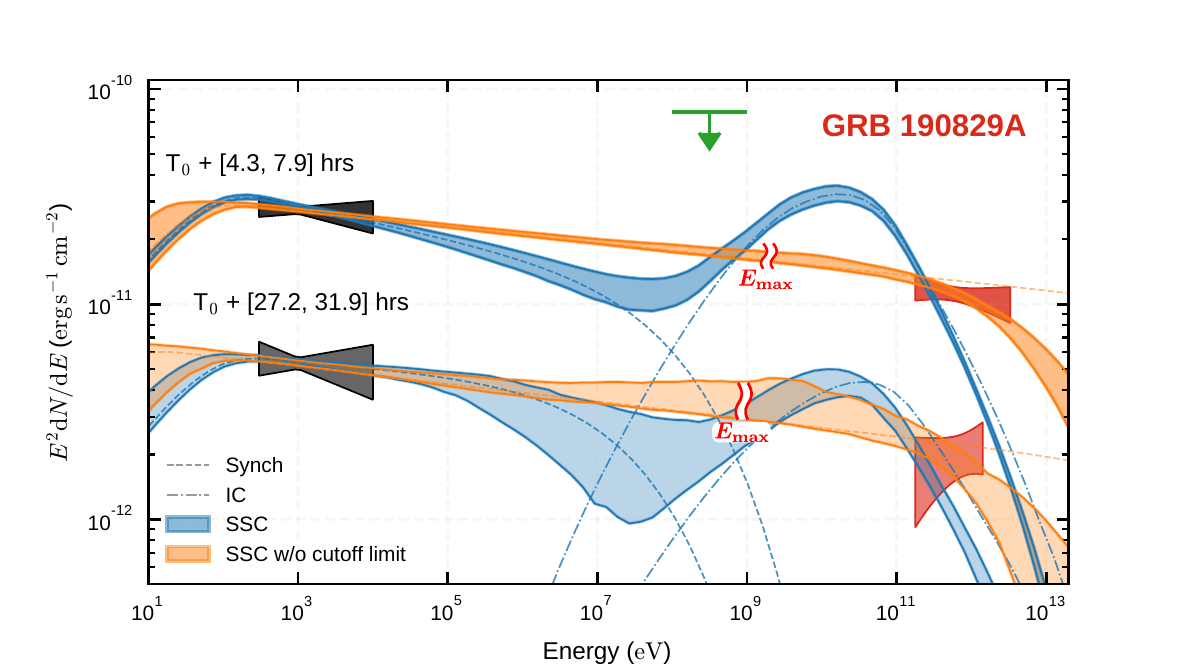}
    \caption{GRB\,190829A: modeling of X-ray, LAT and H.E.S.S. data proposed by the H.E.S.S. collaboration for the two time intervals with VHE and X-ray detections. Two scenarios are investigated for the TeV emission: synchrotron and SSC. H.E.S.S. flux contours are displayed considering the statistical uncertainty. The synchrotron and SSC component are shown in dashed and dash-dotted lines respectively.  The shaded areas represent the 68$\%$ confidence intervals determined from the posterior probability distribution of the MCMC parameter fitting for the standard SSC model (light blue) and for the
model without maximum energy for synchrotron emission (orange). From \cite{190829A_hess}.}
    \label{fig:hess_modeling_190829A}
\end{figure}
%--------------------------------

The H.E.S.S. Collaboration \cite{190829A_hess} investigated both the synchrotron and the SSC emission in the external forward shock as responsible radiation mechanism of the observed TeV component. Multi-wavelength data collected simultaneously with H.E.S.S. observations in the first two nights were modeled separately with a time-independent numerical code using the Markov-chain Monte Carlo (MCMC) approach to explore the parameter space. The results of the fitting show that the SSC mechanism fails to explain the VHE emission. The low Lorentz bulk factor predicted by the observations ($\Gamma \lesssim 10$) implies that the SSC emission occurs in KN cross-scattering regime. As a result, a steep spectrum, inconsistent with the observational VHE data, is obtained {(see Figure~\ref{fig:hess_modeling_190829A}, light blue shaded area)}. Possible improvements between the data and the model foresee a higher $\Gamma$, which is in contrast with the observations, or the presence of an additional hard component in the distribution of the accelerated electrons. However, this latter solution implies extreme assumptions on the density of the circumburst medium ($n_0 = 10^{-5}$ cm$^{-3}$ in case of strong magnetic field or $n_0 = 10^{5}$ cm$^{-3}$ for weak magnetic field) and a SED strongly dominated by the SSC component which is inconsistent with the data. 
A better fitting of the observational data can be obtained when considering an alternative model where the maximum electron energy set by the radiative losses is ignored. In such scenario, the synchrotron emission is able to extend up to TeV energies and the observational broad-band data are described by a single synchrotron component {(see Figure~\ref{fig:hess_modeling_190829A}, orange shaded area).} The SSC contribution is negligible while the $\gamma-\gamma$ absorption shapes the VHE spectrum. The single synchrotron component scenario provides better fit ($> 5 \sigma$) to the multi-wavelength data. On the other hand, this interpretation requires unknown acceleration processes or non-uniform magnetic field strength in the emission region as described for GRB\,180720B {(see Section~\ref{subsec:180720B})}.

%--------------------------------
\begin{figure}[!ht]
    \centering
    {
    \includegraphics[width=0.6\textwidth]{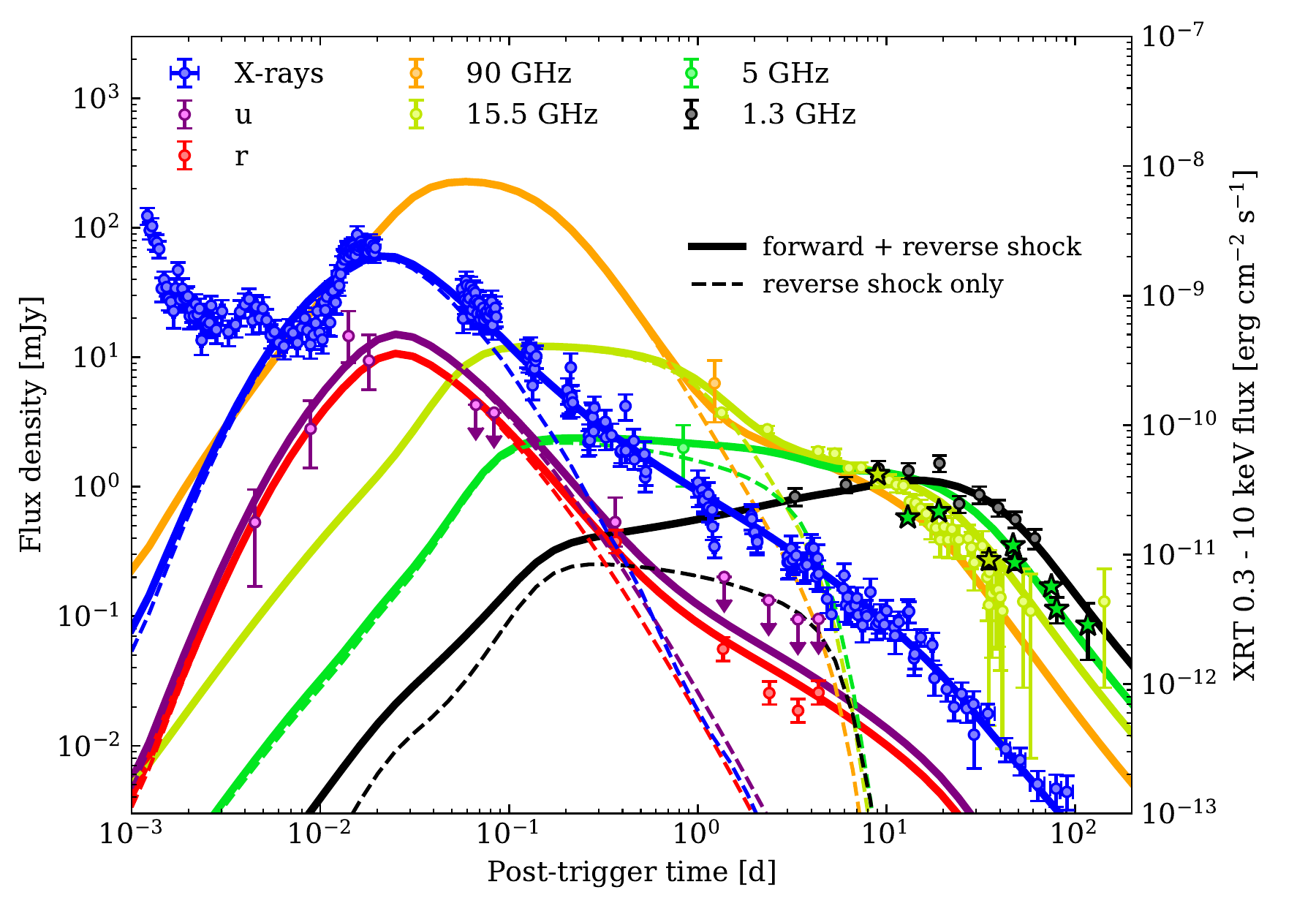}
    \includegraphics[width=0.6\textwidth]{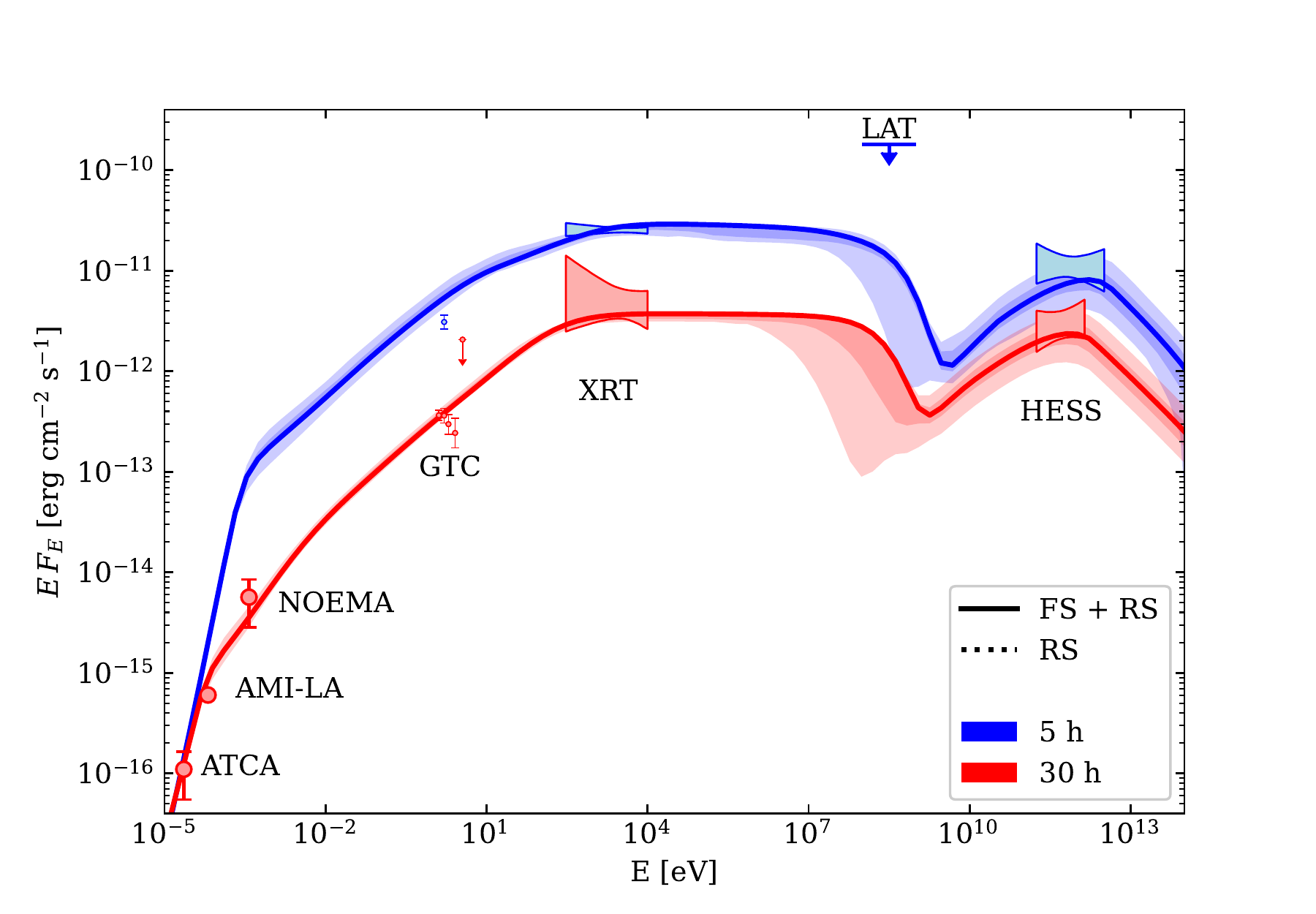}}
    \caption{GRB\,190829A: modeling of multi-wavelength light curves (upper panel) and SED (bottom panel) proposed by \cite{190829A_salafia}. The 90$\%$ and 50$\%$ credible intervals from the fit are shown in lighter shades.}
    \label{fig:om_modeling_190829A}
\end{figure}
%--------------------------------

A complete multi-wavelength modeling of the GRB\,190829A data including contribution of the synchrotron and SSC emission for both the forward and reverse shocks considering a constant-density environment is presented in \cite{190829A_salafia}. The predicted broad-band light curves and the SED at the time of the H.E.S.S. detection are shown in Figure~\ref{fig:om_modeling_190829A}. A MCMC approach was adopted in order to estimate the best-fit parameters for the multi-wavelength modeling. The resulting values of the parameters related with the forward shock scenario are shown in Table~\ref{tab:190829A_modeling}. In contrast with the H.E.S.S. Collaboration results, the VHE emission is well reproduced with the SSC external forward shock scenario. The usual simplified assumption that $\xi_e = 1$ is excluded by the fit which provides acceptable solutions only for $\xi_e \lesssim 6.5 \times 10^{-2}$. Moreover, an isotropic-equivalent kinetic energy at the afterglow onset $E_{k} = 2.5^{+1.9}
_{-1.3} \times 10^{53}$ erg is estimated. Considering the observed GBM prompt energy, such value implies that the prompt efficiency is $\eta = 1.2^{+1.0}_{-0.5} \times 10^{-3}$ which is much lower than the typical values derived from previous GRB studies. The other parameters ($n_0$, $\epsilon_e$ and $\epsilon_B$) are found to be similar to the ones estimated for GRB\,190114C.

A two-component off-axis jet model has also been investigated \cite{190829A_twocomponentjet}. Such model proposes that the GRB jet is seen off-axis ($\theta_{view} = 1.78^{\circ}$) and it consists of a narrow ($\theta_{jet} = 0.86^{\circ}$) fast ($\Gamma = 350$) jet and a slow ($\Gamma = 20$) co-axial jet. The former jet component is responsible for the emission of SSC photons in the VHE band. The calculation of the SSC flux at the time of the H.E.S.S. detection is done following the prescriptions of \cite{sariesin} considering only the Thompson scattering regime.

An EIC plus SSC scenario has also been proposed for the production of the VHE component \cite{190829A_EIC}. The seed photons belong to the long-lasting X-ray flare seen for GRB\,190829A which can be up-scattered to TeV energies. A numerical calculation of the afterglow dynamics and radiative processes have been used to model the observational data. For $t \sim 10^{3} - 10^{4}$ s the EIC component dominates the VHE emission, while for later times ($t \gtrsim 3 \times 10^{4}$ s) the EIC gradually decays and the SSC component becomes relevant. The initial afterglow kinetic energy used for the modeling ($E_k = 10^{52}$ erg) suggests that GRB\,190829A is not a typical low-luminosity GRB but it may have much higher kinetic energy.

%--------------------------------
\begin{table}[!ht]
\begin{center}
\begin{adjustbox}{width=0.72\textwidth}
\begin{tabular}{lccccccc}
\toprule
    & $E_k$ & $\epsilon_e$ & $\epsilon_B$ & n  & $p$ & $\xi_e$ & $\theta_j$ \\
    & erg & & & cm$^{-3}$ & & & rad \\
\midrule
 \small{Hess Coll. (SSC)}   & $2.0 \times 10^{50}$ & 0.91 & $5.9-7.7 \times 10^{-2}$ & 1. & 2.06-2.15 & 1. & / \\
 \small{Hess Coll. (Sync)}   & $2.0 \times 10^{50}$ & 0.03-0.08 & $\approx 1$ & 1. & 2.1 & 1. & / \\
  \small{Salafia + 2021}   & $1.2-4.4 \times 10^{53}$ & 0.01-0.06 & $1.2-6.0 \times 10^{-5}$ & 0.12-0.58 & 2.01 & $< 6.5 \times 10^{-2}$ & 0.25-0.29 \\
  \small{Zhang + 2021}  & $9.8 \times 10^{51}$ & 0.39 & $8.7 \times 10^{-5}$ & 0.09 & 2.1 & 0.34 & 0.1 \\
  \bottomrule
\end{tabular}
\end{adjustbox}
\end{center}
\caption{Parameters for modeling of GRB\,190829A}
\label{tab:190829A_modeling}
\end{table}
%--------------------------------

%-------------------------------------------------------
\subsection{GRB\,201015A}\label{subsec:201015A}
GRB\,201015A is a long GRB at $z=0.426$ detected by the Swift-BAT \cite{swift_bat_201015A} on 15 October 2020, at $T_0=22:50:13$\,UT. The Fermi-GBM instrument did not trigger the event but the targeted search revealed a transient source consistent with the Swift-BAT location \cite{fermi_gbm_201015A}.
MAGIC observations show a possible detection with a significance of $\sim3.5\sigma$.

\subsubsection{General properties and multi-wavelength observations}
The (isotropic equivalent) prompt emission energy inferred from
spectral analysis of Fermi-GBM data is $E_{\gamma,iso} = (1.1 \pm 0.2) \times 10^{50}$\,erg \cite{e_iso_201015A}. The prompt duration is $T_{90} = 9.78 \pm 3.47$\,s ($15 - 350$ keV band). The BAT time-average spectrum in the time interval $0-10$\,s is well fitted by a power-law model with photon index $\beta_T = -3.03 \pm 0.68$ suggesting a low peak energy $E_p < 10$\,keV \cite{swift_bat_201015A_refined}. 

Swift-XRT \cite{swift_xrt_201015A} follow-up the event starting only 3214\,s after $T_0$ due to observational constraints. The light curve up to almost 1 day is well described by a power-law with decay index $\alpha = - 1.49^{+0.24}_{-0.21}$. Late-time observations performed by the Chandra X-ray Observatory \cite{chandra_201015A} and Swift-XRT \cite{swift_xrt_latetime_201015A} from $\sim 8$ days up to $\sim 21$ days showed a flattening of the X-ray light curve, i.e. a flux level inconsistent (higher) with the extrapolation of the power-law decay rate at early times. 
Optical observations confirmed the presence of an afterglow counterpart from around 168\,s \cite{optical_master_net_201015A}. The optical light-curves showed a clear initial rise with a peak around $200$\,s followed by a decay \cite{optical_nutella_201015A}. A bright radio counterpart (flux density $\sim 1.3 \times 10^{-4}$\,Jy at 6\,GHz 1.4 days after the burst) was also detected by several instruments \cite{radio_vla_201015A,radio_emerlin_201015A,radio_evn_201015A}. Late-time optical observations identified an associated supernova rising from 5 days after the burst reaching its maximum flux around 12-20 days after $T_0$ \cite{supernova_201015A,supernova_lbt_201015A}.
The measurement of the redshift was reported by the GTC ($z = 0.426$) \cite{gtc_redshift_201015A} and then confirmed by the NOT ($z = 0.423$) \cite{not_redshift_201015A} instrument. 

\subsubsection{VHE observations and results}

%-----------------------------------
\begin{figure}[!ht]
\begin{center}
    \includegraphics[width=0.5\textwidth]{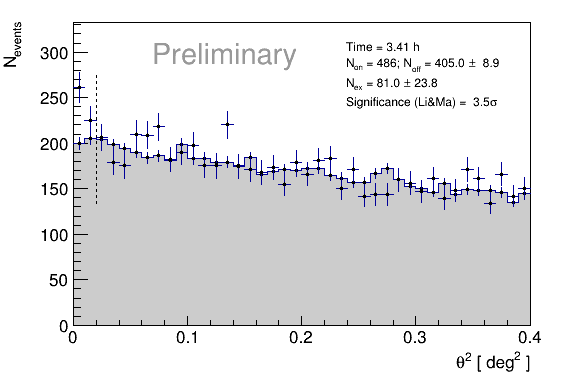}
    \caption{GRB\,201015A: distribution of the angular distance $\theta^2$ between the reconstructed event arrival directions and the nominal source position.  {The gray histogram represents background events while the black point with blue crosses are the $\gamma$-like events.} The vertical dashed line describes the $\theta^2$ cut value and defines the region in which excess events and signal significance are calculated. From \cite{201015A_magic_proc}.}
    \label{fig:201015A_theta2plot}
\end{center}
\end{figure}
%-----------------------------------
Final results from VHE data analysis of GRB\,201015A have not been published yet. Preliminary information reported here have been released in \cite{magic_201015Agcn} and \cite{201015A_magic_proc}.
Observations of GRB\,201015A were performed by the MAGIC telescopes starting 33\,s after the trigger time, under dark conditions, with a zenith angle ranging from 24$^{\circ}$ up to 48$^{\circ}$, and lasted for about 4\,h. In the second half of the data taking the presence of passing clouds affected the observation for $\sim$0.45\,h. These data were removed and the remaining ones were analyzed with the standard MAGIC analysis software. Offline analysis showed a possible excess with a $3.5 \sigma$ significance at the GRB position (see Figure~\ref{fig:201015A_theta2plot}) and a significant spot in the sky map. The energy threshold of the analysis is calculated to be 140\,GeV from Monte Carlo simulated $\gamma$-ray data.

\subsection{GRB\,201216C}\label{subsec:201216C}
GRB\,201216C is a long GRB at $z=1.1$ triggered by the Swift-BAT at $T_0=23:07:31$\,UT on 16 December 2020 \cite{swift_bat_201216C}. Fermi-GBM also detected the event with a slightly different trigger time (6 second before the Swift-BAT) \cite{fermi_gbm_201216C}. 
MAGIC detected GRB\,201216C with a significance of $\sim6\sigma$.

\subsubsection{General properties and milti-wavelength observations}
The duration is estimated as $T_{90} = 48 \pm 16$\,s in the 15-350\,keV band by Swift-BAT \cite{swift_bat_refined_201216C} and around $29.9$\,s in the 50-300\,keV band by Fermi-GBM. The light curve shows a multiple peak structure with a main peak around 20\,s after the trigger time. The time-averaged GBM spectrum in the first 50\,s is best fit by a Band function with $E_{p} = 326 \pm 7$\,keV,
$\alpha = -1.06 \pm 0.01$ and $\beta_T = -2.25 \pm 0.03$. 
The isotropic equivalent energy $E_{\gamma,iso}$ in the 10-1000\,keV band is
$(4.71 \pm 0.16) \times 10^{53}$\,erg, as calculated from the fluence measured by Fermi-GBM.

Fermi-LAT observed the GRB starting from around 3500\,s and up to 5500\,s. No significant emission was reported \cite{fermi_lat_201216C}.
Swift-XRT began the observation at $t=2966.8$\,s due to an observational constrain. A fading source was detected following a broken power-law behaviour with decay indices of $1.97^{+0.10}_{-0.09}$ and $1.07^{+0.15}_{-0.10}$ and a break at 9078\,s \cite{swift_xrt_201216C_refined}. Optical observations were also performed by several instruments. The $r$-band light curve made along with VLT data point \cite{optical_vlt_201216C} and inferred data from FRAM-ORM \cite{optical_fram_orm_201216C} show a power-law decay in flux with index equal to 1. The Liverpool Telescope observations, performed around 177\,s after the trigger time, seems to be around the peak of the optical afterglow \cite{optical_liverpool_201216C}. The HAWC observatory followed-up the event but no significant detection was identified in the TeV band \cite{hawc_201216C}.
Redshift estimation of $z = 1.1$ was performed by the ESO VLT \cite{optical_redshift_vlt_201216C}. 

\subsubsection{VHE observations and results}

%----------------------------------
\begin{figure}[!ht]
\begin{center}
    \includegraphics[width=0.5\textwidth]{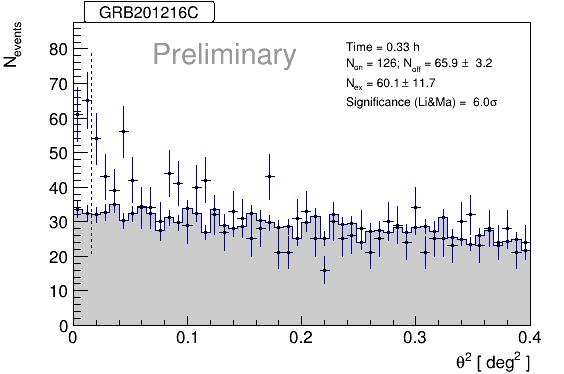}
    \caption{$\theta^2$ angular distance distribution between reconstructed event arrival directions and nominal source position for GRB\,201216C.  {The gray histogram represents background events while the black point with blue crosses are the $\gamma$-like events.} The vertical dashed line describes the $\theta^2$ cut value and defines the region in which excess events and signal significance are calculated. From \cite{201216C_magic}.}
    \label{fig:201216C_theta2plot}
\end{center}
\end{figure}
%----------------------------------
Final results from VHE data analysis of GRB\,201216C have not been published yet. Preliminary information reported here have been released in \cite{magic_201216C_gcn} and \cite{201216C_magic}.
MAGIC observations and data taking of GRB\,201216C started with a delay of 56\,s after the Swift-BAT trigger time. The observation lasted for 2.2\,h and was performed in optimal atmospheric condition and in absence of Moon. The zenith angle ranged from 37$^{\circ}$ to 68$^{\circ}$. The low level of night sky background allowed to retain also the low energy events and therefore obtain a low energy threshold with compared to the other GRBs observed. To keep as many low-energy events as possible, an image cleaning method
to extract dimmer Cherenkov showers initiated by gamma rays than the standard method was adopted.

The signal significance was calculated to be $6.0 \sigma$ pre-trial ($5.9 \sigma$ post-trial\endnote{A trial factor of 2 is considered due to the two sets of analysis cuts used for MAGIC data analysis.}) for the first 20 minutes of observation (see Figure~\ref{fig:201216C_theta2plot}). A preliminary time-integrated spectrum for the first 20 minutes of observation was produced. Due to the strong absorption effect by EBL a very steep power-law decay was found for the observed spectrum, especially for the events with energies higher than a few hundreds of GeV. The intrinsic spectrum, corrected for the EBL absorption was found to be consistent with a flat single power-law until 200 GeV above which no significant spectral points have been derived. A preliminary light curve in the time interval 56\,s - 2.2\,h was also calculated. After 50\, min only upper limits on the emitted flux have been derived since no significant emission was found after this time. The preliminary results are consistent with a monotonically decaying light curve fitted with a power-law. 

\section{The new TeV spectral window: discussion}\label{chap:discussion}

After decades of searches, MAGIC and H.E.S.S. observations have unequivocally proved that (long) GRBs can be accompanied by a significant amount of TeV emission during the afterglow phase. 
Table~\ref{tab:summary} summarizes the main properties of the GRBs detected by IACTs, and presented in detail in the previous section.
The list includes also two events (namely GRB\,160821B and GRB\,201015A) where only a hint of excess (i.e., with a significance at $\sim 3-4 \sigma$) was found. For the other four events, namely GRB\,180720B, GRB\,190114C, GRB\,190829A and GRB\,201216C the detections are robust ($> 5 \sigma$).
The table lists several properties, such as duration $T_{90}$ and total emitted energy $E_{ {\gamma,iso}}$ of the prompt emission, redshift, and information on the IACT detection (the starting time $T_{delay}$ of observations elapsed since the trigger time $T_0$, the energy range where photons have been detected, the name of the telescope and the significance of the excess). 
GRB\,160821B is the only one belonging to the short class, the other five being long GRBs. 

In this section we address the question why these GRBs have been detected, whether they have peculiar properties and whether they show some common behaviour which may be at the basis of the production of TeV radiation.
To do that, one should be careful, since these GRBs have been followed-up under very different observational conditions and with very different time delays after the trigger time, and they span a quite large range of redshifts (from 0.078 to 1.1). 
Keeping in mind these differences, which have a strong impact on the detection capabilities of IACTs, we compare the observed and intrinsic properties of the population of GRBs at VHE, highlighting their similarities and differences, and discuss how they compare to the whole population. 

%-----------------------------------------------------------
\begin{table}[!ht]
\begin{center}
\begin{adjustbox}{width=0.72\textwidth}
\begin{tabular}{lcccccc}
\toprule
    & $T_{90}$ & $E_{\gamma,iso}$ & z & $T_{delay}$ & $E_{range}$  & IACT (sign.)  \\
    & s & erg & & s & TeV & \\
\midrule
 \small{160821B}   & 0.48 & $1.2 \times 10^{49}$ & 0.162 & 24 & 0.5-5 & MAGIC ($ 3.1\sigma$)  \\
  \small{180720B}  & 48.9 & $6.0 \times 10^{53}$ & 0.654 & 3.64$ \times 10^{4}$ & 0.1-0.44 & H.E.S.S. ($ 5.3\sigma$)   \\
  \small{190114C}  & 362 & $2.5 \times 10^{53}$ & 0.424 & 57 & 0.3-1 & MAGIC ($> 50\sigma$)  \\
  \small{190829A}  & 58.2 & $2.0 \times 10^{50}$ & 0.079 & 1.55 $ \times 10^{4}$ & 0.18-3.3 & H.E.S.S. ($ 21.7\sigma$)  \\
  \small{201015A}  & 9.78 & $1.1 \times 10^{50}$ & 0.42 & 33 & 0.14 & MAGIC ($ 3.5\sigma$)   \\
  \small{201216C}  & 48 & $4.7 \times 10^{53}$ & 1.1 & 56 & 0.1 & MAGIC ($ 6.0\sigma$)   \\
  \bottomrule
\end{tabular}
\end{adjustbox}
\end{center}
\caption{List of the GRBs observed by IACTs with a firm detection (significance $>5\sigma$) or a hint of detection ($3-4\sigma$) above 100\,GeV. The $T_{90}$ and $E_{\gamma,iso}$ refer to the duration and total emitted energy of the prompt emission; the redshift is listed in column 3; $T_{delay}$ is the time delay between the trigger time $T_0$ and the time when IACT observations started; $E_{range}$ defines the energy range of the detected photons. The name of the telescope which made the observation and the significance of the detection are listed in the last column.}
\label{tab:summary}
\end{table}
%-----------------------------------------------------------

\subsection{Observing conditions}
Low zenith angles, fast repointing, dark nights, low redshift, and highly energetic events have always been considered as optimal, if not necessary, conditions to have some chances for GRB detections with IACTs.
On the other hand, these first VHE GRBs have demonstrated that GRBs can have a level of TeV emission large enough to be detected by the current generation of IACTs, even under non-optimal conditions.
GRB\,190114C was observed with a zenith angle $> 55^{\circ}$ and in presence of the Moon. Both conditions imply a higher energy threshold (typically $\gtrsim$ 0.2 TeV) and require a dedicated data analysis. 
Another example is GRB\,160821B, that was observed with a NSB 2-8 times higher than the standard dark night conditions. Moreover, significant VHE excess was found not only in case of short delays (less than hundreds of seconds) from the burst trigger but, somewhat surprisingly, also at quite late times, i.e. with delays of several hours or even days, as in the case of GRB\,180720B and GRB\,190829A, respectively. 
This showed the importance of pointing a GRB also at relatively late times, in cases fast follow-up observations are not feasible. 

Optimal observing conditions and short delays remain however crucial to detect GRBs at higher redshift, for which the impact of EBL is large already at a few hundreds GeV.
This explains how it has been possible the detection of a GRB at $z=1.1$ (GRB\,201216C): in this case, optimal observing conditions  allowed to reach a low energy threshold of the sensitivity window ($\sim70$\,GeV). The excess of signal was indeed found only below 200\,GeV (more precisely, between 70 and 200\,GeV) where the attenuation by the EBL is still limited.

\subsection{Redshift and the impact of EBL}
The redshift of the detected GRBs covers a broad range, from $z = 0.079$ (GRB\,1908\-29A) to $z = 1.1$ (GRB\,201216C). The impact of the EBL attenuation on the spectrum is severely changing, depending on the redshift value and on the photon energy. For redshift $z \sim 0.4$ the impact becomes relevant for energies $\gtrsim 0.2$ TeV with a flux attenuation of $\sim 50\%$ for $0.2$ TeV and almost $\sim 99.5\%$ for $1$ TeV \cite{dominguez_ebl}. 
For nearby events ($z \lesssim 0.1-0.2$) the effect of EBL is less severe and becomes relevant only for energies $ \gtrsim 0.3$\,TeV, reaching an attenuation factor of an order of magnitude only for energies $\gtrsim 2$ TeV. 
As a result, the GRB observed photon indices and the energy range of detected TeV photons differs significantly between the events. 
GRBs with redshift $z > 0.4$ such as GRB\,190114C or GRB\,180720B have very steep photon indices and they are detected in the lower energy range up to 0.44\,TeV for GRB\,180720B and 1.0 TeV for GRB\,190114C. Spectral analysis from GRB\,201216C are not yet public but preliminary results indicate that the emission is concentrated in the lower energy band between 0.1-0.2 TeV. Nearby GRBs with redshift $z \lesssim 0.1-0.2$ such as GRB\,160821B or GRB\,190829A show a less steep photon spectrum (around $-2.5$) and the TeV detection range extends above 1\,TeV. The detection of several GRBs with significant value of redshift ($z > 0.4$) is a robust proof that IACTs can overcome the limitations due to the EBL absorption and can expand the VHE detection horizon at the current stage up to $z = 1.1$. On the other hand, it is evident that detection of nearby GRBs is fundamental in order to explore more robustly the spectral shape, unbiased by the EBL effect which is a non-negligible source of uncertainty for higher redshifts.

%----------------------------------
\begin{figure}[!ht]
\begin{center}
    \includegraphics[width=0.5\textwidth,trim={0 0 0.8cm 0},clip]{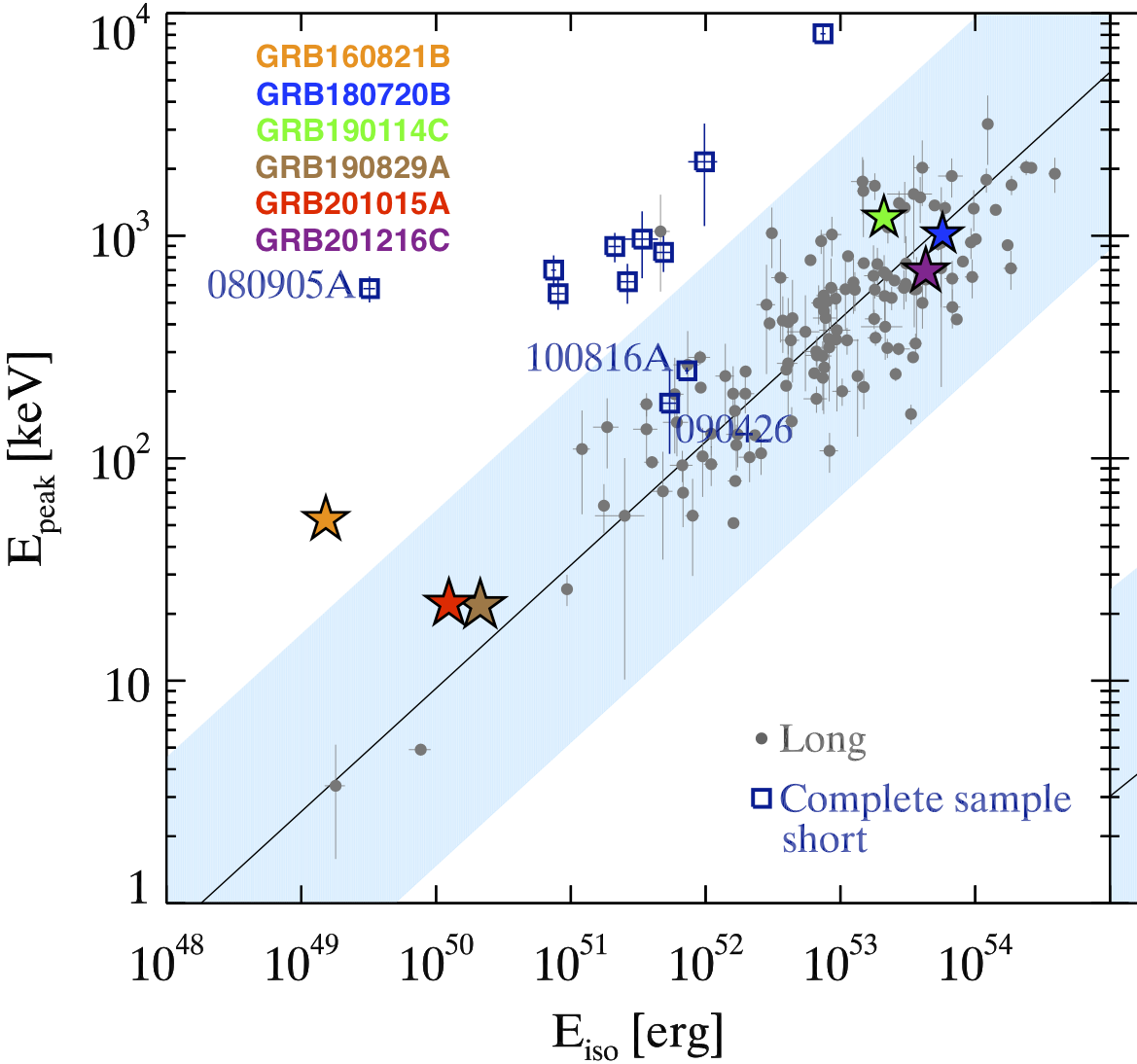}
    \caption{The Amati ($E_{peak}$ - $E_{iso}$) correlation for a sample of 136 long GRBs (grey dots, from \cite{nava12}) and a sample of 11 short GRBs (empty blue squares) detected by Swift. The corresponding power-law fit for the sample of long GRBs and the $3\sigma$ scatter of the distribution of points around the best fits are shown. The six GRBs detected in VHE are also added in the plot. Adapted from \cite{davanzo14}.}
    \label{fig:amati}
\end{center}
\end{figure}
%----------------------------------

\subsection{Energetics}
In terms of $E_{\gamma,iso}$ the VHE GRB sample spans more than three orders of magnitude from $\sim 10^{49}$ erg up to $\sim 6 \times 10^{53}$ erg. {The five long GRBs detected follows the Amati correlation as shown in Figure~\ref{fig:amati}. GRB\,160821B, the only short GRB of the sample, is consistent with the existence of a possible Amati-like correlation for short GRBs, with this event falling in the weak-soft part of the correlation. The} detections of GRB\,190829A and GRB\,201015A show that an event does not need to be extremely energetic in terms of isotropic-equivalent prompt energy in order to produce a TeV emission with (intrinsic) luminosity comparable to the X-ray luminosity. As a result, sources with $E_{\gamma,iso} \sim 10^{50-51}$ erg are not excluded as possible TeV emitters, even though their detection is possible only for relatively low redshift.
This reduces the available volume, and hence the detection rate of similar events. 
In any case, this is relevant also for short GRBs which are less energetic than long ones, with typical isotropic energies falling within the $\sim 10^{49-52}$\,erg range \cite{berger_review_short_grb}.  

\subsection{X-ray lightcurves}
The comparison between TeV and X-ray light-curves suggests an intimate connection between the emission in these two bands, both in terms of emitted energy and luminosity decay rate. 
In Figure~\ref{fig:grb_luminosity} the XRT afterglow light-curves (luminosity versus rest-frame time) in the $0.3-10$ keV energy range are compared with the VHE light-curves (integrated over different energy ranges, depending on the detection window, see Table~\ref{tab:summary}). Different colors refer to the six different GRBs. The VHE luminosity is shown with empty circles.

Considering the X-ray luminosity, the GRB sample can be divided into two groups: GRB\,190114C, GRB\,180720B and GRB\,201216C display large and clustered X-ray luminosity (at $t \sim 10^4$\,s their luminosity is around 1-5 $\sim 10^{47}$ erg s$^{-1}$) and their light curves almost overlap for the entire afterglow phase. The other three GRBs (GRB\,190829A, GRB\,201015A and GRB\,160821B) are much fainter in terms of X-ray luminosity (at least two order of magnitude at $t \sim 10^4$ s). This is consistent with the fact that they also have a smaller $E_{\gamma,iso}$. The correlation between X-ray afterglow luminosity and prompt $E_{\gamma,iso}$ is found in the bulk of the long GRB population and these GRBs make no exception.

Observations in the VHE band (empty circles in Figure~\ref{fig:grb_luminosity}) reveal that the VHE luminosities observed in the afterglow phase are in general smaller but comparable to the simultaneous X-ray luminosity, implying that almost an equal amount of energy is emitted in the two energy bands.
Any theory aimed at explaining the origin of the TeV radiation should explain the origin of these similarity. 
Concerning the decay rate, observations are still not conclusive. The decay rate of the TeV emission is available only for two events. For GRB~190829A the temporal indices in X-ray and VHE are very similar, while for GRB~190114C the VHE clearly decays faster than the X-ray emission.

For GRB\,190114C at $t \sim 380$ s the VHE luminosity $L_{VHE}$ is $\sim 1.5 - 2.5 \times 10^{48}$ erg s$^{-1}$ and the X-ray one $L_X$ is $\sim 0.6 - 1.0 \times 10^{49}$ erg s$^{-1}$. As a result, the power radiated in the VHE band is about $\sim 25\%$ of the X-ray one. Similarly for GRB\,190829A at $t \sim 4.5$ hrs the VHE luminosity $L_{VHE}$ is $\sim 4.0 - 8.5 \times 10^{44}$ erg s$^{-1}$ which is around $\sim 15-20\%$ of the corresponding X-ray one ($L_X$ $\sim 2.0 - 5.0 \times 10^{45}$ erg s$^{-1}$). For GRB\,180720B at $t \sim 2\times 10^{4}$ s the VHE luminosity $L_{VHE}$ is $\sim 9 \times 10^{47}$ erg s$^{-1}$ and the X-ray one $L_X$ is $\sim 1.5 - 2.5 \times 10^{49}$ erg s$^{-1}$. In this case the power radiated in the VHE band is around $\sim 35-60\%$ of the X-ray one.

%---------------------------------------------------
\label{subsec:summary}
\begin{figure}[!ht]
\begin{center}
    \includegraphics[width=0.8\textwidth,trim={2cm 0 1cm 2cm},clip]{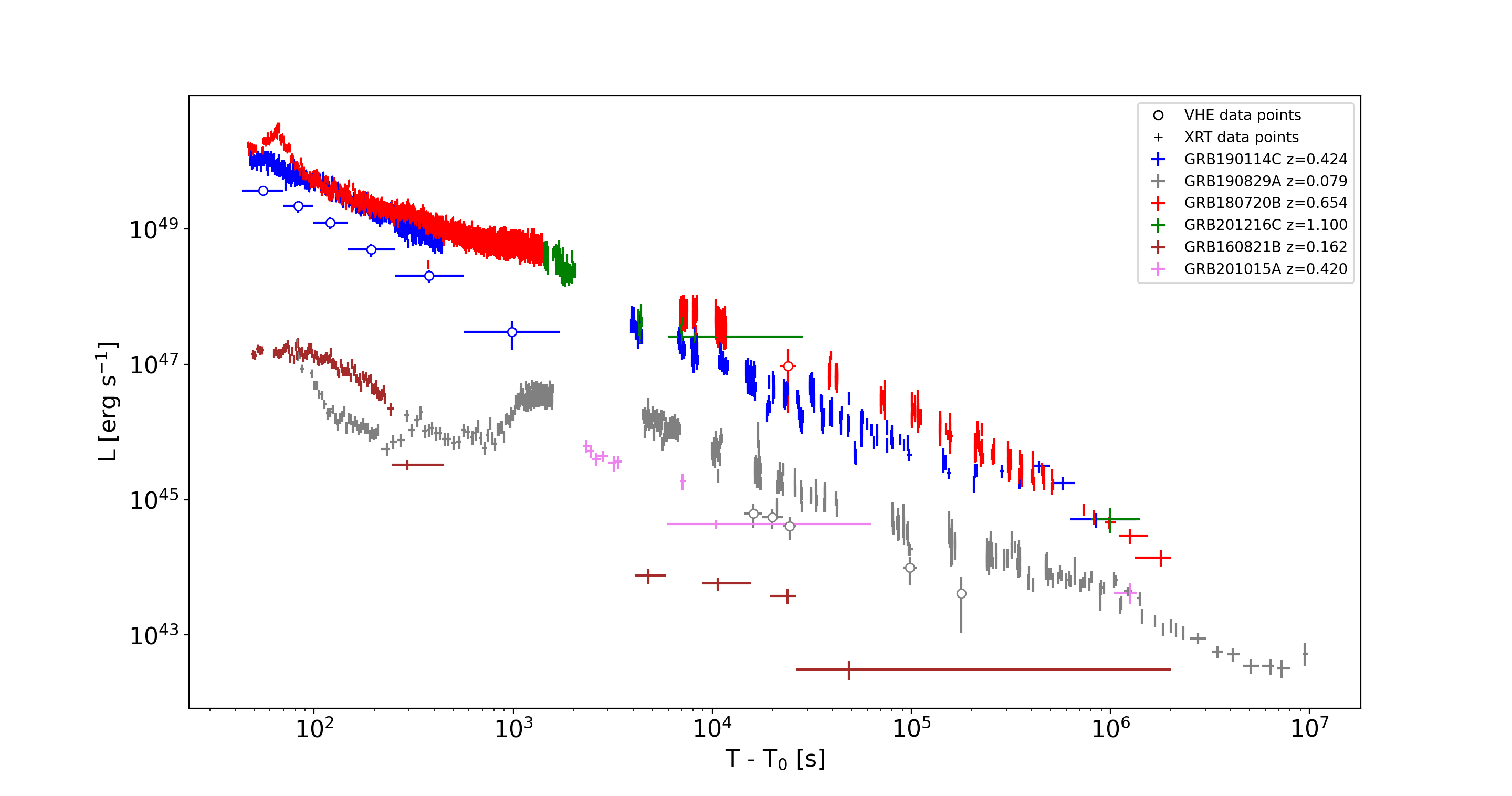}
    \caption{X-ray and VHE luminosity versus rest-frame time for the six GRBs detected in the TeV domain. The X-ray light-curve in the 0.3-10 keV energy range is taken from the Swift Burst Analyzer webpage \protect\endnotemark.}
    \label{fig:grb_luminosity}
\end{center}
\end{figure}
%---------------------------------------------------
\endnotetext{https://www.swift.ac.uk/burst\_analyser/}

\subsection{The TeV contribution to the multi-wavelength modeling}

Modeling of multi-wavelength afterglow data provide important insights concerning the GRB afterglow physics. In particular, the VHE data were crucial to investigate (i) the radiation mechanisms responsible for the production of photons between 10-100\,GeV already detected by LAT; (ii) the environmental conditions at the GRB site; (iii) the free parameters which describe the shock micro-physics, and in particular the self-generated magnetic field.

The modelings proposed so far in literature to explain the VHE component have considered two different radiation processes at the origin of the TeV emission: SSC and synchrotron. In the first case, the VHE emission is interpreted as a distinct spectral component from the synchrotron radiation dominating from radio to $\sim$\,GeV energies, which provides the seed photons that are upscattered at higher energies by the same electron population.
In the second scenario, the VHE emission is seen as the extension of the synchrotron spectrum up to TeV energies. 

In principle, a simultaneous SED covering the X-ray, HE and VHE range should be sufficient to discriminate among these two different possibilities. An hardening of the spectrum from GeV to TeV energies should be the smoking gun for the presence of a distinct component. In reality, the uncertainties in the spectral slope at VHE (caused by the uncertainty on the EBL and on the narrow energy range of TeV detection) can make the distinction hard to perform. In this case, LAT observations are of paramount importance to reveal the presence of a dip in the SED, which would also prove the need to invoke a different origin for the VHE emission.
This seems to be the case for GRB~190114C, for which at least in one SED the LAT flux strongly suggests a dip in the GeV flux and hence the presence of the characteristic double bump observed for a synchrotron-SSC emission (Figure~\ref{fig:sed_magic_all}).
For GRB~190829A, LAT provides only an upper limit which is not constraining for modeling the shape of the SED (Figure~\ref{fig:hess_modeling_190829A}). In this GRB an interpretation of the whole SED in terms of synchrotron radiation can not be excluded, although a modeling as SSC radiation has been proved to be successful \cite{190829A_salafia} (Figure~\ref{fig:om_modeling_190829A}). 
For the other events detected at VHE, either the data do not allow to build a proper SED with simultaneous multi-wavelength observations, or they are not yet public.
Despite this, the SSC emission seems to be the most viable mechanism able to explain the TeV data. 
A firm conclusion on the responsible radiation mechanism has not been reached yet and future detections will be crucial for deeper investigations.

Assuming one of the two scenarios, TeV data coupled with broad band observations at lower energies can be exploited to give additional information on the details of the afterglow external forward shock scenario. 

Concerning the shock micro-physics, several modeling have suggested the possibility that the fraction of electrons accelerated in non-thermal distribution, $\xi_e$, is different from the standard value of 1 which is usually assumed. In a few GRBs modeling, namely for GRB\,190114C \cite{asano_190114C} and GRB\,190829A \cite{190829A_salafia,190829A_twocomponentjet} the introduction of a $\xi_e < 1$ was essential in order to fit consistently the observational data. In particular in \cite{190829A_salafia} the requirement for a low value of $\xi_e \lesssim 6.5\times10^{-2}$ was required to provide acceptable fit of the data. The other modeling assume a greater value of $\xi_e$, around $\sim 0.3$. Further detections will be exploited in order to verify if such indication could be present also in other events.

Some considerations can also be drawn for the equipartition parameters $\epsilon_e$ and $\epsilon_B$. These values, especially the former one, are usually well unconstrained and can span several order of magnitudes. The TeV modeling described so far suggest that around $ \sim 10\%$ of the energy is given to the electrons, while a lower value (from $10^{-5}$ to $10^{-3}$) is given to the magnetic field. Larger values of $\epsilon_B$ such as 0.1-0.01, which are considered in external shock scenario, are excluded. Moreover, some results also can be interpreted as an indication of an evolution in time of these parameters. In Figure~\ref{fig:grb_lc_190114c} the modeling of the broad band light curves of GRB\,190114C is shown. Two different modeling are presented: one optimized for the early time X-ray, HE and VHE observations (solid line) and one optimized for the late time lower energy bands (dotted line). This is due to the fact that the model which reproduce the early time data over-predict the late time optical and radio observations. This result point towards the possibility that some of the fixed parameters of the afterglow theory (e.g. the electron and magnetic field equipartition parameters) may evolve in time. A further clue of the presence of time-dependent shock micro-physics parameters is derived from the low frequency multi wavelength modeling of GRB\,190114C presented in \cite{misra_time_evolution_parameters}. In order to model the optical and the radio data it is required that the micro-physical parameters evolve with time as $\epsilon_e \propto t^{-0.4}$ and $\epsilon_B \propto t^{0.1}$ in the ISM case and $\epsilon_B \propto t^{0.76}$ for the stellar wind scenario.

An issue that still is not solved by TeV observations is the discrimination between constant and wind-like profile for the ambient density. It is expected that long GRBs occur in wind-like environments. Nevertheless, at the current stage there seems to be no preference between such environment and a constant ISM one, which is able to well reproduce the observational data. Therefore conclusive answers on the topic cannot be drawn yet.

In conclusion, the current population of GRBs at VHE already show quite broad properties, spanning more than three order of magnitude in $E_{\gamma,iso}$ and more than two order of magnitude in terms of afterglow luminosity and ranging in redshift between 0.079-1.1. The afterglow X-ray and VHE emission have comparable fluxes and decay slopes. The afterglow emitted power in the VHE band seems to constitute from $15\%$ up to $60\%$ of the X-ray one. Data modeling suggest that the responsible VHE radiation mechanism is the SSC emission although different mechanisms (e.g. synchrotron radiation, EIC) cannot be completely excluded and a conclusive answer cannot be given yet. Multi-wavelength modeling show no preferences concerning the GRB environments between ISM or wind-like scenario and indicate that shock micro-physics parameters which seems to be able to reproduce VHE emission are $\varepsilon_e \sim 0.1$ and $\varepsilon_B \sim 10^{-5} - 10^{-3}$. Such features can be an indication of the universality of TeV emission in GRBs. It is then expected that a larger sample of GRBs than the current one will be detected in the VHE band, including also short GRBs for which, at the current stage, there are no confirmed detection except for the hint of excesses seen for GRB\,160821B.

\section{Conclusions and future prospects}\label{chap:conclusions}

The recent discoveries performed by current generation of Cherenkov telescopes in the VHE band have opened a new observational spectral window on GRBs. The presence of a TeV afterglow component has been unequivocally proven and the studies on the currently available sample have shown the potential that such detections have in probing several long-standing open questions in the GRB field. 
These first studies have focused on the identification of the responsible radiation mechanism, which is the first issue to address, and the comparison of the energetics, luminosity, and temporal behaviour of the VHE component with respect to emission at lower frequencies. 
Modeling of multi-wavelength data covering from radio up to TeV energies were performed giving interesting insights on the shock micro-physics conditions. 

Limitations to the robust use of VHE data for afterglow modeling are imposed by the severe modification of the intrinsic spectrum cased by the energy-dependent flux-attenuation induced by EBL.
GRBs with redshift $z > 0.4$, four out of six in the current VHE sample, are strongly affected by EBL absorption starting from hundreds of GeV. 
This implies large uncertainties on the shape and photon index of the intrinsic VHE spectrum. As a result, firm conclusions on the origin and spectral regime of the TeV component cannot be drawn yet. The low-energy extension of the range of sensitivity of IACTs is then fundamental for reaching a larger rate of detections and a more robust determination of the spectral index of the TeV component.

%-----------------------------------
\begin{figure}[!ht]
\begin{center}
    \includegraphics[width=0.6\textwidth]{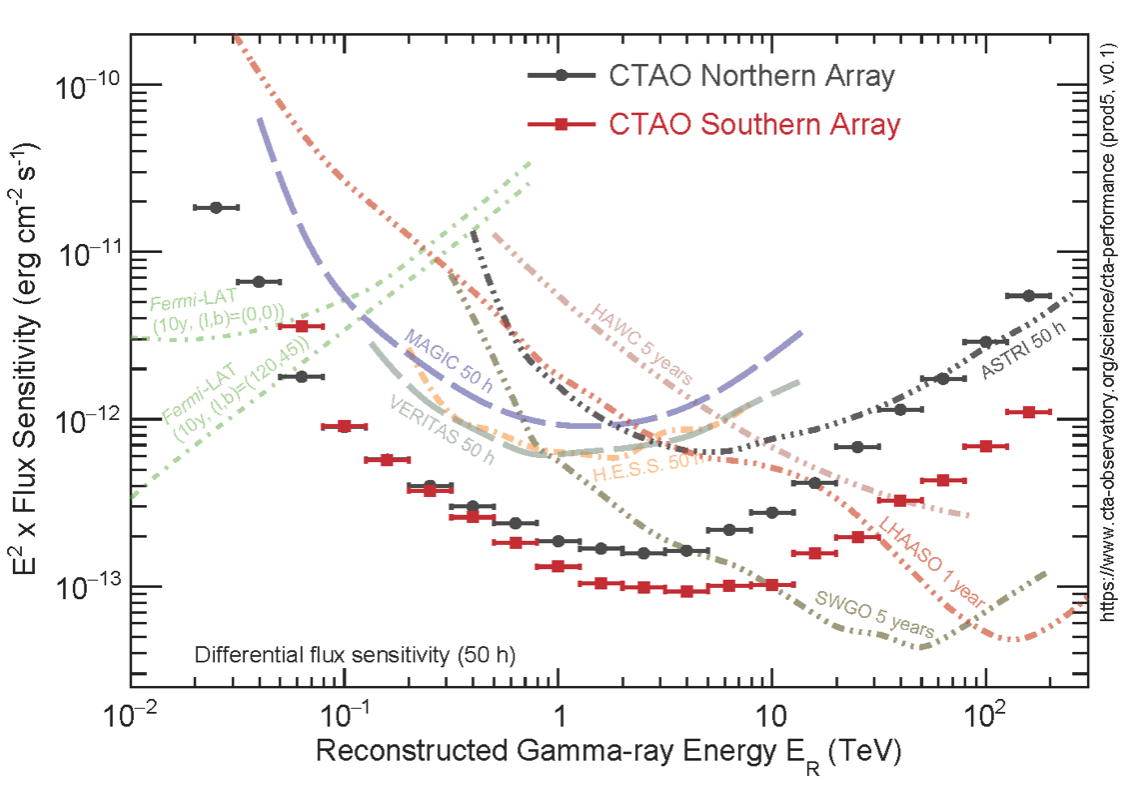}
    \caption{CTAO differential sensitivity\protect\endnotemark (defined as the minimum flux needed to obtain a 5-standard-deviation detection of a point-like source) for 50\,h of observations with the Northern and Southern array compared to the sensitivity of several other Cherenkov telescopes and with Fermi-LAT (1 year).}
    \label{fig:CTA_sensitivity}
\end{center}
\end{figure}
%-----------------------------------
\endnotetext{https://www.cta-observatory.org/science/ctao-performance/}
A full comprehension and exploitation of TeV data is expected to be reached thanks to the next generation of Cherenkov telescopes. The Cherenkov Telescope Array (CTA) will be a huge step forward for the detection of GRBs in the VHE band. The major upgrades with respect to the current generation of Cherenkov telescopes that will impact GRB observations are: i) a lower energy threshold ($\lesssim 30$ GeV), ii) a larger effective area at multi-GeV energies ($\sim 10^{4}$ times larger than Fermi-LAT at 30\,GeV) and iii) a rapid slewing capability (180 degrees azimuthal rotation in 20\,seconds). Moreover, its planned mixed-size array of large, medium and small size telescopes (called LST, MST and SST, respectively) situated at two sites in the northern and southern hemispheres will provide a full sky coverage from few tens of GeV up to hundreds of TeV. CTA will have a much better sensitivity and a broader energy range with respect to current ground-based facilities. A comparison is shown in Figure~\ref{fig:CTA_sensitivity}.
At the present stage, the first prototype of the LSTs has been built and is operative under commissioning phase\endnote{\url{https://www.lst1.iac.es/}} in the northern site at the Roque de los Muchachos Observatory in La Palma. 
Despite these performance improvements, the expected CTA detection rates of GRBs will be anyway influenced by the relatively low duty cycle affecting IACTs and by the synergies with other instruments. Indeed, Cherenkov telescopes repointing relies on external triggers coming from space satellites. 
Assuming that currently operating space telescopes will be still operative,
GRB alerts will be mostly provided by Swift-BAT and partially by the Fermi-GBM, and in the future by the French-Chinese mission Space-based multi-band astronomical Variable Objects Monitor (SVOM \cite{svom}). 

BAT observes around $92$ GRBs per year with a typical localization error of a few arcmin \cite{lien16}. The good localisation (later refined by XRT to a few arcsec) is fundamental for Cherenkov telescopes, given their limited field of view (e.g., about $4^{\circ}$ for the LSTs and $7^{\circ}$ for the MSTs). The GBM provides a much higher number of alerts, around 250 per year but with a larger localization error, from $1-3^{\circ}$ up to $10^{\circ}$, which makes follow-up with IACTs very challenging.
In case of such a large localization errors, CTA can exploit the so-called divergent mode for observations, which is currently under study \cite{divergent}. 
In this pointing strategy, each telescope points to a position in the sky that is slightly offset to extend the field of view. Concerning future instruments, SVOM is expected to provide Swift-like alerts at a rate of $\sim 60-80$ GRBs/yr with a localization error $< 1^{\circ}$, including also 10 GRBs/yr with redshift $z < 1$. 

Available estimates of CTA detection rate of GRBs are reported in \cite{susumu_cta}. These studies were performed before the discovery of TeV emission.
They are based on Swift-like alerts (triggered by Swift-BAT or SVOM) and Fermi-GBM alerts. The predicted detection rate is around a few GRBs per year, depending on the energy threshold of the observation and on the observation delay \cite{susumu_cta}. An updated study that considers current knowledge of TeV emission in the afterglow of GRBs is in progress \cite{cta_icrc}.

Despite for decades GRBs' hunting by Cherenkov telescopes has been primarily focused on reaching low energy thresholds in order to explore the multi-GeV band, these first detections have shown that photons above TeV energies can be produced in GRBs and can be detected. This is mostly valid only for nearby events of redshift below $0.1-0.2$ where EBL attenuation is not too severe. The exploration of the GRB emission component above 1 TeV can be of potential interest for SSTs and for the ASTRI Mini-Array. 
The ASTRI Mini-Array, currently under construction, will be an array of nine imaging atmospheric dual-mirror Cherenkov telescopes at the Teide Observatory site, expected to deliver the first scientific results in 2023. 
After the detection of GRB\,190114C, the capabilities of the ASTRI Mini-Array in detecting and performing spectral studies of an event similar to the MAGIC GRB have been explored \cite{astri}. GRB\,190114C has been taken as a template to simulate possible GRB emission from 
few seconds to hours, and has been extrapolated to 10\,TeV on the bases of model predictions. 
%-----------------------------------
\begin{figure}[!ht]
\begin{center}
    \includegraphics[width=0.6\textwidth]{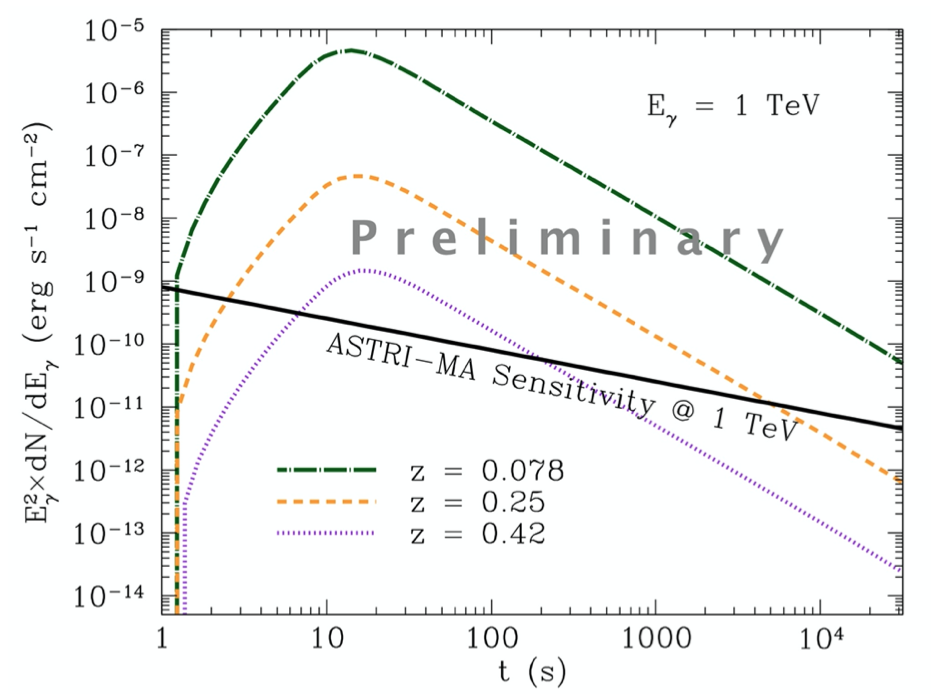}
    \caption{Light-curve of GRB\,190114C at 1\,TeV (dotted purple curve) compared to the sensitivity of the ASTRI Mini-Array. The yellow dashed and green dot-dashed curves show how it would appear GRB\,190114C rescaled at redshift $z=0.25$ and $z=0.078$, respectively. From \cite{astri}.}
    \label{fig:astri}
\end{center}
\end{figure}
%-----------------------------------
The results show that the instrument will be able to detect afterglow TeV emission from an event like GRB 190114C up to $\sim200$\,s (see the comparison between the GRB observed flux at 1\,TeV and the differential ASTRI Mini-Array sensitivity in Figure~\ref{fig:astri}). By moving the GRB at smaller redshift (down to $z=0.078$, the redshift of the TeV GRB\,190829A), the time for which the GRB is detectable increases up to $\sim10^5$\, (however, the light-curve in this case should be re-scaled by the lower energetics of nearby events).
Nearby GRBs are then potential target of interests for the ASTRI Mini-Array.
These are certainly rare events, but their detection will provide a wealth of information, with spectra that can be characterised up to several TeV \cite{astri}.

In conclusion, after decades of huge efforts, current ground-based VHE facilities have started a new era in the comprehension and study of GRB physics. Their breakthrough detections allow unprecedented studies. As discussed in this review, many open questions in afterglow physics can largely benefit from the inclusion of TeV data. The first detections are providing glimpses of such a huge potential.
Luckily, we are at the dawn of the VHE era thanks to the upcoming CTA observatory, which will assure major upgrades in sensitivity, energy range, temporal resolution, sky coverage.
Future observations, if complemented by simultaneous observations in X-rays and at $\sim$\,GeV energies, will play a paramount role to improve our knowledge on the physics of GRB during the afterglow phase and hopefully also in the prompt phase. In particular, the afterglow SSC one-zone model will be tested to understand if it can grasp the main properties of the VHE emission or if a revision of our comprehension on the particle acceleration processes, shock micro-physics and radiation mechanisms is needed.

%%%%%%%%%%%%%%%%%%%%%%%%%%%%%%%%%%%%%%%%%%

\authorcontributions{All authors, D.M. and L.N., have contributed to~writing. All authors have read and agreed to the published version of the manuscript.}
\funding{{This research received no external funding}}
\dataavailability{
This work made use of data supplied by the UK Swift Science Data Centre at the University of Leicester (\url{https://www.swift.ac.uk/xrt\_curves/})}
\conflictsofinterest{The authors declare no conflict of~interest.} 
%%%%%%%%%%%%%%%%%%%%%%%%%%%%%%%%%%%%%%%%%%%

%=====================================

\begin{adjustwidth}{0cm}{0cm}
\printendnotes
\end{adjustwidth}}}}
\end{paracol}

\reftitle{References}
\externalbibliography{yes}
\bibliography{bibliography.bib}

\end{document}